\documentclass[english]{article}
\usepackage[T1]{fontenc}
\usepackage[latin9]{inputenc}
\usepackage[letterpaper]{geometry}
\usepackage{graphicx, hyperref}
\usepackage{amssymb,amsmath}
\usepackage{esint}
\usepackage{epstopdf}

\numberwithin{equation}{section}

\newcommand{\be}{\begin{equation}}
\newcommand{\ee}{\end{equation}}
\newcommand{\ba}{\begin{eqnarray}}
\newcommand{\ea}{\end{eqnarray}}

\newcommand{\nn}{\nonumber}

\def\a {\alpha}

\makeatletter

\providecommand{\tabularnewline}{\\}

\renewcommand\[{\begin{equation}}
\renewcommand\]{\end{equation}} 

\makeatother

\usepackage{babel}

\begin{document}

\begin{titlepage}

\title{TBA, NLO L\"uscher correction, and double wrapping\\
in twisted AdS/CFT}
\maketitle


\author{Changrim Ahn \footnote{
Department of Physics and Institute for the
Early Universe, Ewha Womans University, DaeHyun
11-1, Seoul 120-750, S. Korea; ahn@ewha.ac.kr},
Zoltan Bajnok \footnote{
Theoretical Physics Research Group,
Hungarian Academy of Sciences, 1117 Budapest, P\'azm\'any s. 1/A
Hungary; bajnok@elte.hu}, 
Diego Bombardelli \footnote{Centro de F\'isica do Porto and Departamento de F\'isica e Astronomia
Faculdade de Ci\^encias da Universidade do Porto, Rua do Campo Alegre 687, 4169-007 Porto, 
Portugal; diego.bombardelli@fc.up.pt} and 
Rafael I. Nepomechie \footnote{
Physics Department, P.O. Box 248046,
University of Miami, Coral Gables, FL 33124, USA;
nepomechie@physics.miami.edu}}

\vspace{-8cm}

\hspace{11cm}UMTG-271

\vspace{8cm}

\begin{abstract}
The ground-state energy of integrably-twisted theories is analyzed in
finite volume.  We derive the leading and next-to-leading order (NLO)
L\"uscher-type corrections for large volumes of the vacuum energy for integrable
theories with twisted boundary conditions and twisted S-matrix.  We
then derive the twisted thermodynamic Bethe ansatz (TBA) equations to
describe exactly the ground state, from which we obtain an untwisted
Y-system.  The two approaches are compared by expanding 
the TBA equations to NLO, and exact agreement is found.  We give
explicit results for the $O(4)$ model and for the three-parameter
family of $\gamma$-deformed (non-supersymmetric)
planar AdS/CFT model, where the ground-state energy can be nontrivial and
can acquire finite-size corrections. The NLO corrections, which correspond to 
double-wrapping diagrams, are explicitly evaluated for the latter 
model at six loops. 
\end{abstract}

\end{titlepage}

\setcounter{footnote}{0}

\section{Introduction}

The AdS/CFT correspondence in the planar limit can be described by a
two-dimensional integrable quantum field theory.  The
finite-volume energy levels of this integrable theory correspond on one side to the
string energies in the curved $AdS_{5}\times S^{5}$ background, while
to the anomalous dimensions of gauge-invariant single-trace operators
on the other side.  Integrability provides tools to solve the
finite-volume spectral problem exactly.  (For recent reviews with
many references, see \cite{Tseytlin:2009zz, Beisert:2010jr}.)

For large volume, $L$, (long operators of size $L$), the asymptotic
Bethe ansatz \cite{Staudacher:2010jz, Ahn:2010ka} determines the
spectrum including all polynomial corrections in $L^{-1}$.  In the
weak-coupling limit, this result is exact up to $L$ loops; but over
$L$ loops, wrapping diagrams start to contribute \cite{Sieg:2010jt}.
In the integrable quantum field theory, they show up as exponentially 
small vacuum polarization
effects: virtual particles circling around the space-time modifies the
energy levels \cite{Janik:2010kd}.  These effects have a systematic
expansion which counts how many times virtual particles encircle the
space-time cylinder (or diagrams wrap around).  The leading-order (LO)
L\"uscher correction corresponds to a single circle or wrapping.
Together with the asymptotic Bethe ansatz, they provide an exact
result up to $2L$ loops.  The next-to-leading (NLO) L\"uscher
correction corresponds to two circles and double wrapping.  Including
their contribution describes the energy levels/anomalous dimensions
exactly up to $3L$ loops.

For an exact description, valid for any number of loops, one has to
sum up all virtual processes.  For the ground state, this is done by
the thermodynamic Bethe ansatz (TBA), which evaluates the saddle point
of the partition function for large Euclidean times in the mirror 
(space-time rotated) description 
\cite{Ambjorn:2005wa,Arutyunov:2007tc, Bajnok:2010ke, Gromov:2009tv, Bombardelli:2009ns, 
Gromov:2009bc, Arutyunov:2009ur, Arutyunov:2009ux}.  
The TBA provides coupled integral
equations for infinitely-many unknown functions, whose solutions
determine the exact ground-state energy and 
satisfy the so-called Y-system relations, which is 
characteristic for the model
 and are the same for all the excited states
\cite{Gromov:2010kf}.  What is different for the excited states is the
analytical structure of these Y-functions
\cite{Arutyunov:2009ax,Arutyunov:2010gb,Arutyunov:2011uz}.
Using additional inputs, such as discontinuity relations
\cite{Cavaglia:2010nm, Cavaglia:2011kd} and analytical structure, the
Y-system can be turned into integral equations for excited states
\cite{Balog:2011nm, Balog:2011cx}, which provide the solution of the
finite-volume spectral problem.  An ultimate solution would be to
replace the infinite Y-system with a finite T-Q system (see attempts
\cite{Hegedus:2009ky,Gromov:2010km, Gromov:2010vb,Suzuki:2011dj,Balog:2011cx} 
in this direction), 
which would lead to nonlinear integral equations (NLIE) for only 
finitely-many unknowns.

In the present paper, we would like to analyze the ground state of the
three-parameter family of $\gamma$-deformed planar AdS/CFT theories
\cite{Lunin:2005jy, Frolov:2005dj, Beisert:2005if, Frolov:2005iq, Zoubos:2010kh}, 
for which we refer as $\gamma$-deformed theory from now on.
Contrary to the undeformed or $\beta$-deformed theories, in the most
general case, no supersymmetry is preserved, so the ground state
is indeed nontrivial and affected by wrapping corrections. The
planar gauge theory is nevertheless ultraviolet finite and scale-invariant
\cite{Ananth:2007px}.  This is an ideal laboratory to test ideas
directly on the ground state, which actually contains all information
about the theory.

The $\gamma$-deformation can be implemented in several distinct ways:
in \cite{Arutyunov:2010gu} it was described as an operatorial twisted
boundary condition (the twist depends on the particle number); in
\cite{Ahn:2010yv, Ahn:2010ws} as a (c-number) twisted boundary condition and a
twisted scattering matrix; finally in \cite{Gromov:2010dy} the authors
showed that the untwisted Y-system with twisted asymptotic
conditions is consistent with the LO L\"uscher (single wrapping)
correction as calculated on the gauge-theory side.  In this paper,
based on our previous work \cite{Ahn:2010ws}, we choose twisted
boundary condition and twisted S-matrix.

We begin by analyzing in Sec \ref{sec:finitesizegen} the effect of a
twisted boundary condition on the ground state in general.  We derive
exact expressions for the LO and NLO L\"uscher corrections valid for
any integrable theory with a twisted boundary condition.  The LO
correction contains information about the spectrum of the (mirror) theory,
while the NLO contains the logarithmic derivative of the scattering
matrix.  We show that a Drinfeld-Reshetikhin type twist
\cite{Reshetikhin:1990ep} of the scattering matrix does not affect the
ground-state energy.  We then demonstrate the effect of the twist in
the TBA equations in general.  These equations provides the exact
description of the ground state for any finite size.  By expanding the
result for large sizes, we must recover the LO and NLO L\"uscher
corrections.  This is explicitly elaborated in the examples that 
follow.

As a warm up in a simpler case, we analyze in Sec.  \ref{sec:O4} the $O(4)$
model with twisted boundary conditions.  After calculating the LO and
NLO L\"uscher corrections, we derive the so-called raw (canonical) TBA
equations, which contain the twist as chemical potentials.
Interestingly, the twist does not show up in the simplified TBA
equations except in the asymptotic behavior of the Y-functions.  As a
consequence, the Y-system is the same as the untwisted one.  We solve
the simplified TBA equations at NLO and compare with the NLO L\"uscher
correction.  We find complete agreement.

We turn in Sec.  \ref{sec:AdSCFT} to the $\gamma$-deformed AdS/CFT
model.  We calculate first the LO L\"uscher correction.  In
calculating the NLO correction, we determine the determinant of the
two-particle S-matrix $S^{Q_{1} Q_{2}}$ in all the $su(2)_{L} \otimes
su(2)_{R}$ sectors for the generic $Q_{1}$ and $Q_{2}$ bound-state
case.  We then derive the raw TBA equations from first principles by
evaluating exactly the chemical potentials originating from the
twisted boundary condition. (For the untwisted case, the TBA 
equations were formulated in \cite{Gromov:2009tv, Bombardelli:2009ns, 
Gromov:2009bc, Arutyunov:2009ur, Arutyunov:2009ux}.)
The twist disappears from the simplified
equations, just as it does in the O(4) case.  (See \cite{deLeeuw:2011rw} for a general
argument on this.)  The twist nevertheless reappears in the
asymptotic boundary conditions for the Y-functions.  Since the
simplified equations are not twisted, neither is the Y-system, as was
anticipated by the authors of \cite{Gromov:2010dy, Beccaria:2011qd}.  Our derivation
confirms their assumption.  We then expand the TBA equations to NLO
and compare with the result of the NLO L\"uscher correction.  We find
complete agreement again.

We evaluate in Sec.  \ref{sec:weakcoupling} the weak-coupling
expansion of the NLO L\"uscher correction, which corresponds to
double-wrapping diagrams. We explicitly compute this correction for 
$L=3$, thereby obtaining the anomalous dimension of the operator 
${\rm Tr} Z^{3}$ in the twisted gauge theory up to six loops.

Finally, Sec. \ref{sec:discuss} contains our conclusion and outlook.

\section{Finite-size corrections of the vacuum 
energy}\label{sec:finitesizegen}

In this section we analyze the finite-size corrections for the ground state
with a twisted boundary condition. We consider an integrable 
$(1+1)$-dimensional quantum field theory
that possesses just one multiplet of particles with the same 
dispersion relation. The particles are labeled
by $\alpha$, and their interaction is
described by the two-particle scattering matrix 
$S_{\alpha\beta}^{\delta\gamma}(p_{1},p_{2})$, which does not admit 
any bound states.\footnote{With a view to
later applying this formalism to AdS/CFT, we do not assume
relativistic invariance; hence, the two-particle S-matrix need not be
a function of the difference of the particles' momenta.}
We are interested in the ground-state energy of a system of size
$L$ with a $c$-number twisted boundary condition in terms of the scattering
data.
The twisted boundary condition is defined by means of a conserved charge
$J$, which commutes with the scattering matrix $[J,S]=0$. The twists
are implemented by introducing a so-called defect line on the circle. It
has the effect that, whenever a particle of type $\alpha$ crosses
the defect line from the left to the right, it picks up the transmission
phase $e^{i\gamma J_{\alpha}}$, where $\gamma$ is the twist angle
supposed to be real. 
If the particle moves oppositely, then it picks
up the inverse phase $e^{-i\gamma J_{\alpha}}$. This ensures that
if we formulate the  Bethe-Yang equation by moving one
particle around the circle and scattering with all the other particles
and with the defect line in both directions, then we obtain equivalent
equations. 

\begin{figure}
\begin{centering}
\includegraphics[height=4cm]{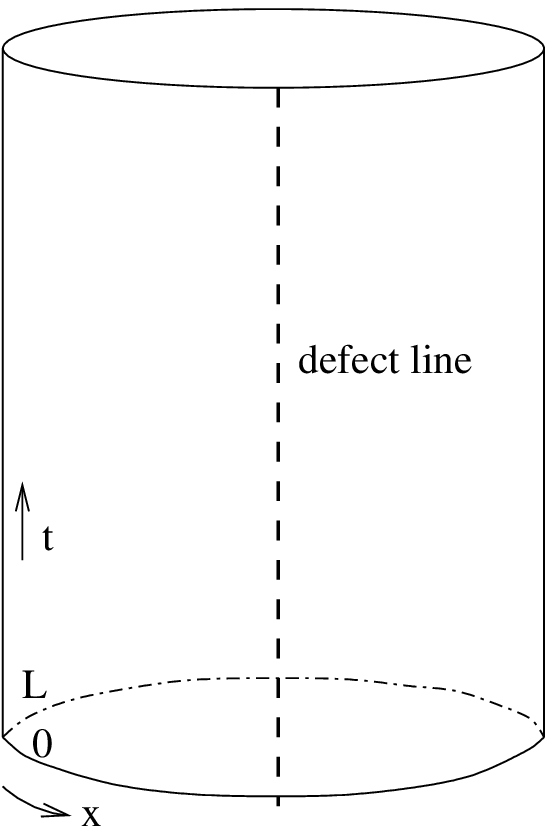}
~~~~~~~~~~~~~~~~\includegraphics[width=4cm]{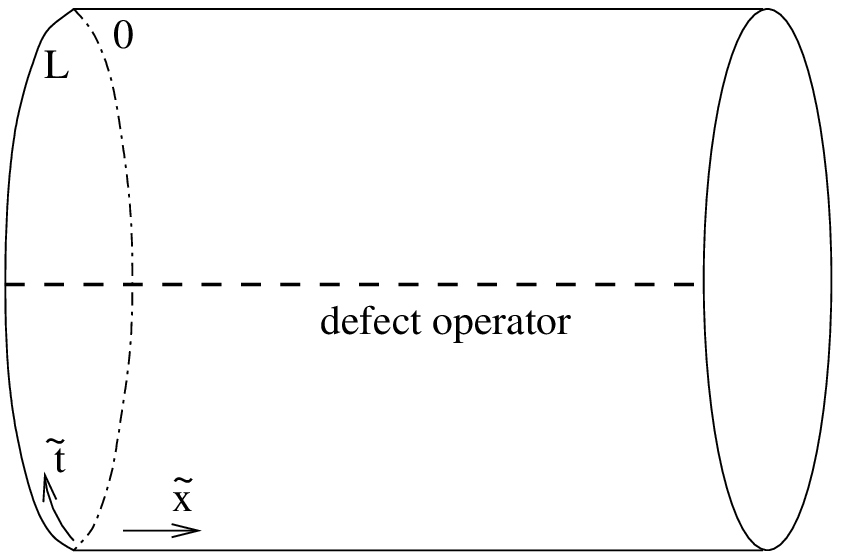}
\par\end{centering}
\caption{Two possible locations of a defect. On the left it is 
located in space,
and it introduces a twisted boundary condition. On the right it is located
in (Euclidean) time, and it acts as an operator on the periodic Hilbert
space.}
\label{Flo:dflinevsop}
\end{figure}

In deriving the finite-size energy of the vacuum with the defect
line, $E_{0}^{d}(L)$, we analyze the twisted Euclidean torus partition
function from two different perspectives, see Figure \ref{Flo:dflinevsop}.

\noindent By compactifying the time-like direction with period $R$
and taking the $R\to\infty$ limit, the ground-state energy of the
twisted system can be extracted from the twisted partition function
as 
\[
\lim_{R\to\infty}Z^{d}(L,R)=\lim_{R\to\infty}Tr\left(e^{-H^{d}(L)R}\right)
=e^{-E_{0}^{d}(L)R}+\dots \,.
\]
In the alternative description in which the role of Euclidean time, $\tilde x= -it$,
and space, $x$, are exchanged, the defect will be localized at a constant
imaginary time $\tilde t= -i x $ of the mirror model. It acts as an operator of the periodic
Hilbert space of the mirror model defined by the configurations on
a fixed-$\tilde t $ slice. The action of this operator can be calculated
from the transmission phase \cite{Bajnok:2004jd}. In the present
case, the operator is simply $e^{i\gamma J}$, and we can evaluate the
twisted partition function alternatively as
\[
Z^{d}(L,R)=\mbox{Tr}(e^{-\tilde{H}(R)L}e^{i\gamma J}) \,,
\]
where we use a tilde $\tilde{}$ to help distinguish quantities in the 
mirror model.
In the first subsection, we suppose that the volume $L$ is large and
expand the partition function at leading and next-to-leading orders.
In this way, we derive the LO and NLO L\"uscher-type corrections for the
ground state energy of the twisted system. Then, in the second 
subsection, we comment on how one can evaluate the partition function in the 
saddle-point approximation to obtain the twisted thermodynamic Bethe 
ansatz (TBA) equations.

\subsection{Large-volume expansion}

In this subsection, we evaluate the twisted partition function at LO
and NLO for large volumes (i.e., $L$ is large, and $R \rightarrow 
\infty$). This means that we keep the first two nontrivial
terms in the expansion of the twisted partition function
\begin{equation}
\lim_{R\rightarrow\infty}\mbox{Tr}(e^{-\tilde{H}(R)L}e^{i\gamma J})=1+\sum_{k,\alpha}e^{i\gamma J_{\alpha}
-\tilde{\epsilon}(\tilde{p}_{k})L}+\sideset{}{'}\sum_{k,l,(\alpha,\beta)}e^{i\gamma
J_{(\alpha,\beta)}
-(\tilde{\epsilon}(\tilde{p}_{k})+\tilde{\epsilon}(\tilde{p}_{l}))L}
+\dots \,,
\label{eq:ZlargeL}
\end{equation}
where $k,l$ are the labels of the allowed mirror momenta $\tilde{p}$;
$\alpha$ is the color index of the one-particle and $(\alpha,\beta)$
is that of the two-particle state. The sum $\sum\nolimits'$ is taken over the 
distinct two-particle states.
$J$ is the conserved charge such
that $J_{\alpha}$ denotes its eigenvalue on the one particle, while
$J_{(\alpha,\beta)}$ is its eigenvalue on the two-particle state. Finally,
$\tilde{\epsilon}(\tilde{p})$ denotes the energy of the mirror particle.
Clearly, the defect does not affect the energy levels, but 
nevertheless modifies the twisted partition function. 
Calculations based on the expansion of the partition function for large volumes can be found 
for boundary entropies in \cite{Dorey:2004xk}, while for the boundary ground state energy
in \cite{Bajnok:2004tq}.

\subsubsection{Leading-order calculation}

In evaluating the twisted partition function at LO, we analyze
the one-particle contributions. In a finite but large volume, $R$,
the momentum is quantized as 
\begin{equation}
e^{i\tilde{p}_{k}R}=1\qquad\to\qquad\frac{R}{2\pi}\tilde{p}_{k}=k\in\mathbb{Z}
\,,
\label{eq:1ptBY}
\end{equation}
which is independent of the color index $\alpha=1,\dots,N$. In the
$R\to\infty$ limit, the allowed momenta become dense, and the summation
can be turned into integration. The change from the discrete label
$k$ to the continuous momentum variable $\tilde{p}$ is dictated by
the Bethe-Yang equation above as
\begin{equation}
\sum_{k}\to R\int\frac{d\tilde{p}}{2\pi} \,.
\label{eq:1ptJacobi}
\end{equation}
Taking the logarithm of the twisted partition function, the ground-state
energy can be obtained
\begin{equation}
E_{0}^{d}(L)=-\lim_{R\to\infty}R^{-1}\log\left[\mbox{Tr}(e^{-\tilde{H}(R)L}e^{
i\gamma J})
\right] \,.
\label{eq:E_0^d}
\end{equation}
Expanding the log as $\log(1+x)=x+O\left(x^{2}\right)$ and keeping
the first term, we obtain
\[
E_{0}^{d}(L)=E_{0}^{(1)}(L) +O(e^{-2\tilde{\epsilon}(0)L}) \,, \qquad
E_{0}^{(1)}(L) = -\mbox{Tr}(e^{i\gamma
J})\,\int\frac{d\tilde{p}}{2\pi}e^{-\tilde{\epsilon}
(\tilde{p})L} \,,
\label{eq:E1gen}
\]
where the color summation gives 
$\sum_{\alpha}e^{i\gamma J_{\alpha}}=\mbox{Tr}(e^{i\gamma J})$,
which is basically the character of the particles' representation.
The physical meaning of this formula is clear: The finite-volume vacuum
contains virtual particles, and they modify the vacuum energy by virtual
processes. The leading volume-dependent process is when a 
particle and anti-particle pair appears from the vacuum, and then the particle
travels around the world and annihilates with the anti-particle on the
other side. Clearly, in so doing, it crosses the defect line and picks
up the phase which, when summed up for the multiplet, results in the
character.

\subsubsection{Next-to-leading order calculation}

At the NLO energy correction, we have to expand the logarithm of the
partition function (\ref{eq:E_0^d}) to second order: 
$\log(1+x)=x-\frac{x^{2}}{2}+O\left(x^{3}\right)$.
This will include the square of the one-particle term and the
two-particle term. The former, however, contains a factor $R^{2}$
which would lead to a divergence in the $R\to\infty$ limit, and has
to be canceled against a similar part of the two-particle term. We
evaluate now the two-particle contribution and see the needed cancellation.
From the remaining terms, we obtain the NLO energy correction. 

In calculating the two-particle term, we must first determine the allowed
momenta. In very large volume $R$, the momentum quantization
conditions are given by the Bethe-Yang (or, in other terminology, 
the asymptotic Bethe ansatz) equations. As the scattering mixes the
color indices, we begin by diagonalizing the two-particle S-matrix:
\[
e^{iR\tilde{p}_{k}}S_{\mu}^{\nu}(\tilde{p}_{k},\tilde{p}_{l})\psi_{\nu}=
\psi_{\mu}\to 
e^{iR\tilde{p}_{k}}e^{i\delta_{\mu}(\tilde{p}_{k},\tilde{p}_{l})}=1 \,. 
\label{eq:2ptBY}
\]
The two-particle S-matrix has $N^{2}$ eigenvalues, and we denote
their phases by $\delta_{\mu}(\tilde{p}_{k},\tilde{p}_{l})$ for
$\mu=1,\dots,N^{2}$.
Unitarity implies $\delta_{\mu}(\tilde{p}_{k},\tilde{p}_{l})=
-\delta_{\mu}(\tilde{p}_{l},\tilde{p}_{k})\ {\rm mod}\ 2\pi$.
We assume that the particles are fermionic: 
$S(\tilde{p},\tilde{p})=-\mathbb{I}$,
thus $\delta_{\mu}(\tilde{p},\tilde{p})=\pi$. Taking the logarithm
of the equations (\ref{eq:2ptBY}) for a given eigenvalue, we 
arrive at the Bethe-Yang equations
\begin{eqnarray}
\frac{R}{2\pi}\tilde{p}_{k}+\frac{1}{2\pi}\delta_{\mu}(\tilde{p}_{k},\tilde{p}_{
l})
 & = & k\nonumber \,, \\
\frac{R}{2\pi}\tilde{p}_{l}-\frac{1}{2\pi}\delta_{\mu}(\tilde{p}_{k},\tilde{p}_{
l}) 
& = & l \,.
\end{eqnarray}
The fermionic nature of the particles excludes $k=l$; and in summing
up over two-particle states, $k>l$ is understood. In changing to momentum
integration, it is better to reorganize the sum as 
$\sum_{k>l}f(k,l)=\frac{1}{2}\sum_{k,l}f(k,l)-\frac{1}{2}\sum_{k}f(k,k)$,
since the summand 
$f(k,l)=e^{i\gamma
J-(\tilde{\epsilon}(\tilde{p}_{k})+\tilde{\epsilon}(\tilde{p}_{l}))L}$
is symmetric. The diagonal part, $-\frac{1}{2}\sum_{k}f(k,k)$, has
the one-particle quantization rule (\ref{eq:1ptBY}); thus, changing
to integration as in (\ref{eq:1ptJacobi}) the contribution to the energy
turns out to be: 
\[
E_{0}^{(2,1)}(L)=\frac{1}{2}\mbox{Tr}(e^{i\gamma
J})^{2}\int\frac{d\tilde{p}}{2\pi}
e^{-2\tilde{\epsilon}(\tilde{p})L} \,,
\label{eq:E21gen}
\]
where we used that $\sum_{(\alpha,\beta)}e^{i\gamma J_{(\alpha,\beta)}}
=\sum_{\mu}e^{i\gamma J_{\mu}}=\mbox{Tr}(e^{i\gamma J})^{2}$. 

We now transform $\frac{1}{2}\sum_{k,l}f(k,l)$ into a double integral.
To this end, we compute the Jacobian for the change of variables
$(k,l)\to(\tilde{p}_{k},\tilde{p}_{l})$:
\[
\left| \begin{array}{cc}
\frac{\partial k}{\partial \tilde{p}_{k}} & \frac{\partial 
k}{\partial \tilde{p}_{l}} \\\\
\frac{\partial l}{\partial \tilde{p}_{k}} & \frac{\partial 
l}{\partial \tilde{p}_{l}}
\end{array}\right|=
\frac{1}{(2\pi)^{2}}\left|\begin{array}{cc}
R+\delta_{\mu, k} & \delta_{\mu, l}\\
-\delta_{\mu, k} & R-\delta_{\mu, l}
\end{array}\right|=\frac{1}{(2\pi)^{2}}\left[R^{2}+R(\delta_{\mu,k} 
-\delta_{\mu,l}) \right] \,,
\label{eq:Jacobian}
\]
where $\delta_{\mu, k}=\partial_{\tilde{p}_{k}}\delta_{\mu}
(\tilde{p}_{k},\tilde{p}_{l})$ and $\delta_{\mu, 
l}=\partial_{\tilde{p}_{l}}\delta_{\mu}
(\tilde{p}_{k},\tilde{p}_{l})$. As already mentioned,
the terms which contribute to the ground-state energy have to be proportional
to $R$. Indeed the dangerous $R^{2}$ term \[
\frac{R^{2}}{2}\mbox{Tr}(e^{i\gamma J})^{2}\int\frac{d\tilde{p}_{1}}{2\pi}
\int\frac{d\tilde{p}_{2}}{2\pi}e^{-(\tilde{\epsilon}(\tilde{p}_{1})+
\tilde{\epsilon}(\tilde{p}_{2}))L}\]
will cancel against the $-\frac{x^{2}}{2}$ term of the expansion
of the logarithm of the one-particle contribution. 
The second term of the Jacobi determinant (\ref{eq:Jacobian}) is proportional to
the volume
$R$, and contributes to the ground-state energy as
\[
E_{0}^{(2,2)}(L)=-\int\frac{d\tilde{p}_{1}}{2\pi}\, e^{-\tilde{\epsilon}
(\tilde{p}_{1})L}\int\frac{d\tilde{p}_{2}}{2\pi}e^{-\tilde{\epsilon}
(\tilde{p}_{2})L}\sum_{\mu}e^{i\gamma 
J_{\mu}}\partial_{\tilde{p}_{1}}\delta_{\mu}(\tilde{p}_{1},\tilde{p}_{2}) \,,
\label{eq:E22gen}
\]
where we have used that $\delta_{\mu}(\tilde{p}_{1},\tilde{p}_{2})$
is antisymmetric in its arguments; and that, as the twist commutes with
the scattering matrix $[e^{i\gamma J},S]=0$, both can be diagonalized
in the same basis. We note that
\[
\sum_{\mu}e^{i\gamma
J_{\mu}}\partial_{\tilde{p}_{1}}\delta_{\mu}(\tilde{p}_{1},\tilde{p}_{2})=
-i\partial_{\tilde{p}_{1}}\mbox{Tr}(e^{i\gamma 
J}\log[S(\tilde{p}_{1},\tilde{p}_{2})]) \,.
\label{eq:eigtomat}
\]
In particular, this implies that if the S-matrix is twisted (\`a la 
Drinfeld-Reshetikhin \cite{Reshetikhin:1990ep})
with another
conserved charge $\tilde{S}=FSF$, such that $[e^{i\gamma J},F]=0$,
then the finite-size correction is the same as in the undeformed case:
\begin{eqnarray}
\partial_{\tilde{p}_{1}}\mbox{Tr}(e^{i\gamma J}\log(\tilde{S})) & = & 
\partial_{\tilde{p}_{1}}\sum_{\alpha}e^{i\gamma J_{\alpha}} 
\mbox{Tr}_{\alpha} \log( F_{\alpha} S_{\alpha} F_{\alpha})  
= \partial_{\tilde{p}_{1}} \sum_{\alpha}e^{i\gamma J_{\alpha}} 
\log\det ( F_{\alpha} S_{\alpha} F_{\alpha}) \nonumber \\
& = &
\partial_{\tilde{p}_{1}} \sum_{\alpha}e^{i\gamma J_{\alpha}} 
\log\det S_{\alpha} =  \partial_{\tilde{p}_{1}}\mbox{Tr}(e^{i\gamma 
J}\log (S) ) \,,
\label{eq:DrinfeldReshetikhin}
\end{eqnarray}
where we have denoted by $F_{\alpha}$ ($S_{\alpha}$) the matrix $F$  
($S$) in the subspace where $J$ has eigenvalue $J_{\alpha}$, 
respectively; and we have used the fact that 
$\det ( F_{\alpha} S_{\alpha} F_{\alpha}) =  \det S_{\alpha} \det 
F_{\alpha}^{2}$.

We conclude that the LO and NLO corrections to the finite-volume 
vacuum energy in the twisted theory come only from the twisted boundary
condition, and are given by 
\begin{eqnarray}
E_{0}^{d}(L) & = & E_{0}^{(1)}(L)+E_{0}^{(2,1)}(L)+E_{0}^{(2,2)}(L)\nonumber \\
 & = & -\mbox{Tr}_{1}(e^{i\gamma J})\int\frac{d\tilde{p}}{2\pi}\, 
e^{-\tilde{\epsilon}(\tilde{p})L}+\frac{1}{2}\mbox{Tr}_{1}(e^{i\gamma J})^{2}
\int\frac{d\tilde{p}}{2\pi}e^{-2\tilde{\epsilon}(\tilde{p})L}\nonumber \\
 &  & +\int\frac{d\tilde{p}_{1}}{2\pi}\, e^{-\tilde{\epsilon}(\tilde{p}_{1})L}
\int\frac{d\tilde{p}_{2}}{2\pi}e^{-\tilde{\epsilon}(\tilde{p}_{2})L}i
\partial_{\tilde{p}_{1}}\mbox{Tr}_{2}(e^{i\gamma J}\log[S(\tilde{p}_{1},
\tilde{p}_{2})]) \,,
\label{eq:Luscher}
\end{eqnarray}
where the omitted terms are of order of 
$O(e^{-3\tilde{\epsilon}(0)L})$,
and $\mbox{Tr}_{i}()$ for $i=1, 2$ means that the trace is taken over the one-
or two-particle states, respectively. 
This derivation is an alternative formulation of the virial expansion
of the partition function in statistical physics.  (See also the
result for the $O(n)$ case \cite{Balog:2001sr}.)

Our result (\ref{eq:Luscher}) can also be used to make the connection 
between the scattering description
and other descriptions of the theory. Indeed, given an integral
equation for the ground-state energy, we can extract from it the S-matrix
by expanding for large volume to NLO.

\subsection{Twisted TBA}\label{sec:genTBA}

We have so far supposed that the physical volume $L$ is large, and we 
have calculated
the LO and NLO energy corrections. If the volume is not large and
we are interested in the exact description of the vacuum, we have to
evaluate the contributions of multiparticle states. This, in the untwisted
case, is done by the TBA; and we shall now see
how the derivations are modified in the presence of the twist. 

The first step in calculating the partition function is the determination
of the momentum quantization of multiparticle states. This is done
by solving the Bethe-Yang equations by means of the asymptotic Bethe 
ansatz (BA).
Here, in addition to the physical momentum-carrying particles, one
has to introduce so-called magnonic particles that take care of the non-diagonal
nature of the scattering. They are useful objects, since in terms of them
the scattering can be regarded as diagonal.  One then analyzes
the various ``diagonal'' scattering matrices and looks for bound states:
i.e., complex string-like solutions of the asymptotic BA equations.
The scattering matrices of the bound states are determined from the
scattering matrices of their constituents. Let us label the particles
(momentum-carrying, magnonic and their bound states) by a multilabel
$n$; and their scattering matrices by $S_{nm}(u_{1}^{n},u_{2}^{m})$,
where $u_{i}^{n_{i}}$ is some generalized rapidity of a particle
of type $n_{i}$. Greek indices such as $\alpha$ will denote magnons only.
The asymptotic BA equations for large particle numbers (thermodynamic
limit) takes the generic form
\[
-1=e^{i\tilde{p}_{n}(u_{k}^{n})R}\prod_{m}\prod_{l}S_{nm}(u_{k}^{n},u_{l}^{m})
\,,
\label{eq:BA}
\]
where the mirror momentum vanishes for magnons
$\tilde{p}_{\alpha}(u^{\alpha})=0$,
and $S_{nn}(u_{k}^{n},u_{k}^{n})=-1$. We note that not only the momentum,
but also the energy vanishes for magnons, 
$\tilde{\epsilon}_{\alpha}=0$.
Thus, the magnonic equations can be inverted, without changing their
physical meaning. We have to choose such equations which give rise
to positive particle densities in the thermodynamic limit. In this
limit, the partition function is dominated by finite-density configurations.
The density of the particles (holes) of type $n$ can be introduced
as $\rho_{n}=\frac{\Delta N_{n}}{R\Delta\tilde{p}}$, ($\bar{\rho}_{n}=
\frac{\Delta\bar{N}_{n}}{R\Delta\tilde{p}}$),
where $\Delta N_{n}$($\Delta\bar{N}_{n}$ ) denotes the number of particles (holes)
in the interval $(\tilde{p},\tilde{p}+\Delta\tilde{p}$), respectively.
In terms of these densities, the energy of the configuration is 
\[
\tilde{E}[\rho]=R\sum_{n}\int d\tilde{p}\, \rho_{n}(\tilde{p})\,
\tilde{\epsilon}_{n}(\tilde{p})=R\sum_{n}\int du\,\rho_{n}(u)\,
\tilde{\epsilon}_{n}(u) \,,
\]
while the entropy is 
\[
S[\rho,\bar{\rho}]=R\sum_{n}\int du\left[(\rho_{n}+\bar{\rho}_{n})\log(\rho_{n}
+\bar{\rho}_{n})-\rho_{n}\log\rho_{n}-\bar{\rho}_{n}\log\bar{\rho}_{n}\right]\,.
\]
The particle and the hole densities are not independent, and the derivative
of the logarithm of the asymptotic BA (\ref{eq:BA}) connects them as 
\[
\rho_{n}+\bar{\rho}_{n}-\frac{1}{2\pi}\partial_{u}\tilde{p}_{n}=
\int du'\sum_{m}K_{nm}(u,u')\rho_{m}(u')=:K_{nm}\star\rho_{m} \,,
\]
where $K_{nm}(u,u')=\frac{1}{2\pi i}\partial_{u}\log S_{nm}(u,u')$. If we had
inverted
any of the asymptotic BA equations, then we would have obtained the
sign-changed kernel here. By choosing the proper signs of the kernels 
for the magnons, we
can ensure the positivity of all the densities. If we had started 
instead with the Drinfeld-Reshetikhin-twisted S-matrix, then $S_{nm}$ in (\ref{eq:BA}) 
would be replaced by $\tilde S_{nm}$, which differs from $S_{nm}$ by 
constant phases; and these phases would disappear from the kernel $K_{nm}$. 
Consequently, the TBA equations are independent of twists of the 
S-matrix, as is the L\"uscher correction (\ref{eq:DrinfeldReshetikhin}).

We have seen that the twist does not change the energy levels of the
periodic mirror system, but nevertheless modifies the partition function. 
Since the twist commutes with the scattering matrix, the
particles of the asymptotic BA equations which diagonalize the multiparticle
scatterings will have diagonal twist eigenvalues, too. Let us denote
the eigenvalue of $i\gamma J$ on a particle with label $n$ by $\mu_{n}$.
The total contribution of the twist on the multiparticle state is
\[
\mu[\rho]=R\sum_{n}\int du\,\rho_{n}(u)\, \mu_{n} \,.
\]
In terms of these quantities, the partition function can be written as 
\[
Z^{d}(L,R)=\mbox{Tr}(e^{-\tilde{H}(R)L}e^{i\gamma J})=\int\prod_{n}
d[\rho_{n},\bar{\rho}_{n}]e^{S[\rho,\bar{\rho}]+\mu[\rho]-L\tilde{E}[\rho]} \,.
\label{eq:partitionfunction}
\]
Evaluating the integrals in the saddle-point approximation, the minimizing
condition
for the pseudo-energies $\epsilon_{n}=\log\frac{\bar{\rho}_{n}}{\rho_{n}}$ 
turns out to be
\[
\epsilon_{n}+\mu_{n}=\tilde{\epsilon}_{n}L-\log(1+e^{-\epsilon_{m}})\star K_{mn}
\,.
\label{eq:TBA}
\]
Once we have calculated the pseudo-energies, the ground-state energy can
be extracted from the saddle-point value as 
\[
E_{0}^{d}(L)=-\sum_{n}\int\frac{du}{2\pi}\partial_{u}\tilde{p}_{n}
\log(1+e^{-\epsilon_{n}}) \,.
\]
Clearly the only difference compared with the untwisted case is 
the appearance in the TBA equations (\ref{eq:TBA}) of the
chemical potential $\mu_{n}$, which is proportional to the charge
of the particle. (TBA equations with chemical potentials have been studied 
previously; see e.g. \cite{Klassen:1990dx}.)

As the determination of the magnons and their charges is model
dependent, we work out the details in the following for the $O(4)$
model, and then for twisted planar AdS/CFT.

\section{Case study: $O(4)$ model}\label{sec:O4}

In this section, as a warm-up, we elaborate explicitly the simpler 
case of the twisted $O(4)$ model, also known as the $su(2)$ principal chiral
model.  We calculate the LO and NLO L\"uscher corrections, derive the
twisted TBA equations, and compare the two approaches by expanding the
TBA equations up to second order.

The $O(4)$ model is a relativistic theory containing one multiplet
of particles with mass $m$. The dispersion relation $E(p)=\sqrt{m^{2}+p^{2}}$
can be parameterized in terms of the rapidity as 
\[
E(\theta)=m\cosh\pi\theta\,, \qquad 
p(\theta)=m\sinh\pi\theta \,.
\]
The particles transform under the bifundamental representation of
$su(2)$. The two-particle S-matrix is the simplest $su(2)\otimes su(2)$
symmetric, unitary and crossing-invariant scattering matrix 
\cite{Zamolodchikov:1977nu, Zamolodchikov:1992zr}
\[
S(\theta)=\frac{S_{0}^{2}(\theta)}{(\theta-i)^{2}}\hat{S}(\theta)
\otimes\hat{S}(\theta)\,, \qquad
\hat{S}(\theta)=\theta\,\mathbb{I}-i\,\mathbb{P} \,,
\]  
where $\theta=\theta_{1}-\theta_{2}$, and the scalar factor 
\[
S_{0}(\theta)=i\frac{\Gamma(\frac{1}{2}-\frac{i\theta}{2})\Gamma(\frac{i\theta}{
2})}
{\Gamma(\frac{1}{2}+\frac{i\theta}{2})\Gamma(-\frac{i\theta}{2})}
\]
does not have any poles in the physical strip, showing the absence of
physical bound states. 

We analyze this theory on a circle of size $L$ with a twisted boundary
condition. We twist the theory with independent twist angles 
$\gamma_{\mp}$ for the left 
and right $su(2)$ factors, respectively:
\[
e^{i\gamma J}=e^{i\gamma_{-}J_{0}\otimes\mathbb{I}+i\gamma_{+}\mathbb{I}\otimes
J_{0}}=
e^{i\gamma_{-}J_{0}}\otimes e^{i\gamma_{+}J_{0}}=\mbox{diag}(\dot {q},\dot{q}^{-1})
\otimes\mbox{diag}(q,q^{-1})\,,
\]
where $J_{0}$ has eigenvalues $\pm 1$ on the two components of the
doublet, and $\dot {q}=e^{i\gamma_{-}}$, $q=e^{i\gamma_{+}}$. We could
also twist the S-matrix, i.e. change $S\to FSF$,
but this would have no effect on the ground-state energy, as 
explained in (\ref{eq:DrinfeldReshetikhin}).

\subsection{L\"uscher corrections}

We now proceed to evaluate the L\"uscher correction for the vacuum
(\ref{eq:Luscher}).
As the theory is relativistically invariant, the mirror dispersion
relation is $\tilde{\epsilon}(\tilde{p})=\sqrt{m^{2}+\tilde{p}^{2}}$,
which we parameterize in terms of the rapidity as above:
$\tilde{p}(\theta)=m\sinh\pi\theta$.
In this parameterization, the leading-order result for the ground-state
energy is
\[
E_{0}^{(1)}(L) =-[2]_{q}[2]_{\dot{q}}m
\int\frac{d\theta}{2}\cosh\pi\theta\, e^{-mL\cosh\pi\theta} \,,
\label{O4LuscherE1}
\]
where we used that 
\[
\mbox{Tr}(e^{i\gamma J})=\mbox{Tr}(e^{i\gamma_{-}J_{0}})
\mbox{Tr}(e^{i\gamma_+J_{0}})=(\dot{q}+\dot{q}^{-1})(q+q^{-1})=[2]_{q}[2]_{
\dot{q}}\,.
\]
It is useful to introduce the $q$-numbers 
\[
[n]_{q}=\frac{q^{n}-q^{-n}}{q-q^{-1}}=q^{n-1}+q^{n-3}+\dots+q^{3-n}+q^{1-n}\,,
\]
for which $[n]_{q}\to n$ in the untwisted limit $q\to1$. 

In the second-order correction, we have the term without the scattering
matrix 
\[
E_{0}^{(2,1)}(L)=\frac{1}{2}[2]_{q}^{2}[2]_{\dot{q}}^{2}m
\int\frac{d\theta}{2}\cosh\pi\theta\, e^{-2mL\cosh\pi\theta}\,.
\label{O4LuscherE21}
\]
In the other term, we have to diagonalize the two-particle S-matrix
\[
S(\theta)=S_{0}^{2}(\theta)S_{1}(\theta)\otimes S_{2}(\theta)\,,
\qquad S_{1}(\theta)=S_{2}(\theta)= \frac{1}{\theta-i}\hat S(\theta)
=\left(\begin{array}{cccc}
1 & 0 & 0 & 0\\
0 & \frac{\theta}{\theta-i} & \frac{-i}{\theta-i} & 0\\
0 & \frac{-i}{\theta-i} & \frac{\theta}{\theta-i} & 0\\
0 & 0 & 0 & 1\end{array}\right) \,. \label{eq:O4Smatrix}
\]
The twist matrix acts on the two-particle states as 
\[
e^{i\gamma J}=e^{i\gamma_{-}J_{0}}\otimes e^{i\gamma_{+}J_{0}}=\dot{A}\otimes A=
\mbox{diag}(\dot{q}^{2},1,1,\dot{q}^{-2})\otimes\mbox{diag}(q^{2},1,1,q^{-2})\,,
\]
and commutes with the scattering matrix. The twist and the S-matrix
can be diagonalized in the same basis, where the S-matrix eigenvalues
take the form 
\[
S=S_{0}^{2}\,\Lambda\otimes\Lambda=S_{0}^{2}\,\mbox{diag}(1,1,
\frac{\theta+i}{\theta-i},1)\otimes\mbox{diag}(1,1,\frac{\theta+i}{\theta-i},1)\
,.
\]
For the L\"uscher correction, we need to calculate 
$\mbox{Tr}(e^{i\gamma J}(-i\partial_{\theta})\log S)$.
As the scattering matrix has the specific tensor product structure 
(\ref{eq:O4Smatrix}), we
can write 
\begin{eqnarray}
\mbox{Tr}(e^{i\gamma J}\log S) & = & \mbox{Tr}((\dot{A}\otimes A)\left(2\log 
S_{0}\mathbb{I}\otimes\mathbb{I}+\log
S_{1}\otimes\mathbb{I}+\mathbb{I}\otimes\log S_{2}\right))\\
 & = & \mbox{Tr}(\dot{A})\mbox{Tr}(A)2\log S_{0}+\mbox{Tr}(A)
\mbox{Tr}(\dot{A}\log S_{1})+\mbox{Tr}(\dot{A})\mbox{Tr}(A\log S_{2})\nonumber
\\
 & = & 2\mbox{Tr}(\dot{A})\mbox{Tr}(A)\log S_{0}+\mbox{Tr}(A)\sum_{i}\dot{A}_{i}
\log\Lambda_{i}+\mbox{Tr}(\dot{A})\sum_{i}A_{i}\log\Lambda_{i}\,.
\nonumber 
\end{eqnarray}
In Fourier space, the logarithmic derivatives take a particularly
simple form: 
\begin{eqnarray}
K_{00}(\theta)=\frac{1}{2\pi i}\partial_{\theta}\log S_{0}^{2}(\theta) & \to &
 \tilde{K}_{00}(\omega)=\frac{2t}{t+t^{-1}}\,, \nonumber \\
K(\theta)=\frac{1}{2\pi i}\partial_{\theta}\log\frac{\theta+i}{\theta-i} & \to
& 
\tilde{K}(\omega)=-t^{2}\,,
\end{eqnarray}
where we have indicated the Fourier transform by tilde, and
$t=e^{-\frac{\vert\omega\vert}{2}}$.
The integrand of the second order L\"uscher correction is finally 
\[
\frac{1}{2\pi}\mbox{Tr}(e^{i\gamma J}(-i\partial_{\theta})\log
S)=[2]_{q}^{2}[2]_{
\dot{q}}^{2}K_{00}+\left([2]_{q}^{2}+[2]_{\dot{q}}^{2}\right)K\,.
\]
In terms of these quantities, the second part of the L\"uscher correction is
\begin{eqnarray}
E_{0}^{(2,2)}(L) & = & -[2]_{q}^{2}[2]_{\dot{q}}^{2}\, \frac{m}{2} 
\int d\theta_{1}\, e^{-mL\cosh\pi\theta_{1}}
\int d\theta_{2}\cosh\pi\theta_{2}\, e^{-mL\cosh\pi\theta_{2}}\nonumber \\
 &  & \times \left\{ K_{00}(\theta_{1}-\theta_{2})+([2]_{q}^{-2}
+[2]_{\dot{q}}^{-2})K(\theta_{1}-\theta_{2})\right\} \,.
\label{eq:E22}
\end{eqnarray}

\subsection{Twisted TBA}

Following the general procedure outlined in section \ref{sec:genTBA}, in 
order to formulate the twisted TBA equations, we need to classify the
particles: 
momentum-carrying, magnons and their bound states. We also have to calculate
their
scattering matrices; and, additionally to the untwisted case, we also
must identify the twist charge on all the excitations.

\subsubsection{Raw twisted TBA }

In order to derive the mirror nested asymptotic BA equations,
we start with an $N$-particle state consisting of down-spin particles
only. We label these particles by $0$. They scatter on each other
as
\[
S_{00}(\theta)=S_{0}(\theta)^{2} \,,
\]
and they have the dispersion relation $\tilde \epsilon_{0}(\tilde p) 
=\tilde \epsilon(\tilde p)$.
As the $J_{0}$ eigenvalue of the lower component is $-1$ on both
$su(2)$ sides, the chemical potential is $\mu_{0}=-i\gamma_{-}-i\gamma_{+}$.
We can now introduce up-spins in the sea of down-spins. These are
the magnons, which do not change the energy and momentum, rather describe
the polarization degrees of freedom. We label them by $1$ for the
right $su(2)$ factor, and by $-1$ for the left $su(2)$ factor. Let us 
first focus on the positive (right)
part, and denote magnon rapidities by $u$. The magnons scatter on the
massive particles and on themselves as 
\[
S_{01}(\theta-u)=\frac{\theta-u+\frac{i}{2}}{\theta-u-\frac{i}{2}}\,,
\qquad S_{11}(u-u')=\frac{u-u'-i}{u-u'+i} \,,
\]
respectively.
The magnons do not have any energy and momentum 
$\tilde{\epsilon}_{1}(u)=\tilde{p}_{1}(u)=0$,
but they do have chemical potential. Since a magnon swaps a spin from
down to up, it changes the charge by $2$: $\mu_{1}=2i\gamma_{+}$.
This means that a state with $m$ up-spins and $N-m$ down-spins, which
contains $N$ type-$0$ particles and $m$ type-$1$ particles, has $J_{0}$
charge $-N+2m$. Inspecting the magnon scattering matrices, we can
conclude that a magnon and a massive particle cannot form bound states.
In contrast, magnons among themselves can bound. Bound states in the
thermodynamic limit consist of strings of any length $M \in\mathbb{N}$: 
\[
u_{j}=u+i\frac{M+1-2j}{2}\,, \qquad j=1,\dots,M \,.
\]
We label this string as $M$. Clearly, the $M=1$ string is the magnon
itself. The scattering of the $M$-string and the massive particle
can be calculated from the bootstrap,
\[
S_{0M}(\theta-u)=\prod_{j=1}^{M}S_{01}(\theta-u_{j})=\frac{\theta-u
+\frac{i}{2}M}{\theta-u-\frac{i}{2}M}\,.
\]
As $S_{0M}(\theta -u)S_{M0}(u-\theta)=1$, we conclude that $S_{0M}(u)=
S_{M0}(u)$. 
Similarly, the magnon-magnon scatterings are given by
\begin{eqnarray}
S_{MM'}(u-u') & = & \prod_{j=1}^{M}\prod_{j'=1}^{M'}S_{11}(u_{j}-u'_{j'})\\
 & = & 
\left(\frac{u-u'-\frac{i}{2}\vert M-M'\vert}
{u-u'+\frac{i}{2}\vert M-M'\vert}\right)
\left(\frac{u-u'-\frac{i}{2}(\vert M-M'\vert+2)}
{u-u'+\frac{i}{2}(\vert M-M'\vert+2)}\right)^{2} \nn \\
& & \times\ldots
\left(\frac{u-u'-\frac{i}{2}(M+M'-2)}{u-u'+\frac{i}{2}(M+M'-2)}\right)^{2}
\left(\frac{u-u'-\frac{i}{2}(M+M')}{u-u'+\frac{i}{2}(M+M')}\right) \,.
\nonumber 
\end{eqnarray}
These bound states have no energy and momentum 
$\tilde{\epsilon}_{M}(u)=\tilde{p}_{M}(u)=0$,
while their chemical potential is the sum of their constituents':
$\mu_{M}=2Mi\gamma_{+}.$

Similar considerations apply to the left excitations, which are denoted
by $-M$. They scatter only on themselves and on the massive particle,
such that the scattering is independent of the sign of $M$. The only
difference is in the chemical potential, as the twists are different
on the two sides: $\mu_{-M}=2Mi\gamma_{-}$. 

Summarizing, we have particles for any $M\in\mathbb{Z}$. The only
massive excitation that has nontrivial energy and momentum has the
label $0$; all others are magnons. The scattering kernels in Fourier
space have the form
\[
\tilde{K}_{00}=\frac{2t}{(t+t^{-1})}\,, \qquad 
\tilde{K}_{0n}=\tilde{K}_{n0}=-t^{n}\,,\qquad 
\tilde{K}_{nm}=
\frac{t+t^{-1}}{t-t^{-1}}(t^{n+m}-t^{\vert n-m\vert})-\delta_{nm}\,,
\]
where $t=e^{-\frac{\vert\omega\vert}{2}}$ and $n,m>0$. For the other
values, we have $K_{0n}=K_{0-n}$, $K_{n0}=K_{-n0}$, $K_{-n-m}=K_{n\, m}$ and $K_{-n\,
m}=K_{n\,-m}=0$. 

In the general procedure, one has to invert the magnonic equations
before introducing the magnon densities. In so doing, one obtains the 
``raw'' (canonical) twisted TBA equations
\begin{eqnarray}
\epsilon_{0}+\mu_{0} & = & L\tilde{\epsilon}_{0}-\log(1+e^{-\epsilon_{0}})
\star K_{00}+\sum_{M\neq0}\log(1+e^{-\epsilon_{M}})\star K_{M0} \,, 
\label{eq:O4TBA0} \\
\epsilon_{M}+\mu_{M} & = & -\log(1+e^{-\epsilon_{0}})\star K_{0M}+
\sum_{M'\neq0}\log(1+e^{-\epsilon_{M'}})\star K_{M'M} \,, \quad M \ne 
0 \,.
\label{eq:O4TBAM} 
\end{eqnarray}
These equations for the untwisted ($\mu =0  $) case reduce to those in \cite{Gromov:2008gj}, 
although in slightly different convention. 

\subsubsection{Universal TBA and Y-system}

Using identities among the kernels, we now bring the TBA equations 
(\ref{eq:O4TBA0}), (\ref{eq:O4TBAM}) to
a universal local form. This means that the pseudo-energies can be 
associated with vertices of a two-dimensional lattice, such that only
neighboring sites couple
to each other with the following universal kernel
\begin{equation}
s\, I_{MN}=\delta_{MN}-(K+1)_{MN}^{-1}\,, \qquad 
s(\theta)=\frac{1}{2\cosh\pi\theta} \,,
\label{eq:Ksop}
\end{equation}
where $I_{MN}=\delta_{M+1,N}+\delta_{M-1,N}$ and 
$(K+1)_{MN}^{-1}\star(K_{NL}+\delta_{NL})=\delta_{ML}$.
We also have $(K_{n1}+\delta_{n1})\star s=-K_{0n}$, which can be easily
seen in Fourier space where $\tilde{s}=\frac{1}{t+t^{-1}}$. 

Let us introduce the Y-functions:
\[
Y_{0}=e^{-\epsilon_{0}}\,,\qquad 
Y_{M}=e^{\epsilon_{M}}\,, \quad M\neq0 \,.
\]
We take the equations (\ref{eq:O4TBAM}) for $Y_{M}$, act with the operator 
$\delta_{MN}-s\, I_{MN}=(K+1)_{MN}^{-1}$
from the right, and use the kernel identity 
$K_{0N}\star(K+1)_{NM}^{-1}=-s\,\delta_{M,1}$.
Since the chemical potentials are annihilated by the discrete Laplacian
\[
\mu_{M}\star(sI_{MN}-\delta_{MN})=\frac{1}{2}(\mu_{N-1}+\mu_{N+1})-\mu_{N}=0 \,,
\]
they completely disappear from the equations, and we arrive at
\begin{eqnarray}
\log Y_{M} & = & I_{MM'}\log(1+Y_{M'})\star s
 \,, \quad M \neq 0 \,.
\label{eq:O4simpleTBAM}
\end{eqnarray}
Finally, we take the equations for $M=\pm1$ and convolute
them with the kernel $s$. We combine these equations with the massive
equation (\ref{eq:O4TBA0}). Using the magic property of the kernel 
$K_{00}=-2s\star K_{01}$,
and exploiting that $\mu_{0}+\frac{1}{2}(\mu_{1}+\mu_{-1})=0$, we
obtain the equation for the massive node
\[
\log Y_{0}+mL\cosh\pi\theta=
\left(\log(1+Y_{1})+\log(1+Y_{-1})\right)\star s\,.
\label{eq:O4simpleTBA0}
\]
Thus, the twists completely disappear from the ``simplified'' equations 
(\ref{eq:O4simpleTBAM}), (\ref{eq:O4simpleTBA0}).
Nevertheless, they enter in the asymptotics of the $Y$-functions as 
\be
\lim_{M\to\infty}\frac{1}{M}\log Y_{\pm M}=-2i\gamma_{\pm} \,,
\label{eq:O4Yasymptotics}
\ee
since the kernels in (\ref{eq:O4TBAM}) vanish in this limit. 
After all, it should not come 
as a surprise that the $Y$-system is not twisted,
\[
Y_{M}^{+}Y_{M}^{-}=(1+Y_{M-1})(1+Y_{M+1})\,,\qquad 
Y^{\pm}(\theta)=Y(\theta\pm\frac{i}{2})\,.
\]
The ground-state energy contains the contribution of the only massive
node,
\begin{equation}
E_{0}(L)=-\frac{m}{2}\int d\theta\cosh\pi\theta\,\log(1+Y_{0})\,.
\label{eq:O4E}
\end{equation}

\subsubsection{Asymptotic expansion}

We now make a LO and NLO asymptotic expansion of the simplified TBA
equations (\ref{eq:O4simpleTBAM}), (\ref{eq:O4simpleTBA0}) for $L\to\infty$. 

At leading order, $Y_{0}$ is exponentially small and the other $Y$
functions are constant. Let us expand the $Y$-functions as 
\[
Y_{M}=\mathcal{Y}_{M}(1+y_{M})+\dots \,,
\label{eq:Yexpansion}
\]
and determine all functions iteratively. The Y-system at leading order
will be split into two independent constant $Y$-systems. The solutions
with the correct initial and asymptotic behaviors will determine
the exponentially small leading-order $\mathcal{Y}_{0}$ in terms of
$\mathcal{Y}_{\pm1}$.
Then, in calculating the NLO $y_{M}$ functions, we can proceed independently
for the two parts. Again, the initial condition is provided by 
$\mathcal{Y}_{0}$,
which appears as a multiplicative factor; while uniqueness is provided
by the vanishing asymptotics $\lim_{M\to\infty}y_{M}=0$. The
$y_{\pm1}$ obtained in this way will determine the NLO correction $y_{0}$, which
is needed
for the energy correction. 

Let us now carry out these calculations. Using the fact that $s\star
f=\frac{1}{2}f$
if $f$ is constant, we see from (\ref{eq:O4simpleTBA0}) that 
\[
\log\mathcal{Y}_{0}=-mL\cosh\pi\theta+\frac{1}{2}\log(1+
\mathcal{Y}_{1})+\frac{1}{2}\log(1+\mathcal{Y}_{-1})\,,
\]
where the LO constant $Y$-functions satisfy the relations
\[
(\mathcal{Y}_{M})^{2}=(1+\mathcal{Y}_{M-1})(1+
\mathcal{Y}_{M+1})\,, \quad M \ne 0\,,
\]
as follows from (\ref{eq:O4simpleTBAM}).
The solution with the correct asymptotics (\ref{eq:O4Yasymptotics}) 
is \footnote{The twists 
$\gamma_{\pm}$ have small positive imaginary parts in order to suppress 
large-$M$ magnonic contributions to the 
partition function (\ref{eq:partitionfunction}).}
\[
\mathcal{Y}_{M}=[M]_{q}[M+2]_{q}\,,\qquad
\mathcal{Y}_{-M}=[M]_{\dot{q}}[M+2]_{\dot{q}}\,.
\label{eq:calYO4}
\]
Clearly, the twist dependence reenters through the asymptotic solution.
This means that at leading non-vanishing order 
\[
Y_{0}\approx\mathcal{Y}_{0}=\sqrt{(1+\mathcal{Y}_{1})(1+
\mathcal{Y}_{-1})}e^{-mL\cosh\pi\theta}=[2]_{q}[2]_{\dot{q}}e^{-mL\cosh\pi\theta
}\,,
\]
which, when substituted back into the energy formula (\ref{eq:O4E}),
reproduces the leading-order L\"uscher correction 
(\ref{O4LuscherE1}). Actually, expanding
the log in the energy formula (\ref{eq:O4E}) to second order 
$\log(1+\mathcal{Y}_{0})=\mathcal{Y}_{0}-\frac{1}{2}\mathcal{Y}_{0}^{2}$
reproduces also $E_{0}^{(2,1)}$ in (\ref{O4LuscherE21}). Thus, we need to expand
the $Y$-functions
to NLO to obtain the remaining $E_{0}^{(2,2)}$ in (\ref{eq:E22}). 

We see from (\ref{eq:O4simpleTBA0}) and (\ref{eq:Yexpansion}) that the
massive node has the NLO expansion
\begin{equation}
Y_{0}=\mathcal{Y}_{0}\left(1+s\star\left(\frac{\mathcal{Y}_{1}}{1
+\mathcal{Y}_{1}}y_{1}+\frac{\mathcal{Y}_{-1}}{1+\mathcal{Y}_{-1}}y_{-1}
\right)\right)+\dots \,.
\label{eq:Y0NLO}
\end{equation}
We need to calculate $y_{\pm 1}$. We expand the TBA equations
(\ref{eq:O4simpleTBAM}),
keeping only the linear terms in $y$,
\[
y_{k}=s\star\left(\frac{\mathcal{Y}_{k+1}}{
1+\mathcal{Y}_{k+1}}y_{k+1}+\frac{\mathcal{Y}_{k-1}}
{1+\mathcal{Y}_{k-1}}y_{k-1}\right)\,, \quad k \ne 0\,.
\]
We solve this equation by Fourier transform
\[
(t+t^{-1})\tilde{y}_{k}=\frac{[k+1]_{q} [k+3]_{q}}
{[k+2]_{q}^{2}}\tilde{y}_{k+1}
+\frac{[k-1]_{q} [k+1]_{q}}
{[k]_{q}^{2}}\tilde{y}_{k-1}\,,
\]
where we have also used the result (\ref{eq:calYO4}) and the identity 
$1+[k-1]_{q} [k+1]_{q} = [k]_{q}^{2}$. 
Being a second-order difference equation, the generic solution contains
two parameters. These parameters can be fixed by demanding that
$\lim_{k\to\infty}\tilde{y}_{k}=0$ and
$\tilde{\mathcal{Y}}_{0}=\lim_{k\to 0}\mathcal{Y}_{k}\tilde{y}_{k}$.
The result is  
\[
\tilde{y}_{k}=t^{k}\frac{[k+1]_{q}}{[2]_{q}[k]_{q}[k+2]_{q}}([k+2]_{q}-[k]_{q}t^
{2})
\tilde{\mathcal{Y}}_{0}\,,\qquad
\tilde{y}_{-k}=\tilde{y}_{k}(q\to\dot{q})\,,
\]
which is just the deformed version of the $O(4)$ solution \cite{Gromov:2008gj}.
Thus, for the needed $y_{\pm 1}$, we have 
\[
\tilde{y}_{1}=\left(t^{1}-\frac{t^{3}}{[3]_{q}}\right)\tilde{\mathcal{Y}}_{0}\,,\qquad
\tilde{y}_{-1}=
\left(t^{1}-\frac{t^{3}}{[3]_{\dot{q}}}\right)\tilde{\mathcal{Y}}_{0}
\,.
\]
Performing inverse Fourier transform,
\[
y_{1}=- \left(K_{01}-\frac{K_{03}}{[3]_{q}} \right)\star\mathcal{Y}_{0}\,,\qquad 
y_{-1}=-\left(K_{01}-\frac{K_{03}}{[3]_{\dot{q}}} \right)\star\mathcal{Y}_{0}\,.
\]
Substituting back into (\ref{eq:Y0NLO}), we obtain
\[
Y_{0}=[2]_{q}[2]_{\dot{q}}e^{-mL\cosh\pi\theta}\left(1+s\star\left[(K_{03}-[3]_{q}K_{01})
[2]_{q}^{-2}
+(K_{03}-[3]_{\dot{q}}K_{01})[2]_{\dot{q}}^{-2}\right]\star[2]_{q}[2]_{\dot{q}}e^{
-mL\cosh\pi\theta} \right) \,.
\]
Comparing the double-convolution term with $E_{0}^{(2,2)}$ in (\ref{eq:E22})
in Fourier space, we obtain complete agreement.

\section{Twisted AdS/CFT}\label{sec:AdSCFT}

In this section, we apply the previous methodology to the twisted
AdS/CFT model. After defining the model by its scattering matrix,
dispersion relation and twist matrix, we derive the LO and NLO L\"uscher
corrections. As the model has infinitely many massive bound states
$Q\in\mathbb{N}$, in the NLO L\"uscher correction we have a sum of
the form $\sum_{Q_1,Q_2=1}^{\infty}$. We first elaborate the summand $Q_1=Q_2=1$
in detail, and we then treat the general case, which entails detailed 
knowledge of all scattering matrices $S^{Q_{1} Q_{2}}$. We next derive the twisted TBA
equations by evaluating
the charges of the magnons and their bound states in the thermodynamic
limit of the mirror asymptotic BA. The twist, just as in
the $O(4)$ model, disappears from the universal equations, which
lead to the untwisted $Y$-system. We expand the TBA equations
to NLO and compare to the L\"uscher correction, and again find 
perfect agreement.

The AdS/CFT integrable model has an $su(2\vert2)\otimes su(2\vert2)$
symmetry. The elementary particle transforms under the bifundamental
representation of $su(2\vert2)$. For one copy of $su(2\vert2)$, 
Latin indices $a=1,2$ label the bosonic, while Greek indices $\alpha=3,4$
label the fermionic components of the four-dimensional representation. 
We will introduce twist in the bosonic subspace by the generator $L_0$, which 
has nonvanishing diagonal matrix elements: $(L_0)_1^1=1$  and $(L_0)_2^2=-1$. 

The symmetry completely determines the left/right scattering matrix, which
has the nonvanishing amplitudes
\begin{equation}
S_{aa}^{aa}=S_{ab}^{ab}+S_{ab}^{ba}=a_{1}=\frac{x_{2}^{-}-x_{1}^{+}}{x_{2}^{+}
-x_{1}^{-}}
\sqrt{\frac{x_{2}^{+}}{x_{2}^{-}}}\sqrt{\frac{x_{1}^{-}}{x_{1}^{+}}}\,,
\qquad S_{ab}^{ab}-S_{ab}^{ba}=a_{2}\,,
\label{eq:a1def}
\end{equation}
\[
S_{\alpha\alpha}^{\alpha\alpha}=S_{\alpha\beta}^{\alpha\beta}+
S_{\alpha\beta}^{\beta\alpha}=a_{3}=-1\,, \qquad
S_{\alpha\beta}^{\alpha\beta}-S_{\alpha\beta}^{\beta\alpha}=a_{4}\,,
\]
\[
S_{ab}^{\alpha\beta}=-\frac{1}{2}\epsilon_{ab}\epsilon^{\alpha\beta}a_{5}\,, \qquad
S_{\alpha\beta}^{ab}=-\frac{1}{2}\epsilon_{\alpha\beta}\epsilon^{ab}a_{6}\,,
\]
\[
S_{a\alpha}^{a\alpha}=a_{7}\,,\qquad  
S_{a\alpha}^{\alpha a}=a_{8}\,, \qquad
S_{\alpha a}^{a\alpha}=a_{9}\,,\qquad 
S_{\alpha a}^{\alpha a}=a_{10}\,,
\]
where $a, b \in \{1, 2\}$ with $a\ne b$; $\alpha, \beta \in \{3, 4\}$ 
with $\alpha\ne \beta$;
and the various coefficients can be extracted from 
\cite{Arutyunov:2006yd}.\footnote{Indeed, $a_{1}, 
\ldots, a_{10}$ are given by the coefficients of the ten terms in 
Eq. (8.7) in \cite{Arutyunov:2006yd}, respectively.}
For $Q_1=Q_2=1$ we shall need explicitly only $a_{1}$, since -- as a consequence of 
some identities among the various
coefficients -- we shall be able to express the L\"uscher corrections purely
in terms of it. The scattering matrix depends independently on the
momenta of the particles $p_{1}$ and $p_{2}$ via 
\[
\frac{x^{+}}{x^{-}}=e^{ip}\,, \qquad 
x^{+}+\frac{1}{x^{+}}-x^{-}-
\frac{1}{x^{-}}=\frac{2i}{g}\,,
\]
where $g=\sqrt{\lambda}/(2\pi)$ and $\lambda = g^{2}_{YM} N$ is the 't Hooft coupling.
The full scattering matrix has the form 
\[
\mathcal{S}_{11}(p_{1},p_{2})=S_{sl(2)}^{11}(p_{1},p_{2})\left[S_{su(2
\vert2)}^{11}(x_{1}^{\pm},x_{2}^{\pm})\otimes 
S_{su(2\vert2)}^{11}(x_{1}^{\pm},x_{2}^{\pm})\right]^{-1}\,,
\]
where $S_{sl(2)}^{11}(p_{1},p_{2})$ is the scalar factor
\[
S_{sl(2)}^{11}(u,u')=\frac{u-u'+\frac{i}{g}}{u-u'-\frac{i}{g}}\,
\Sigma_{11}^{-2}\,, \qquad
\Sigma_{11}=\frac{1-\frac{1}{x_{1}^{+}x_{2}^{-}}}
{1-\frac{1}{x_{1}^{-}x_{2}^{+}}}\sigma\,,
\label{S11}
\]
with $\sigma$ being the dressing factor.
We remark that $\mathcal{S}_{11}$ denotes actually the inverse of the AFZ $S$-matrix \cite{Arutyunov:2006yd}, since we are using the relativistic convention $1=e^{ipL}\prod_j S(p,p_j)$ , as in Section \ref{sec:finitesizegen}, instead of $e^{ipL}=\prod_j S(p,p_j)$.

The dispersion relation can be easily expressed in terms of $x^{\pm}$
as 
\[
E=-\frac{ig}{2}\left(x^{+}-\frac{1}{x^{+}}-x^{-}+\frac{1}{x^{-}}\right)\,.
\label{eq:dispersion}
\]
In analogy with the $O(4)$ model, we introduce different twists for
the two $su(2|2)$ factors, which we label by $\alpha=\pm$,
\[
e^{i\gamma J}=e^{i\gamma_{-}L_{0}}\otimes e^{i\gamma_{+}L_{0}}=
\mbox{diag}(\dot{q},\dot{q}^{-1},1,1)\otimes\mbox{diag}(q,q^{-1},1,1)\,,
\]
where again $q=e^{i\gamma_{+}}$, $\dot{q}=e^{i\gamma_{-}}$; and 
$\gamma_{\pm}$ are related to the deformation parameters $\gamma_{i}$ 
used in \cite{Beisert:2005if, Frolov:2005iq} by 
$\gamma_{\pm}=(\gamma_{3}\pm\gamma_{2})\frac{L}{2}$.

The scattering matrix has poles, which signal the existence of
bound states. These states transform under the $4Q$-dimensional totally
\emph{symmetric} representation of $su(2\vert2)$ for any $Q\in\mathbb{N}$.
The dispersion relation of the bound states can be obtained from 
(\ref{eq:dispersion}) by changing the shortening condition to 
\begin{equation}
x^{+}+\frac{1}{x^{+}}-x^{-}-\frac{1}{x^{-}}=\frac{2iQ}{g} \,.
\label{eq:shortening}
\end{equation}
The matrix part of the scattering matrix can be fixed
\cite{Arutyunov:2009mi} from the Yangian symmetry
\cite{Beisert:2007ds}, while the scalar factor can be determined  
\cite{Arutyunov:2009kf} from the bootstrap principle.

The mirror model has the analytically-continued scattering matrix:
$x^{\pm}(p)\to x^{\pm}(\tilde{p})$, where $\tilde{p}=-iE$. Since
the physical domains of $p$ and $\tilde{p}$ are different, the bound states
are different, too. The mirror bound states transform under the 
$4Q$-dimensional totally \emph{antisymmetric} representation of 
$su(2\vert2)$,
and the twist charge acts as 
\[
e^{i\gamma_{+}L_{0}}=\mbox{diag}(\mathbb{I}_{Q-1},\mathbb{I}_{Q+1},q
\mathbb{I}_{Q},q^{-1}\mathbb{I}_{Q})\,.
\label{eq:AdSCFTtwist1particle}
\]
The scattering matrix of the antisymmetric bound states are related 
to those of the symmetric ones by changing the labels $1\leftrightarrow 3$, 
$2 \leftrightarrow 4$ and simultaneously flipping $x^{\pm} \leftrightarrow x^{\mp}$ inside the matrix part.
Combining this with the previously mentioned notational differences, we can use the following scattering matrices to calculate the L\"uscher 
correction:

\[
\mathcal{S}=S_{sl(2)}^{Q_1Q_2}(S_{su(2|2)}^{Q_1Q_2}\otimes 
S_{su(2|2)}^{Q_1Q_2})\,,
\label{SQQ}
\]
where 
\[
S_{sl(2)}^{Q_{1}Q_{2}}(u_{1},u_{2})=
\prod_{j_{1}=1}^{Q_{1}}\prod_{j_{2}=1}^{Q_{2}}S_{11}(u^1_{j_{1}}
,u^2_{j_{2}}) \,,\quad u^n_{j_{n}}=u_{n}+(Q_{n}+1-2j_{n})\frac{i}{g}\,.
\label{Ssl2QQ}
\]
and $S_{su(2\vert2)}^{Q_1Q_2}$ denotes the symmetric-symmetric bound state scattering matrix in 
the conventions of \cite{Arutyunov:2009mi}.

\subsection{L\"uscher corrections}

The derivation of Section \ref{sec:finitesizegen} is not general
enough to describe the AdS/CFT problem.  We have to incorporate two
new features: the existence of fermions, and of multiple species of
particles that are labeled by the charge $Q$.  The fermionic nature can
be taken into account by changing the trace to the supertrace.  This
is equivalent to imposing antiperiodic boundary conditions on the
fermions, which can be implemented by an $e^{i\pi F}$ twist, where $F$
is the fermion number operator:
\[
\mbox{STr}_{Q}(e^{i\gamma J})=\mbox{Tr}_{Q}((-1)^{F}e^{i\gamma J})=
\mbox{Tr}_{Q}(e^{i(\pi F+\gamma J)})=\mbox{STr}_{Q}(e^{i\gamma_{-}L_{0}})
\mbox{STr}_{Q}(e^{i\gamma_{+}L_{0}})=([2]_{q}-2)([2]_{\dot{q}}-2)Q^{2} \,.
\]
Clearly, the supertrace vanishes in the untwisted $q\to1$ limit.  The
generalization of the derivation of Section \ref{sec:finitesizegen}
will contain the scattering matrices $\mathcal{S}^{Q_{1}Q_{2}}.$ They
arise from two-particle states with charges $Q_{1}$ and $Q_{2}$.  As
the species are different, we should not constrain the summation on
the quantization numbers $\sum_{k<l}$, and must keep all $\sum_{k,l}$,
as they label distinct two-particle states. One can verify that the 
dangerous $R^{2}$ terms from the determinant cancel against the cross 
terms coming from the square of the one-particle contribution.
Otherwise the derivation goes along the same lines as before.  As a
final result, we obtain the LO and NLO L\"uscher correction as
follows:
\begin{eqnarray}
E_{0}^{(1)} & = & -\sum_{Q}\mbox{STr}_{Q}(e^{i\gamma J})\int
\frac{d\tilde{p}}{2\pi}\, 
e^{-\tilde{\epsilon}_{Q}(\tilde{p})L}\,, \label{eq:AdSCFTLuscherE1}\\
E_{0}^{(2,1)} & = & \frac{1}{2}\sum_{Q}\mbox{STr}_{Q}(e^{i\gamma J})^{2}
\int\frac{d\tilde{p}}{2\pi}\, e^{-2\tilde{\epsilon}_{Q}(\tilde{p})L}
\,, \label{eq:AdSCFTLuscherE21}\\
E_{0}^{(2,2)} & = & \sum_{Q_{1},Q_{2}=1}^{\infty}\int
\frac{d\tilde{p}_{1}}{2\pi}e^{-L\tilde{\epsilon}_{Q_{1}}(\tilde{p}_{1})}\int
\frac{d\tilde{p}_{2}}{2\pi}e^{-L\tilde{\epsilon}_{Q_{2}}(\tilde{p}_{2})}i
\partial_{\tilde{p}_{1}}\mbox{STr}_{Q_{1}Q_{2}}(e^{i\gamma J}\log
\mathcal{S}^{Q_{1}Q_{2}}(\tilde{p}_{1},\tilde{p}_{2}))
\,, \label{eq:AdSCFTLuscherE2}
\end{eqnarray}
cf. Eqs. (\ref{eq:E1gen}),  (\ref{eq:E21gen}) and  
(\ref{eq:E22gen}),  (\ref{eq:eigtomat}), 
respectively. 
Here and below it is understood that ${\tilde p}_{i}={\tilde p}_{Q_{i}}$.

In evaluating these expressions, we note that the mirror dispersion
relation is defined via $\tilde{p}=-iE$ and $\tilde{\epsilon}=-ip$.
This dispersion relation can be then encoded into 
\[
e^{\tilde{\epsilon}_{Q}(\tilde{p})}=\frac{x^{-}}{x^{+}}\,,
\qquad\frac{2\tilde{p}}{g}=x^{-}-\frac{1}{x^{-}}-x^{+}+\frac{1}{x^{+}} \,,
\label{eq:mirrordispersion}
\]
where the shortening condition (\ref{eq:shortening}) is satisfied.

The leading L\"uscher correction for the vacuum (\ref{eq:AdSCFTLuscherE1})
receives contributions from each particle 
\[
E_{0}^{(1)}=-([2]_{q}-2)([2]_{\dot{q}}-2)
\sum_{Q}Q^{2}\int\frac{d\tilde{p}}{2\pi}\, 
e^{-\tilde{\epsilon}_{Q}(\tilde{p})L} \,.
\label{eq:E1LuscherAdS}
\]

The simple part of the NLO correction (\ref{eq:AdSCFTLuscherE21}) is
also straightforward to compute
\[
E_{0}^{(2,1)}=\frac{1}{2}([2]_{q}-2)^{2}([2]_{\dot{q}}-2)^{2}\sum_{Q}Q^{4}
\int\frac{d\tilde{p}}{2\pi}e^{-2\tilde{\epsilon}_{Q}(\tilde{p})L} \,.
\label{eq:E21LuscherAdS}
\]
In order to calculate the $E_{0}^{(2,2)}$-part of the NLO correction, we
need the supertrace of the logarithmic derivative of the mirror $S$-matrix (\ref{SQQ}):
\[
\mbox{STr}\left(e^{i\gamma_{-}L_{0}}\otimes 
e^{i\gamma_{+}L_{0}}\log\mathcal{S}\right)\,.
\]

We now diagonalize the twist matrix and the scattering matrix on the
same basis,
\begin{eqnarray}
e^{i\gamma_{-}L_{0}}\otimes e^{i\gamma_{+}L_{0}} & = & \dot{A}\otimes A=\mbox{diag}
(\dot{A}_{1},\dots,\dot{A}_{n})\otimes\mbox{diag}(A_{1},\dots,A_{n}) 
\,, \nonumber \\
\mathcal{S} & = & \Lambda\otimes\Lambda=\mbox{diag}(\Lambda_{1},\dots,
\Lambda_{n})\otimes\mbox{diag}(\Lambda_{1},\dots,\Lambda_{n})\,,
\end{eqnarray}
where $\Lambda_{i}$ are the eigenvalues of $S_{su(2|2)}^{Q_1Q_2}$,
and $n=16 Q_{1} Q_{2}$.
Calculation similar to the one in the $O(4)$ model gives
\begin{eqnarray}
\mbox{STr}(e^{i\gamma J}\log\mathcal{S}) & = & \mbox{STr}\left(\dot{A}\otimes 
A(\log S_{sl(2)}^{Q_1Q_2}\mathbb{I}\otimes\mathbb{I}+\log S_{su(2|2)}^{Q_1Q_2}
\otimes\mathbb{I}+\mathbb{I}\otimes\log S_{su(2|2)}^{Q_1Q_2})\right)\\
 & = & \mbox{STr}(\dot{A})\,\mbox{STr}(A)\log
S_{sl(2)}^{Q_1Q_2}+\sum_{i}(-1)^{F_{i}}
\left(\mbox{STr}(A)\dot{A}_{i}+\mbox{STr}(\dot{A})A_{i}\right)\log\Lambda_{i}\,.
\nonumber 
\end{eqnarray}
Using the derivative of this expression, we can express the NLO
L\"uscher correction (\ref{eq:AdSCFTLuscherE2}) in the following form:
\begin{eqnarray}
E_{0}^{(2,2)} & = &
\sum_{Q_1,Q_2=1}^{\infty}Q_1 Q_2 \int\frac{d\tilde{p}_{1}}{2\pi}
e^{-L\tilde{\epsilon}_{Q_1}(\tilde{p}_{1})}\int\frac{d\tilde{p}_{2}}{2\pi}
e^{-L\tilde{\epsilon}_{Q_2}(\tilde{p}_{2})}i\partial_{\tilde{p}_{1}}
\times\nonumber \\
 &  & \,\,\,\,\,\,\,\,\,\,\,\,\,\, \Biggl[
 Q_1 Q_2(2-[2]_{q})^{2}(2-[2]_{\dot{q}})^{2}
\log S_{sl(2)}^{Q_1Q_2}(\tilde{p}_{1},\tilde{p}_{2})\nonumber \\
 &  & \left.\,\,\,\,\,\,\,\,\,\,\,\,\,\,\,\,\,\,\,+\sum_{i}(-1)^{F_{i}}
\left(\dot{A}_{i}(2-[2]_{q})^{2}+A_{i}(2-[2]_{\dot{q}})^{2}\right)\log
\Lambda_{i}^{Q_1 Q_2}(\tilde{p}_{1},\tilde{p}_{2})\right]\,.
\label{NLOmatrix}
\end{eqnarray}

\subsubsection{NLO L\"uscher correction: the case $Q_{1}=Q_{2}=1$}
To warm up, let us evaluate the NLO L\"uscher correction for the simplest
$Q_1=Q_2=1$
case. We focus on the matrix part in (\ref{NLOmatrix}). Performing 
the calculation explicitly, we obtain
\begin{eqnarray}
 &  & (2-[2]_{\dot{q}})^{2}i\partial_{\tilde{p}_{1}}\left\{ ([3]_{q}-1)\log
a_{1}+
\log\left[a_{1}a_{3}^{3}\left((a_{1}+2a_{2})(a_{3}+2a_{4})-4a_{7}a_{8}
\right)\right]\right.\nonumber \\
 &  & \left.-2[2]_{q}\log(a_{5}a_{6}-a_{10}a_{9})\right\} 
 +(q\leftrightarrow\dot{q}) \,,
\label{Q=Q'=1}
\end{eqnarray}
where $(-1)^{F}=(1,1,-1,-1)$.
Using the explicit expressions for the coefficients found in 
\cite{Arutyunov:2006yd},
we observe the following identities
\begin{equation}
a_{5}a_{6}-a_{10}a_{9}=a_{1}\,, \qquad(a_{1}+2a_{2})(a_{3}+2a_{4})-4a_{7}a_{8}=
-a_{1} \,.
\label{simplifications}
\end{equation}
Substituting these identities into (\ref{Q=Q'=1}), we obtain
a very simple expression for the matrix part of the
NLO L\"uscher correction for $Q_1=Q_2=1$ in terms of only $a_{1}$,
\begin{equation}
(2-[2]_{q})^{2}(2-[2]_{\dot{q}})^{2}\left(\frac{[2]_{\dot{q}}}{2-[2]_{\dot{q}}}
+
\frac{[2]_{q}}{2-[2]_{q}}\right)\int\frac{d\tilde{p}_{1}}{2\pi}e^{-L\tilde{
\epsilon}_{1}
(\tilde{p}_{1})}\int\frac{d\tilde{p}_{2}}{2\pi}e^{-L\tilde{\epsilon}_{1}(\tilde{
p}_{2})}i
\partial_{\tilde{p}_{1}}\log a_{1}(\tilde{p}_{1},\tilde{p}_{2})\,.
\label{NLOLuschermatrix}
\end{equation}

\subsubsection{NLO L\"uscher correction: the general case 
$(Q_{1},Q_{2})$}\label{sec:LuscherAdS}
\vskip0.2in

Although the above approach can also be used for the cases
$(Q_{1},Q_{2})=(1,2), (2,2)$ for which the explicit S-matrices are
available \cite{Arutyunov:2008zt}, it is impractical for
higher-dimensional cases.  Clearly, a more powerful approach is needed
to treat the general case.
Observe from (\ref{NLOmatrix}) that the NLO L\"uscher correction involves the quantity
$\sum_{i} (-1)^{F_{i}}A_{i} \log \Lambda_{i}^{Q_{1}Q_{2}}$, and a
similar quantity with $A_{i}$ replaced by $\dot A_{i}$. 
We exploit the fact that the $su(2\vert 2)$ part of the $\gamma_+$ twist $e^{i\gamma J}={
\mathbb I}
\otimes e^{i\gamma_{+} L_{0}}$  involves nontrivially only the 
$su(2)_{R}$ factor in $ su(2)_{L}\otimes su(2)_{R} \subset su(2\vert 2)$, 
as is evident from (\ref{eq:AdSCFTtwist1particle}). 
Since $su(2)_{L}\otimes su(2)_{R}$ is the symmetry of the scattering matrix, 
we can perform an expansion in the left ($s_L$) and right ($s_R$) 
spins:
\ba
\sum_{i} (-1)^{F_{i}}A_{i} \log \Lambda_{i}^{Q_{1},Q_{2}}
&=& \sum_{(s_L,s_R)} {\rm STr}[ ({ \mathbb I}
\otimes e^{i\gamma_{+} L_{0}}) \log S^{Q_{1} Q_{2}}(s_L,s_R)]\nonumber \\ 
&=& \sum_{(s_L,s_R)} (-1)^{2s_R}
(2s_{L}+1) \left[ 2s_{R}+1\right]_{q} 
\log  \det S^{Q_{1}Q_{2}}(s_{L}, s_{R})  \,,
\label{Luscherid}
\ea
where $S^{Q_{1}Q_{2}}(s_{L}, s_{R})$ is the 2-particle S-matrix in
the sector with left and right $su(2)$ spins $s_{L}$ and $s_{R}$\footnote{In
other words, 
$\det S(s_{L}, s_{R}) =
\prod_{i}\Lambda_{i}(s_{L}, s_{R})$, where $\Lambda_{i}(s_{L}, s_{R})$
are the eigenvalues of the 2-particle S-matrix corresponding to
eigenstates which are also $su(2)_{L} \otimes su(2)_{R}$ 
highest-weight states with given values of $s_{L}$ and $s_{R}$. For 
further details, see appendix \ref{sec:dets}. As usual, the 
spins $s_{L}, s_{R}$ are non-negative integers or half-odd integers.}, 
and we calculated the traces as $ {\rm Tr}_{s_L}({ \mathbb I})
=2s_L+1 $  and   $ {\rm STr}_{s_R}(e^{i\gamma_{+} L_{0}})
=(-1)^{2s_R}[2s_R+1]_q $, respectively.
 The sum is over
all the possible values of $s_{L}$ and $s_{R}$ for the given values of $Q_{1}$ 
and $Q_{2}$.

The problem of computing the NLO L\"uscher correction therefore
reduces to the determination of  $\det S^{Q_{1}Q_{2}}(s_{L}, s_{R})$ for
general values of $Q_{1}$ and $Q_{2}$, and for all the possible
corresponding values of $s_{L}$ and $s_{R}$.  This is a formidable
technical challenge, since individual S-matrix elements -- and 
particularly the eigenvalues --  are not known explicitly enough in general,
and those that are known explicitly enough \cite{Arutyunov:2008zt, Bajnok:2008bm} generally
have very complicated expressions.  Nevertheless, it turns out that --
remarkably -- these determinants have simple compact expressions,
which are constructed from a small number of elementary building
blocks.

We propose that, with both particles in {\em symmetric}
representations and $Q_{1}\,,Q_{2}>1$, the determinants are given by
the expressions in Table \ref{table:general}.  In order to save
writing, we have introduced the following notation
\begin{equation}
U_{0}=\frac{x_{1}^{-}-x_{2}^{+}}{x_{1}^{+}-x_{2}^{-}}\,,\quad 
U_{1}=\sqrt{\frac{x_{1}^{+}}{x_{1}^{-}}}\,,\quad U_{2}=
\sqrt{\frac{x_{2}^{-}}{x_{2}^{+}}}\,,\quad U_{3}=
\frac{x_{1}^{+}x_{2}^{+}-1}{x_{1}^{-}x_{2}^{-}-1}\,,
\quad S_{Q}=\frac{u_{1}-u_{2}-\frac{iQ}{g}}{u_{1}-u_{2}+
\frac{iQ}{g}}\,,\label{detnotation}
\end{equation}
where $x_{j}^{\pm}$ are the parameters of the $Q_{j}$ bound-state 
representation, and $u_{j}\pm\frac{i Q_{j}}{g} =x_{j}^{\pm}+\frac{1}{x_{j}^{\pm}}$. 
Note that there are only three possible values of the right-spin, namely $s_{R} = 0, 
\frac{1}{2}, 1$, as $2s_{R}$ counts the number of fermions in the 
basis of the Hilbert space. If at least one of either $Q_{1}$ or $Q_{2}$ is 1, then
the corresponding results are collected in Table \ref{table:special}. 
A brief account of how these results were obtained is presented in
Appendix \ref{sec:dets}.

\begin{table}
\begin{centering}
\begin{tabular}{c|l|l}
$s_{R}$ & \qquad \qquad  \qquad $\det S^{Q_{1}Q_{2}}(s_{L},s_{R})$ 
&\qquad \qquad \qquad \qquad $s_{L}$\tabularnewline
\hline
& & \tabularnewline
$1$ & $U_{0}U_{1}U_{2}S_{2s_{L}+2}S_{2s_{L}+4}\ldots S_{Q_{1}+Q_{2}-2}$ & 
$\frac{1}{2}\vert Q_{12}\vert,\frac{1}{2}(\vert Q_{12}\vert+2),\ldots,
\frac{1}{2}(Q_{1}+Q_{2}-2)$\tabularnewline
& & \tabularnewline
\hline
& & \tabularnewline
$\frac{1}{2}$ & $U_{0}U_{1}U_{2}$ & $\frac{1}{2}(Q_{1}+Q_{2}-1)$\tabularnewline
$\frac{1}{2}$ & $\left(U_{0}U_{1}U_{2}\right)^{4}S_{2s_{L}+1}^{2}S_{2s_{L}+3}^{4}
\ldots S_{Q_{1}+Q_{2}-2}^{4}$ & $\frac{1}{2}(\vert Q_{12}\vert+1),
\ldots,\frac{1}{2}(Q_{1}+Q_{2}-3)$\tabularnewline
$\frac{1}{2}$ & $\left(U_{0}U_{1}U_{2}\right)^{2}\left(
\frac{U_{2}U_{3}}{U_{1}}\right)^{\delta_{21}}S_{|Q_{12}|+2}^{2}
S_{|Q_{12}|+4}^{2}\ldots S_{Q_{1}+Q_{2}-2}^{2}$ & $\frac{1}{2}
(\vert Q_{12}\vert-1)\ge0$\tabularnewline
& & \tabularnewline
\hline
& & \tabularnewline
$0$ & $1$ & $\frac{1}{2}(Q_{1}+Q_{2})$\tabularnewline
$0$ & $\left(U_{0}U_{1}U_{2}\right)^{3}S_{2s_{L}}$ 
& $\frac{1}{2}(Q_{1}+Q_{2}-2)$\tabularnewline
$0$ & $\left(U_{0}U_{1}U_{2}\right)^{5}S_{2s_{L}}
S_{2s_{L}+2}^{4}$ & $\frac{1}{2}(Q_{1}+Q_{2}-4)$\tabularnewline
$0$ & $\left(U_{0}U_{1}U_{2}\right)^{5}S_{2s_{L}}S_{2s_{L}+2}^{4}S_{2s_{L}+4}^{5}
\ldots S_{Q_{1}+Q_{2}-2}^{5}$ & $\frac{1}{2}(\vert Q_{12}\vert+2),
\ldots,\frac{1}{2}(Q_{1}+Q_{2}-6)$\tabularnewline
$0$ & $\left(U_{0}U_{1}U_{2}\right)^{4}\left(\frac{U_{2}U_{3}}{U_{1}}\right)^{\delta_{21}}
S_{|Q_{12}|+2}^{3}S_{|Q_{12}|+4}^{4}\ldots S_{Q_{1}+Q_{2}-2}^{4}$ & 
$\frac{1}{2}\vert Q_{12}\vert\neq0$\tabularnewline
$0$ & $U_{0}U_{1}U_{2}\left(\frac{U_{2}U_{3}}{U_{1}}\right)^{\delta_{21}}
S_{|Q_{12}|+2}S_{|Q_{12}|+4}\ldots S_{Q_{1}+Q_{2}-2}$ & 
$\frac{1}{2}(\vert Q_{12}\vert-2)\ge0$\tabularnewline
$0$ & $\left(U_{0}U_{1}U_{2}\right)^{3}S_{2}^{2}S_{4}^{3}S_{6}^{3}
\ldots S_{Q_{1}+Q_{2}-2}^{3}$ & $0=Q_{12}$\tabularnewline
& & \tabularnewline
\end{tabular}
\par\end{centering}

\caption{The values of $\det S^{Q_{1}Q_{2}}(s_{L},s_{R})$ for $Q_{1}, 
Q_{2} > 1$ and for all possible
$s_{R}$ and $s_{L}$, where $\delta_{21}=\frac{Q_{21}}{\vert 
Q_{21}\vert}=\pm 1$, 
and $Q_{ij}=Q_{i}-Q_{j}$. See (\ref{detnotation}) for further notations.}
\label{table:general}
\end{table}

\begin{table}
\begin{centering}
\begin{tabular}{c|l|l}
$s_{R}$ & $ \quad \det S^{Q_{1}Q_{2}}(s_{L},s_{R})$ 
&  \qquad  \quad  $s_{L}$\tabularnewline
& & \tabularnewline
\hline
& & \tabularnewline
$1$ & $U_{0}U_{1}U_{2}$ & $\frac{1}{2}(Q_{1}+Q_{2}-2)$\tabularnewline
& & \tabularnewline
\hline
& & \tabularnewline
$\frac{1}{2}$ & $U_{0}U_{1}U_{2}$ & $\frac{1}{2}(Q_{1}+Q_{2}-1)$\tabularnewline
$\frac{1}{2}$ & $\left(U_{0}U_{1}U_{2}\right)^{2}\left(
\frac{U_{2}U_{3}}{U_{1}}\right)^{\delta_{21}}$ & $
\frac{1}{2}(\vert Q_{12}\vert-1)\ge0$\tabularnewline
& & \tabularnewline
\hline
& & \tabularnewline
$0$ & $1$ & $\frac{1}{2}(Q_{1}+Q_{2})$\tabularnewline
$0$ & $\left(U_{0}U_{1}U_{2}\right)^{2}\left(\frac{U_{2}U_{3}}{U_{1}}\right)^{\delta_{21}}$ 
& $\frac{1}{2}\vert Q_{12}\vert\neq0$\tabularnewline
$0$ & $U_{0}U_{1}U_{2}\left(\frac{U_{2}U_{3}}{U_{1}}\right)^{\delta_{21}}$ 
& $\frac{1}{2}(\vert Q_{12}\vert-2)\ge0$\tabularnewline
$0$ & $U_{0}U_{1}U_{2}$ & $0=Q_{12}\quad (Q_{1}=Q_{2}=1) $\tabularnewline
& & \tabularnewline
\end{tabular} 
\par\end{centering}

\caption{The values of $\det S^{Q_{1}Q_{2}}(s_{L},s_{R})$ for all possible
$s_{R}$ and $s_{L}$ if either $Q_{1}$ or $Q_{2}$ is 1. }
\label{table:special}
\end{table}

Substituting the results from Tables \ref{table:general} and 
\ref{table:special} into (\ref{Luscherid}), and 
carefully simplifying the resulting expression, we obtain
\begin{eqnarray}
 \sum_{i}(-1)^{F_{i}}A_{i}\log\Lambda_{i}^{Q_{1}Q_{2}}=&  & [3]_{q}
\bigl(Q_{1}Q_{2}\log U_{0}U_{1}U_{2}+{\cal 
K}^{Q_{1}Q_{2}}\bigr)\label{LuscherQQ'} \\
 &  & -[2]_{q}\bigl( (4Q_{1}Q_{2}-Q_{1}-Q_{2})\log U_{0}+2Q_{2}(2Q_{1}-1)
\log U_{1}  \nonumber \\
&  & \qquad \quad +2Q_{1}(2Q_{2}-1)\log U_{2}+(Q_{2}-Q_{1})\log U_{3}+4{\cal 
K}^{Q_{1}Q_{2}}\bigr) \nonumber \\
 &  & +[1]_{q}\bigl((5Q_{1}Q_{2}-2Q_{1}-2Q_{2})\log U_{0}+Q_{2}(5Q_{1}-4)
\log U_{1}  \nonumber \\ 
&  & \qquad \quad  +Q_{1}(5Q_{2}-4)\log U_{2}+2(Q_{2}-Q_{1})\log U_{3}+5{\cal 
K}^{Q_{1}Q_{2}}\bigr)\,, \nonumber
\end{eqnarray}
where we have defined 
\begin{equation}
{\cal K}^{Q_{1}Q_{2}}=\sum_{j=0}^{Q_{1}-1}(Q_{2}-Q_{1}+2j+1)
\sum_{k=1}^{Q_{1}-j-1}\log S_{Q_{2}-Q_{1}+2j+2k}\,.
\label{calK}
\end{equation}
In deriving the result (\ref{LuscherQQ'}), we have made use of the fact that 
$\log S_{Q}$ is an antisymmetric function of $Q$ (i.e., $\log S_{-Q} 
= -\log S_{Q}$, up to an irrelevant additive constant), which in 
particular implies that ${\cal K}^{Q_{1}Q_{2}}={\cal K}^{Q_{2}Q_{1}}$.
We emphasize that (\ref{LuscherQQ'}) holds for any $Q_{1},Q_{2} \in 
\mathbb{N}$. An analogous result can be derived for 
$\sum_{i}(-1)^{F_{i}}\dot{A}_{i}\log\Lambda_{i}^{Q_{1}Q_{2}}$
by replacing $q\rightarrow\dot{q}$ in (\ref{LuscherQQ'}). 

Thus, in order
to calculate the L\"uscher correction, we have to plug (\ref{LuscherQQ'}) into the formula (\ref{NLOmatrix}):
\ba
E_{0}^{(2,2)}  &=&  
\sum_{Q_{1},Q_{2}=1}^{\infty}Q_{1}Q_{2}\int_{-\infty}^{\infty}\frac{d\tilde{p}_{1}}{2\pi}
e^{-L\tilde{\epsilon}_{Q_{1}}(\tilde{p}_{1})}\int_{-\infty}^{\infty}\frac{d\tilde{p}_{2}}{2\pi}
e^{-L\tilde{\epsilon}_{Q_{2}}(\tilde{p}_{2})}\nonumber \\
&\times& i\partial_{\tilde{p}_{1}}\Biggl\{ (2-[2]_{\dot{q}})^{2}\Big[[3]_{q}
\left(Q_{1}Q_{2}\log U_{0}U_{1}U_{2}+{\cal K}^{Q_{1}Q_{2}}\right) 
\nonumber \\
&& \hspace{3cm} -[2]_{q}\Big((4Q_{1}Q_{2}-Q_{1}-Q_{2})\log U_{0}+2Q_{2}(2Q_{1}-1)
\log U_{1} \nonumber\\
&& \hspace{4cm} +2Q_{1}(2Q_{2}-1)\log U_{2}+(Q_{2}-Q_{1})\log U_{3}+4{\cal 
K}^{Q_{1}Q_{2}}\Big)\nonumber \\
&& \hspace{3cm} +[1]_{q}\Big((5Q_{1}Q_{2}-2Q_{1}-2Q_{2})\log U_{0} 
+Q_{2}(5Q_{1}-4)
\log U_{1}\nonumber \\
&& \hspace{4cm} +Q_{1}(5Q_{2}-4)\log U_{2}+2(Q_{2}-Q_{1})\log U_{3}+5{\cal 
K}^{Q_{1}Q_{2}}\Big)\Big] \nonumber \\
&& \hspace{2cm} +( q \leftrightarrow \dot q ) \nonumber \\
&& \qquad \qquad +Q_{1}Q_{2}(2-[2]_{q})^{2}(2-[2]_{\dot{q}})^{2}
\log S_{sl(2)}^{Q_{1}Q_{2}}(\tilde{p}_{1},\tilde{p}_{2}) \Bigg\}\,.
\label{NLOLuscher}
\ea 
We shall compare this result to the TBA output in Section \ref{sec:TBAQQ'}.

\subsection{Twisted TBA equations}

In \cite{Cavaglia:2010nm, Cavaglia:2011kd, Arutyunov:2009ur}, the
authors derived the TBA equations for the AdS/CFT model with the most
general chemical potentials.  Hence, the TBA equations for the
$\gamma$-deformed theories correspond to some special cases.  However,
since we must determine precisely the charges/chemical potentials of
the various excitations in terms of the deformation parameters, we now
briefly sketch the derivation.

In order to derive the TBA equations, we have to recall the various 
types of excitations (both massive and magnonic) and their scattering 
matrices; and we must calculate
their twist charges. We label the fundamental massive particle as $Q=1$, corresponding to the $(3\dot{3})$ label of the fundamental 
representation. The $S$-matrix of this kind of particles is in fact given by (\ref{S11}) and they can form bound states
for any $Q$ with string-like complex roots defined like in (\ref{Ssl2QQ}).
We label such a massive composite particle by $Q$ and the scattering
matrix of such particles is (\ref{Ssl2QQ}).
Since the twist charge acts
trivially in the $(3,4)$ subspace, the massive particles
are not charged: $\mu_{Q}=0$.

We now focus on the magnonic excitations. They encode the color $su(2\vert2)$
structure of the scattering, and come in independent left and right
copies. We first consider the right part. We label a magnon, which 
introduces label $2$ in the sea of massive $3$-particles, by $y$. It scatters
trivially on itself, but nontrivially on the massive particles 
\[
S_{1y}(u,y)=\frac{x^{-}(u)-y}{x^{+}(u)-y}\sqrt{\frac{x^{+}(u)}{x^{-}(u)}}\,,
\qquad 
S_{Qy}(u,y)=\prod_{j=1}^{Q}S_{1y}(u_{j},y)\,.
\]
The twist charge of the $y$ particles is $\mu_{y}=-i\gamma_{+}$. 

We can also introduce the label $1$ in the sea of $2$-particles. These particles
are labeled by $w$. They scatter nontrivially only on the $y$ particles
and on themselves: 
\[
S_{yw}(y,w)=S_{-1}(v(y)-w)\,, \qquad 
S_{ww}(w,w')=S_{2}(w-w')\,,
\]
where $v(y)=y+y^{-1}$, and $S_{n}(u)$ is defined as in 
(\ref{detnotation}), namely
\be
S_{n}(u)=
\frac{u-\frac{in}{g}}{u+\frac{in}{g}} \,.
\label{eq:Sndef}
\ee 
The twist charge of these particles is $\mu_{w}=2 i\gamma_{+}$.

As the scattering matrix $S_{yw}(y,w)$ has a difference form in the
variable $v(y)=y+y^{-1}$, we might use the parameter $v$ instead
of $y$. The inverse of the relation, however, is not unique. Defining
$y_{-}(v)=\frac{1}{2}(v-i\sqrt{4-v^{2}})$ with the branch cuts running
from $\pm\infty$ to $\pm2$, we can describe any $y$ with $\Im m(y)<0$
for $v\in[-2,2]$. Clearly $y_{+}(v)=y_{-}(v)^{-1}$ describes the
other $\Im m(y)>0$ case; and in the scattering matrices $S_{1y}$
which depend on $y$, and not on $v$, we have to specify which root
is taken. As a consequence, we have two types of $y$ particles $y\vert\delta$
with $\delta=\pm$; and the scattering matrices split as $S_{1y}(u,y)\to 
S_{1y\vert\delta}(u,v):=S_{1y}(u,y_{\delta}(v))$.
Clearly, the $y\vert\delta$ magnons scatter on the momentum bound states
as $S_{Q\, y\vert\delta}(u,v)=\prod_{j}S_{1\, y\vert\delta}(u_{j},v)$. 

Let us now focus on the magnonic bound states. Detailed investigation
showed \cite{Arutyunov:2009zu} that $v$ and $w$ particles can form bound states for any positive
integer $M$. It consist of $2M$ $v$-particles
$v_{\pm(M+2-2j)}=v\pm(M+2-2j)\frac{i}{g}$
for $j=1,\dots,M$ with $y_{j}=(y_{-j}^{-1})^{*}$, and $M$ $w$-particles
with synchronized parameters $w_{j}=v+(M+1-2j)\frac{i}{g}$ for $j=1,\dots,M$.
The scattering matrix of the $v\vert M$ particle with all other particles
is simply the product of the scatterings of each of its individual constituents
\[
S_{v\vert M\, i}(v,q)=\prod_{j=1}^{M+1}S_{y\vert-\, i}(v_{M+2-2j},q)
\prod_{j}^{M}S_{wi}(w_{j},q)\prod_{j=1}^{M-1}S_{y\vert+\, 
i}(v_{M-2j},q)\,.
\]
The twist charge of the bound state simply sums up to 
$\mu_{v\vert M}=2M(- i \gamma_{+})+M2 i \gamma_{+}=0$. 

The $w$-type particles can form bound states among themselves: 
an $N$-string of $w$-particles can be formed as $w_{j}=w+(N+1-2j)\frac{i}{g}$.
The
scattering of the $N$-string with any other particle is
\[
S_{w\vert N\, i}(w,q)=\prod_{j=1}^{N}S_{w\, i}(w_{j},q)\,,
\]
while the twist charge is $\mu_{w\vert N}=2N i \gamma_{+}$. 

We summarize the various scattering matrices and chemical potentials
in Table \ref{Flo:SmatrixTable}. 

\begin{table}[h]
\begin{centering}
\begin{tabular}{|c||c|c|c|c|c|}
\hline 
 & $Q_{2}$ & $v\vert M_{2}$ & $w\vert N_{2}$ & $y\vert\delta_{2}$ & $\mu$\tabularnewline
\hline
\hline 
$Q_{1}$ & $S_{Q_{1}Q_{2}}$ & $S_{Q_{1}\, v\vert M_{2}}$ & $1$ & $S_{Q_{1}\, y\vert\delta_{2}}$ &
$0$\tabularnewline
\hline 
$v\vert M_{1}$ & $S_{v\vert M_{1}\, Q_{2}}$ & $S_{v\vert M_{1}\, v\vert M_{2}}$  
& $1$ & $S_{v\vert M_{1}\, y\vert\delta_{2}}$ & $0$\tabularnewline
\hline 
$w\vert N_{1}$ & $1$ & $1$ & $S_{w\vert N_{1}\, w\vert N_{2}}$ & $S_{w\vert N_{1}\,
 y\vert\delta_{2}}$ & $2N_{1}i\gamma$\tabularnewline
\hline 
$y\vert\delta_{1}$ & $S_{y\vert\delta_{1}\, Q_{2}}$ & $S_{y\vert\delta_{1}\, v\vert M_{2}}$ &
 $S_{y\vert\delta_{1}\, w\vert N_{2}}$ & $1$ & $-i\gamma$\tabularnewline
\hline
\end{tabular}
\par\end{centering}

\caption{Scattering matrices of the various particles and their 
chemical potentials
for any of the two $su(2\vert2)$ wings.}
\label{Flo:SmatrixTable}
\end{table}

Once we know the chemical potentials, we can calculate the kernels
\[
K_{jj'}(u,u')=\frac{1}{2\pi i}\partial_{u}\log S_{jj'}(u,u')\,,
\]
and write the TBA equations one by one. To ensure positive particle
densities, we have to invert the equations for $v\vert M$ and for
$y\vert-$. The equation for the massive nodes then read as 
\[
\epsilon_{Q_{2}}=L\tilde{\epsilon}_{Q_{2}}-\log(1+e^{-\epsilon_{Q_{1}}})\star
 K_{Q_{1}Q_{2}}+\sum_{\alpha=\pm}\log(1+e^{-\epsilon_{v\vert M}^{\alpha}})\star
K_{v\vert M\, Q_{2}}-\delta\log(1+e^{-\epsilon_{y\vert\delta}^{\alpha}})
\star K_{y\vert\delta\, Q_{2}}\,.
\label{eq:AdSTBAQ}
\]
Note that for particles of type $v\vert M$ and $y\vert\delta$, we must
include contributions of the two $su(2\vert2$) copies,
which we denote by $\alpha=\pm$. The remaining equations are valid
for the two $su(2\vert2)$ factors separately, so we omit the $\alpha$
index:
\begin{eqnarray}
\epsilon_{v\vert M} & = & -\log(1+e^{-\epsilon_{Q_{2}}})\star K_{Q_{2}\, v\vert M}+
\log(1+e^{-\epsilon_{v\vert M'}})\star K_{v\vert M'\, v\vert M}-\delta
\log(1+e^{-\epsilon_{y\vert\delta}})\star K_{y\vert\delta\, v\vert 
M}\,, \nonumber \\
\epsilon_{w\vert N} & = &-\mu_{w\vert N} -\log(1+e^{-\epsilon_{w\vert N'}})
\star K_{w\vert N'\, w\vert N}-\delta\log(1+e^{-\epsilon_{y\vert\delta}})
\star K_{y\vert\delta\, w\vert N}\,, \label{eq:AdSTBA}\\
\epsilon_{y\vert\delta} & = & i\pi -\mu_{y\vert\delta} -\log(1+e^{-\epsilon_{Q_{2}}})
\star K_{Q_{2}\, y\vert\delta}+\log(1+e^{-\epsilon_{v\vert M}})\star K_{v
\vert M\, y\vert\delta}-\log(1+e^{-\epsilon_{w\vert N}})\star K_{w\vert N\,
 y\vert\delta}\,.
\nonumber 
\end{eqnarray}
Once these equations are solved, the ground-state energy can be obtained
as
\[
E_{0}(L)=-\sum_{Q_{2}=1}^{\infty}\int\frac{du}{2\pi}\partial_{u}\tilde{p}_{Q_{2}}
\log(1+e^{-\epsilon_{Q_{2}}})\,.
\]

In \cite{Cavaglia:2010nm, Cavaglia:2011kd} the authors analyzed the TBA equations with 
generic chemical potentials, and formulated the requirement under which the Y-system 
remains unchanged. Our chemical potentials, which correspond to 
$\gamma$-deformations, satisfy their requirement.

\subsection{Universal TBA equations and Y-system}

The TBA equations can usually be brought into a local form. As 
already remarked, this means
that the pseudo-energies can be drawn on a two-dimensional lattice, such
that only neighboring sites couple to each other with the universal kernel
\begin{equation}
s\, I_{MN}=\delta_{MN}-(K+1)_{MN}^{-1}\,, \qquad 
s(u)=\frac{g}{4\cosh \frac{g\pi u}{2}} \,,
\label{eq:Ksop2}
\end{equation}
where $I_{MN}=\delta_{M+1,N}+\delta_{M-1,N}$ and $(K+1)_{MN}^{-1}
\star(K_{NL}+\delta_{NL})=\delta_{ML}$.
To simplify the notation, we introduce the following $Y$-functions\footnote{To
compare with \cite{Arutyunov:2009ur, Arutyunov:2009ux},
we note that $Y_{w\vert N}=Y_{N\vert w}^{AF}$,  $Y_{v\vert M}=Y_{M\vert
vw}^{AF}$ and $K_{vx}^{Q_{1} Q_{2}} = K_{vwx}^{Q_{1} Q_{2}\, AF}$.
Also, $K_{n}(u)=\frac{1}{2\pi i} \frac{d}{du}\log S_{n}(u)$, 
where $S_{n}(u)$ is defined in (\ref{eq:Sndef}); its Fourier 
transform is $\tilde K_{n} = {\rm sign}(n) t^{|n|}$, $t=e^{-\frac{\vert \omega \vert}{g}}$.}
\[
Y_{Q}=e^{-\epsilon_{Q}}\,, \qquad 
Y_{v\vert M}=e^{\epsilon_{v\vert M}}\,,\qquad 
Y_{w\vert N}=e^{\epsilon_{w\vert N}}\,,\qquad 
Y_{\delta}= -e^{\epsilon_{y\vert\delta}}\,.
\]
Clearly, we have two copies for the magnonic $Y$-functions: 
$Y_{v\vert M}^{\alpha}, Y_{w\vert N}^{\alpha}, Y_{\delta}^{\alpha}$
where $\alpha=\pm$ refers to the two $su(2\vert2)$ copies. Acting
with the operator (\ref{eq:Ksop2}) on the TBA equations 
(\ref{eq:AdSTBAQ}), (\ref{eq:AdSTBA}), and using
kernel identities such as $(K+1)_{MN}^{-1}\star K_{N}=s\,\delta_{M,1}$
as well as the special properties of the chemical potentials 
$\mu_{w\vert N-1}-2\mu_{w\vert N}+\mu_{w\vert N+1}=0$
and $\mu_{w\vert1}=-2\mu_{y}$, we arrive at their simplified form \cite{Arutyunov:2009ux}.
For later purposes, we write the simplified equations for $v\vert M$
and $w\vert N$ magnons, and a useful combination (hybrid) of the un-simplified
equations for $Q$ and $y$ particles \cite{Arutyunov:2009ax} 
\begin{eqnarray}
\log Y_{Q_{2}} & = & -L\tilde{\epsilon}_{Q_{2}}+\log(1+Y_{Q_{1}})\star
\left(K_{sl(2)}^{Q_{1}Q_{2}}+2s\star K_{vx}^{Q_{1}-1,Q_{2}}\right)\nonumber \\
 &  & +\sum_{\alpha=\pm}\biggr[\log\biggl(1+Y_{v|1}^{\alpha}\biggr)
\star s\,\hat{\star}K_{yQ_{2}}+\log(1+Y_{v|Q_{2}-1}^{\alpha})\star s-\log
\frac{1-Y_{-}^{\alpha}}{1-Y_{+}^{\alpha}}\hat{\star}s\star K_{vx}^{1Q_{2}}\nonumber
\\
 &  & \,\,\,\,\,\,\,\,\,\,\,\,\,\,\,\,\,\,\,\,\,+\frac{1}{2}\log
\frac{1-\frac{1}{Y_{-}^{\alpha}}}{1-\frac{1}{Y_{+}^{\alpha}}}
\hat{\star}K_{Q_{2}}+\frac{1}{2}\log(1-\frac{1}{Y_{-}^{\alpha}})
(1-\frac{1}{Y_{+}^{\alpha}})\hat{\star}K_{yQ_{2}}\biggl]\,, \label{simplTBA1}\\
\log Y_{-}^{\alpha}Y_{+}^{\alpha} & = & -\log(1+Y_{Q_{2}})\star K_{Q_{2}}+
2\log(1+Y_{Q_{2}})\star K_{xv}^{Q_{2}1}\star s+2\log\frac{1+Y_{v|1}^{
\alpha}}{1+Y_{w|1}^{\alpha}}\star s\,,  \label{simplTBA-}\\
\log\frac{Y_{+}^{\alpha}}{Y_{-}^{\alpha}} & = & 
\log(1+Y_{Q_{2}})\star K_{Q_{2}y}\,, \label{simplTBA-+} \\
\log Y_{v|M}^{\alpha} & = & -\log(1+Y_{M+1})\star s+I_{MN}\log(1+Y_{v|N}^{
\alpha})\star s+\delta_{M1}\log\frac{1-Y_{-}^{\alpha}}{1-Y_{+}^{\alpha}}
\hat{\star}s\,,  \label{simplTBAv}\\
\log Y_{w|M}^{\alpha} & = & I_{MN}\log(1+Y_{w|N}^{\alpha})\star s+
\delta_{M1}\log\frac{1-\frac{1}{Y_{-}^{\alpha}}}{1-
\frac{1}{Y_{+}^{\alpha}}}\hat{\star}s \,,
\label{simplTBAw}
\end{eqnarray}
where in the convolution $\hat{\star}$ we integrate over the interval
$[-2,2]$ only. To conform with part of the literature, we have renamed
some kernels $K_{vx}^{MQ}=K_{v\vert M\, Q}$, $K_{xv}^{QM}=K_{Q\, v\vert M}$,
$K_{yQ}=K_{y\vert-\, Q}+K_{y\vert+\, Q}$, $K_{Qy}=K_{Q\, y\vert-}-K_{Q\,
y\vert+}$.
The ground-state energy is given by summing the contributions of the
massive nodes only:
\[
E_{0}(L)=-\sum_{Q=1}^{\infty}\int\frac{du}{2\pi}\partial_{u}\tilde{p}_{Q}
\log(1+Y_{Q})\,.
\label{eq:AdSEngergy}
\]

Evidently, as in the case of the $O(4)$ model, the chemical potentials
and so the twists completely disappear from the simplified equations:
They show up only in the asymptotics of the $Y_{w\vert N}$ functions,
as $\lim_{N\to\infty}\log Y_{w\vert N}=-\mu_{w\vert N}$.  It follows
that the $Y$-system relations are not modified by the twists, as was
supposed in \cite{Gromov:2010dy}.
Equations (\ref{simplTBAw})-(\ref{eq:AdSEngergy}) together with the 
asymptotic prescription give the complete solution for the 
finite-size energy of the twisted 
AdS/CFT model for any coupling $g$. We now  check this solution against
LO and NLO L\"uscher corrections.

\subsection{Asymptotic expansion}

We now expand the simplified TBA equations to leading and next-to-leading order.
We expand any $Y$-functions as 
\[
Y=\mathcal{Y}(1+y+\dots) \,.
\]
We solve iteratively these equations similarly to the $O(4)$ case:
At leading order, all the massive nodes $Y_{Q}$ are exponentially small, which
splits the $Y$-system into two independent subsystems which have constant
asymptotic solutions. These constant values then determine the LO
exponentially small expressions for $\mathcal{Y}_{Q}$.
At NLO, we obtain linear integral equations for the $y$ corrections
of the two subsystems, whose initial values are provided by the asymptotic
$\mathcal{Y}_{Q}$ functions. The solution of the linearized equations
determine the NLO correction for the massive nodes $y_{Q}$, which
provides the NLO energy correction.

\subsubsection{Leading-order expansion}

At LO, the massive $\mathcal{Y}_{Q}$ functions are exponentially
small, and we can neglect the convolutions involving all $\log(1+Y_{Q})$.
The magnonic
$\mathcal{Y}_{\pm}^{\alpha},\,\mathcal{Y}_{v|M}^{\alpha},\,\mathcal{Y}_{w|N}^{
\alpha}$
functions are constants. From (\ref{simplTBA-+}), we see that 
$\mathcal{Y}_{+}^{\alpha}=\mathcal{Y}_{-}^{\alpha}$. It then follows
from (\ref{simplTBAv}) and (\ref{simplTBAw}) that
the equations for
$\mathcal{Y}_{v\vert M}^{\alpha}$ and $\mathcal{Y}_{w\vert N}^{\alpha}$ 
are the same as those for one of the wings of the
$O(4)$ model (\ref{eq:O4simpleTBAM}). From the asymptotic behavior, we see that
the solution for $v\vert M$ is the same as in the undeformed model, while
the solution for $w\vert N$ is that of the deformed model:
\begin{equation}
\mathcal{Y}_{v|M}^{\alpha}=M(M+2)\,, \qquad
\mathcal{Y}_{w|N}^{+}=[N]_{q}[N+2]_{q}\,,\qquad
\mathcal{Y}_{w|N}^{-}=[N]_{\dot{q}}[N+2]_{\dot{q}}\,.
\label{eq:AdSasymptotics1}
\end{equation}
Since $1\star s=\frac{1}{2}$, the equations (\ref{simplTBA-}) for 
$\mathcal{Y}_{\pm}^{\alpha}$ 
can be solved as 
\[
\mathcal{Y}_{+}^{\alpha}=\mathcal{Y}_{-}^{\alpha}=\sqrt{\frac{1+\mathcal{Y}_{v|1
}
^{\alpha}}{1+\mathcal{Y}_{w|1}^{\alpha}}}=
\frac{2}{[2]_{\alpha}}\,, 
\label{eq:AdSasymptotics2}
\]
where we have further streamlined the notation by defining
\[
[n]_{+}=[n]_{q}\,, \qquad
[n]_{-}=[n]_{\dot{q}} \,.
\]
The sign in (\ref{eq:AdSasymptotics2}) can be fixed by the last equation in (\ref{eq:AdSTBA}), 
and is consistent with the vanishing of the ground-state energy 
(\ref{result}) in the undeformed ($q, \dot q \rightarrow 1$) limit.
We now use that $1\hat{\star}K_{yQ}=1$ (see (6.12) in 
\cite{Arutyunov:2009ur}) to write
\begin{equation}
\log\mathcal{Y}_{Q}=-L\,\tilde{\epsilon}_{Q}+\frac{1}{2}\sum_{\alpha=\pm}
\left[\log\biggl(1+\mathcal{Y}_{v|1}^{\alpha}\biggr)+\log(1+\mathcal{Y}_{v|Q-1}^
{
\alpha})+\log\biggl(1-\frac{1}{\mathcal{Y}_{-}^{\alpha}}\biggr)\biggl(1-\frac{1}
{
\mathcal{Y}_{+}^{\alpha}}\biggl)\right]\,.
\label{TBAQLO}
\end{equation}
Using the asymptotic solution (\ref{eq:AdSasymptotics1}), 
(\ref{eq:AdSasymptotics2}), we obtain the leading-order result for $Y_{Q}$
\begin{equation}
    \mathcal{Y}_{Q}=(2-[2]_{q})(2-[2]_{\dot{q}})Q^{2}e^{-L\tilde{
\epsilon}_{Q}(\tilde{p})} \,.
\label{TBAQLOresult}
\end{equation}
Substituting back into the energy formula (\ref{eq:AdSEngergy}), the LO correction
reads as 
\begin{equation}
E_{0}(L)\simeq E_{0}^{(1)}(L) =
-\sum_{Q=1}^{\infty}\int\frac{d\tilde{p}}{2\pi}\mathcal{Y}_{Q}=
-(2-[2]_{q})(2-[2]_{\dot{q}})\sum_{Q=1}^{\infty}Q^{2}\int\frac{d\tilde{p}}{
2\pi}
e^{-L\tilde{\epsilon}_{Q}(\tilde{p})}\,,
\label{result}
\end{equation}
which agrees with the result (\ref{eq:E1LuscherAdS}) that we obtained from the
L\"uscher calculation.

\subsubsection{NLO expansion}
\label{sec:NLOexpansion}

Expanding the energy formula (\ref{eq:AdSEngergy}) to NLO, we obtain
\begin{equation}
E_{0}(L)=-\sum_{Q=1}^{\infty}\int\frac{d\tilde{p}}{2\pi}\log(1+Y_{Q})\simeq-
\sum_{Q=1}^{\infty}\int\frac{d\tilde{p}}{2\pi}\mathcal{Y}_{Q}(1+y_{Q})+
\sum_{Q=1}^{\infty}\int\frac{d\tilde{p}}{2\pi}\frac{1}{2}\mathcal{Y}_{Q}^{2}
\,,
\label{energy}
\end{equation}
The quadratic term nicely reproduces our previous result 
(\ref{eq:E21LuscherAdS}) for $E_{0}^{(2,1)}$, since using again 
(\ref{TBAQLOresult}) gives
\[
E_{0}^{(2,1)}(L)=
\sum_{Q=1}^{\infty}\int\frac{d\tilde{p}}{2\pi}\frac{1}{2}\mathcal{Y}_{Q}^{2}
= \frac{1}{2}(2-[2]_{q})^{2}(2-[2]_{\dot{q}})^{2}
\sum_{Q=1}^{\infty}Q^{4}\int\frac{d\tilde{p}}{2\pi}e^{-2L
\tilde{\epsilon}_{Q}(\tilde{p})}\,.
\label{E021}
\]
In order to evaluate
\[
E_{0}^{(2,2)}(L) = 
-\sum_{Q=1}^{\infty}\int\frac{d\tilde{p}}{2\pi} \mathcal{Y}_{Q} 
y_{Q} \,,
\label{energy2}
\]
we must first calculate $y_{Q}$. This
will be given by the solution of the following linearized set of TBA
equations: \footnote{We note that in \cite{Arutyunov:2010gb} there is 
an erroneous term in Eq. (2.7): $-Y^0_Q\star s$ should be instead 
$-Y^0_Q\star K_{Q}$, as in (\ref{linTBAy++y-}).}
\begin{eqnarray}
y_{Q_{2}} & = & \mathcal{Y}_{Q_{1}}\star\left(K_{sl(2)}^{Q_{1}Q_{2}}+2s\star
K_{vx}^{Q_{1}-1,Q_{2}}\right)
+\sum_{\alpha=\pm}\left[A_{v\vert1}^{\alpha}y_{v|1}^{\alpha}\star
s\,\hat{\star}K_{yQ_{2}}
+A_{v\vert Q_{2}-1}^{\alpha}y_{v|Q_{2}-1}^{\alpha}\star s\right.\nonumber \\
 &  &
\,\,\,\,\,\,\,\,\,\,\,\,\,\,\,\,\,\,\,\,\,\,\,\left.-\frac{y_{-}^{\alpha}-y_{+}^
{
\alpha}}{1-\frac{1}{\mathcal{Y}_{+}^{\alpha}}}\hat{\star}s\star
K_{vx}^{1Q_{2}}+\frac{y_{-}^{
\alpha}-y_{+}^{\alpha}}{2(\mathcal{Y}_{+}^{\alpha}-1)}\hat{\star}K_{Q_{2}}+\frac{y_{
-}^{
\alpha}+y_{+}^{\alpha}}{2(\mathcal{Y}_{+}^{\alpha}-1)}\hat{\star}
K_{yQ_{2}}\right]\,, \label{linTBAQ}\\
y_{+}^{\alpha}+y_{-}^{\alpha} & = &
2\left(A_{v\vert1}^{\alpha}y_{v|1}^{\alpha}-A_{w\vert1}^{\alpha}y_{
w|1}^{\alpha}\right)\star s-\mathcal{Y}_{Q_{2}}\star K_{Q_{2}}+2\mathcal{Y}_{Q_{2}}\star
K_{xv}^{Q_{2}1}
\star s\,, \label{linTBAy++y-}\\
y_{+}^{\alpha}-y_{-}^{\alpha} & = & \mathcal{Y}_{Q_{2}}\star K_{Q_{2}y}\,,
\label{linTBAy+-y-}\\
y_{v|M}^{\alpha} & = & \left(A_{v\vert M-1}^{\alpha}y_{v|M-1}^{\alpha}+A_{v\vert
M+1}^{\alpha}y_{v|M+1}^{
\alpha}\right)\star s-\mathcal{Y}_{M+1}\star
s+\delta_{M1}\frac{y_{-}^{\alpha}-y_{+}^{
\alpha}}{1-\frac{1}{\mathcal{Y}_{+}^{\alpha}}}\hat{\star}s\,, 
~~~~\label{linTBAyvM}\\
y_{w|N}^{\alpha} & = & \left(A_{w\vert N-1}^{\alpha}y_{w|N-1}^{\alpha}+A_{w\vert
N+1}^{\alpha}y_{w|N+1}^{
\alpha}\right)\star s+\delta_{N1}\frac{y_{+}^{\alpha}-y_{-}^{\alpha}}{1-
\mathcal{Y}_{+}^{\alpha}}\hat{\star}s\,,  \label{linTBAywN}
\end{eqnarray}
where 
\[
A_{v\vert M}^{\alpha}=\frac{\mathcal{Y}_{v\vert
M}^{\alpha}}{1+\mathcal{Y}_{v\vert M}^{\alpha}}
=\frac{M(M+2)}{(M+1)^{2}}\,, \qquad
A_{w\vert N}^{\alpha}=\frac{\mathcal{Y}_{w\vert
N}^{\alpha}}{1+\mathcal{Y}_{w\vert N}^{\alpha}}
=\frac{[N]_{\alpha}[N+2]_{\alpha}}{[N+1]^{2}_{\alpha}} \,.
\]

We start with the equation (\ref{linTBAywN}) for $y_{w\vert N}^{\alpha}$. The
difference
between $\alpha=+$ and $\alpha=-$ is only in the asymptotics 
(\ref{eq:AdSasymptotics1}), (\ref{eq:AdSasymptotics2}). 
Since one equation 
can be obtained from the other by interchanging 
$q\leftrightarrow\dot{q}$,
we do not write out explicitly the $\alpha$ index. Replacing
$y_{+}-y_{-}$ in (\ref{linTBAywN}) with the contributions from the 
massive nodes (\ref{linTBAy+-y-}), and using
the explicit form of the asymptotic solution, we obtain an equation
similar to the one for the $O(4)$ case: 
\begin{equation}
y_{w\vert N}=\left(\frac{[N-1][N+1]}{[N]^{2}}y_{w\vert N-1}+\frac{[N+1]
[N+3]}{[N+2]^{2}}y_{w\vert N+1}\right)\star s+\delta_{N1}c_{1}\star s\,,
\end{equation}
where 
\begin{equation}
c_{1}=\frac{[2]}{[2]-2}\mathcal{Y}_{Q}\star\hat{K}_{Qy}\,, \qquad
\hat{K}_{Qy}(u,v)
=K_{Qy}(u,v)\left(\Theta(v+2)-\Theta(v-2)\right)\,,
\label{c1}
\end{equation}
and $\Theta(v)$ is the standard unit step function.
We solve the difference equation in Fourier space. We use that
$\tilde{s}=(2\cosh\frac{\omega}{g})^{-1}
=(t+t^{-1})^{-1}$
where $t\equiv e^{-\frac{|\omega|}{g}}$. The solution which decreases
for large $N$ (to respect the asymptotics of $Y_{w\vert N}$) and
is compatible with the $\delta_{N,1}$ term is 
\begin{equation}
\tilde{y}_{w\vert N}=\frac{\tilde{c}_{1}t}{[2]}\left(\frac{[N+1]}{[N]}t^{N-1}-
\frac{[N+1]}{[N+2]}t^{N+1}\right)\,.
\end{equation}

We now analyze the equation (\ref{linTBAyvM}) for $y_{v\vert 
M}^{\alpha}$. This difference equation
is not the same as for the undeformed $O(4)$ model, as it has inhomogeneous
terms, 
\begin{equation}
y_{v\vert M}=\left(\frac{(M-1)(M+1)}{M^{2}}y_{v\vert M-1}+\frac{(M+1)(M+3)}
{(M+2)^{2}}y_{v\vert M+1}\right)\star s-\mathcal{Y}_{M+1}\star
s+\delta_{M1}c_{2}\star s\,,
\end{equation}
where 
\begin{equation}
c_{2}=\frac{2}{[2]-2}\mathcal{Y}_{Q}\star\hat{K}_{Qy}\label{c2}\,.
\end{equation}
Taking the Fourier transform, we obtain the difference equation
\[
(t+t^{-1})\tilde{y}_{v\vert M}=\frac{(M-1)(M+1)}{M^{2}}\tilde{y}_{v\vert M-1}+
\frac{(M+1)(M+3)}{(M+2)^{2}}\tilde{y}_{v\vert M+1}-\tilde{\mathcal{Y}}_{M+1}+
\delta_{M1}\tilde{c}_{2}\,.
\]
The general solution with two arbitrary parameters $A_{1}$ and $A_{2}$
reads as 
\begin{eqnarray}
\tilde{y}_{v\vert M} & = &
\left(\frac{M+1}{M}t^{M-1}-\frac{M+1}{M+2}t^{M+1}\right)
\left(A_{1}-\sum_{k=1}^{M}\frac{\tilde{\mathcal{Y}}_{k+1}t^{-k-2}\left(t^{-2}
k-k-2
\right)}{\left(t^{-2}-1\right)^{3}(k+1)}\right)\nonumber \\
 &+& \left(\frac{M+1}{M}t^{1-M}-\frac{M+1}{M+2}t^{-1-M}\right)\left(A_{2}-
\sum_{k=1}^{M}\frac{\tilde{\mathcal{Y}}_{k+1}t^{k-2}\left(t^{-2}(k+2)-k\right)}
{\left(t^{-2}-1\right)^{3}(k+1)}\right)\,.
\label{tildeyvM}
\end{eqnarray}
The parameters can be fixed from $\lim_{M\to\infty}\tilde{y}_{v\vert M}=0$
and from the $M=1$ term as 
\begin{equation}
A_{1}=t^{2}\left(\frac{t^{-1}}{2}\tilde{c}_{2}-A_{2}\right)\,, \qquad 
A_{2}=\sum_{k=1}^{\infty}
\frac{\tilde{\mathcal{Y}}_{k+1}t^{k-2}\left(t^{-2}(k+2)-k\right)}
{\left(t^{-2}-1\right)^{3}(k+1)}\,.
\end{equation}


The NLO hybrid equation for $y_{Q_{2}}$ is (\ref{linTBAQ}); we plug into it the equations
(\ref{linTBAy++y-}) and (\ref{linTBAy+-y-}), and obtain
\ba
&&y_{Q_{2}}=
{\cal Y}_{Q_{1}}\star\left(K^{Q_{1}Q_{2}}_{sl(2)}+2s\star
K_{vx}^{Q_{1}-1,Q_{2}}\right)+\sum_{\a=1,2}\left[\frac{A_{v|1}^{\a}}{1-\frac{1}{{\cal
Y}_{+}^{\a}}}y_{v|1}^{\a}\star
s\hat{\star}K_{yQ_{2}}-\frac{A_{w|1}^{\a}}{{\cal
Y}_{+}^{\a}-1}y_{w|1}^{\a}\star s\hat{\star}K_{yQ_{2}}\right.\nn\\
&&\left.+\frac{{\cal Y}_{Q_{1}}\star K_{xv}^{Q_{1}1}}{{\cal Y}_{+}^{\a}-1}\star
s\hat{\star }K_{yQ_{2}}+\frac{{\cal Y}_{Q_{1}}\star K_{Q_{1}y}}{1-\frac{1}{{\cal
Y}_{\pm}^{\a}}}\hat{\star}s\star K_{vx}^{1Q_{2}}-\frac{{\cal Y}_{Q_{1}}\star
K_{Q_{1}y}}{2({\cal Y}_{\pm}^{\a}-1)}\hat{\star}K_{Q_{2}}-\frac{{\cal Y}_{Q_{1}}\star
K_{Q_{1}}}{2({\cal Y}_{+}^{\a}-1)}\hat{\star }K_{yQ_{2}}\right.\nn\\
&&\left.+A_{v|Q_{2}-1}^{\a}y_{v|Q_{2}-1}^{\a}\star s\right]\,,
\label{yQ}
\ea 
Since $y_{v\vert1}$ and $y_{w\vert1}$ can be expressed in terms of $\mathcal{Y}_{Q}$,
we see that the solution for $y_{Q_{2}}$ has the general form 
\[
y_{Q_{2}}=\mathcal{Y}_{Q_{1}}\star 
K_{sl(2)}^{Q_{1}Q_{2}}+\mathcal{Y}_{Q_{1}}\star\mathcal{M}^{Q_{1}Q_{2}}\,.
\label{eq:genform}
\]

Consider the first term $\mathcal{Y}_{Q_{1}}\star K_{sl(2)}^{Q_{1}Q_{2}}$.
It is easy to see that its contribution to the integrand in the 
energy formula (\ref{energy2})
\[
\mathcal{Y}_{Q_{2}}y_{Q_{2}}=\mathcal{Y}_{Q_{2}}\,(\mathcal{Y}_{Q_{1}}\star 
K_{sl(2)}^{Q_{1}Q_{2}})\,,
\]
with $\mathcal{Y}_{Q_{2}}$ given by (\ref{TBAQLOresult}), 
matches with the ``scalar part'' of the integrand of the L\"uscher 
correction $E_{0}^{(2,2)}$ in (\ref{NLOmatrix}). We now proceed to 
analyze the remaining contribution in (\ref{eq:genform}), and show
that it gives the ``matrix part'' of the integrand of the L\"uscher 
correction.

\subsubsection{NLO TBA correction: the case $Q_{1}=Q_{2}=1$}

To warm up, let us evaluate the NLO correction 
for the $Q_{1}=Q_{2}=1$ case; thus, we calculate $\mathcal{M}^{11}$.
In so doing, we can freely put $\mathcal{Y}_{Q_{2}}=0$ for $Q_{2}>1$. The
corresponding solutions read as 
\begin{eqnarray}
y_{w\vert1} & = & \frac{[2]}{[2]-2}\mathcal{Y}_{1}\star K_{1y}
\hat{\star}\left(K_{1}-\frac{1}{[3]}K_{3}\right)\,, \label{eq:M1sltn}\\
y_{v\vert1} & = & \frac{2}{[2]-2}\mathcal{Y}_{1}\star K_{1y}
\hat{\star}\left(K_{1}-\frac{1}{3}K_{3}\right)\,, \nonumber \\
y_{+}-y_{-} & = & \mathcal{Y}_{1}\star K_{1y}\,, \nonumber \\
y_{+}+y_{-} & = & \frac{2}{[2]-2}\mathcal{Y}_{1}\star K_{1y}
\hat{\star}\biggr(\left(\frac{3}{2}-\frac{[3]}{[2]}\right)K_{1}-
\left(\frac{1}{2}-\frac{1}{[2]}\right)K_{3}\biggr)\star s-
\mathcal{Y}_{1}\star K_{1}+2\mathcal{Y}_{1}\star K_{xv}^{11}\star s\,.
\nonumber 
\end{eqnarray}
It is convenient to substitute these solutions directly into (\ref{linTBAQ}), i.e.,
\[
y_{1}=\mathcal{Y}_{1}\star
K_{sl(2)}^{11}+\sum_{\alpha=\pm}\left[A_{v\vert1}y_{v|1}^{
\alpha}\star s\,\hat{\star}K_{y1}-\frac{y_{-}^{\alpha}-y_{+}^{\alpha}}{1-
\frac{1}{\mathcal{Y}_{+}^{\alpha}}}\hat{\star}s\star K_{vx}^{11}+
\frac{y_{-}^{\alpha}-y_{+}^{\alpha}}{2(\mathcal{Y}_{+}^{\alpha}-1)}\hat{\star}K_{1}+
\frac{y_{-}^{\alpha}+y_{+}^{\alpha}}{2(\mathcal{Y}_{+}^{\alpha}-1)}\hat{\star}K_{y1}
\right]\,.
\]
Using the explicit form of the asymptotic solutions, one can see that
the terms involving the convolution with $K_{3}$ completely cancel.
Exploiting further that $K_{1y}\hat{\star}K_{1}=K_{xv}^{11}$ (which 
can be shown using relations from Sec. 6 in \cite{Arutyunov:2009ur}), we arrive
at
\begin{eqnarray}
 &  & y_{1}=\mathcal{Y}_{1}\star K_{sl(2)}^{11}+\sum_{\alpha=\pm}\biggl[\frac{[2]_{
\alpha}}{2([2]_{\alpha}-2)}\left(\mathcal{Y}_{1}\star K_{1y}\hat{\star}K_{1}+
\mathcal{Y}_{1}\star K_{1}\hat{\star}K_{y1}-2\mathcal{Y}_{1}\star K_{xv}^{11}
\star s\hat{\star}K_{y1}\right)\nonumber \\
 &  &
\,\,\,\,\,\,\,\,\,\,\,\,\,\,\,\,\,\,\,\,\,\,\,\,\,\,\,\,\,\,\,\,\,\,\,\,\,\,\,
\,\,\,\,\,\,\,\,\,\,\,\,\,\,\,\,\,\,\,\,\,+\frac{[3]_{\alpha}-3}{([2]_{\alpha}
-2)^{2}}
\mathcal{Y}_{1}\star K_{xv}^{11}\star
s\,\hat{\star}K_{y1}-\frac{2}{[2]_{\alpha}-2}
\mathcal{Y}_{1}\star K_{1y}\hat{\star}s\star K_{vx}^{11}\biggr]\,.
\end{eqnarray}
This expression further simplifies to 
\[
y_{1}=\mathcal{Y}_{1}\star K_{sl(2)}^{11}+\sum_{\alpha=\pm}\biggl[\frac{[2]_{
\alpha}}{2([2]_{\alpha}-2)}\mathcal{Y}_{1}\star\left(K_{1y}\hat{
\star}K_{1}+K_{1}\hat{\star}K_{y1}\right)+\frac{2}{[2]_{\alpha}-2}
\mathcal{Y}_{1}\star\left(K_{xv}^{11}\star s\hat{\star}K_{y1}-K_{1y}
\hat{\star}s\star K_{vx}^{11}\right)\biggr]\,.
\label{eq:y1intermed}
\]
In the second term, using $K_{1y}\hat{\star}K_{1}=K_{xv}^{11}$ and
$K_{vx}^{11}=K_{1}\hat{\star}K_{y1}$, we can write
\[
K_{xv}^{11}\star s\hat{\star}K_{y1}-K_{1y}\hat{\star}s\star K_{vx}^{11}=
K_{1y}\hat{\star}(K_{1}\star s-s\star K_{1})\hat{\star}K_{y1}=0\,,
\]
as both $s$ and $K_{1}$ depend on the differences of their arguments,
and therefore their convolution is commutative. In the previous term 
in (\ref{eq:y1intermed}), we can obtain 
\begin{eqnarray}
K_{1y}\hat{\star}K_{1}+K_{1}\hat{\star}K_{y1} & = & \frac{1}{2\pi i}
\partial_{u_{1}}\log\left(\frac{x_{1}^{-}-x_{2}^{+}}{x_{1}^{-}-1/x_{2}^{+}}
\frac{x_{1}^{+}-1/x_{2}^{-}}{x_{1}^{+}-x_{2}^{-}}\frac{u_{1}-u_{2}-2i/g}
{u_{1}-u_{2}+2i/g}\right)\nonumber \\
 & = & \frac{1}{\pi
i}\partial_{u_{1}}\log\left(\frac{x_{1}^{-}-x_{2}^{+}}{x_{1}^{+}-x_{2}^{-}}
\sqrt{\frac{x_{1}^{+}}{x_{1}^{-}}}\right)=-\frac{1}{\pi i}\partial_{u_{1}}
\log a_{1}(u_{1},u_{2})\,,
\label{a1identity}
\end{eqnarray}
where we have used identities from Sec. 6 
in \cite{Arutyunov:2009ur} and Eq. (3.7) in \cite{Arutyunov:2009ux}, 
and we have recalled the definition in (\ref{eq:a1def}) of $a_{1}$.
The final expression for the $Q=Q'=1$ contribution to the energy
(\ref{energy2}) is therefore given by
\[
y_{1}=\mathcal{Y}_{1}\star
K_{sl(2)}^{11}+\mathcal{Y}_{1}\star\mathcal{M}^{11}\,, \qquad
\mathcal{M}^{11}=\frac{1}{2\pi i}\partial_{u_{1}}\log a_{1}(u_{1},u_{2})
\sum_{\alpha}\frac{[2]_{\alpha}}{2-[2]_{\alpha}}\,,
\]
which completely reproduces the result (\ref{NLOLuschermatrix}) obtained
directly from the
L\"uscher correction.

\subsubsection{NLO TBA correction for any $Q_{1},Q_{2}$}
\label{sec:TBAQQ'}

We now consider the general case.  Let us recall the result (\ref{yQ})
for $y_{Q_{2}}$
\ba
&&y_{Q_{2}}=
{\cal Y}_{Q_{1}}\star\left(K^{Q_{1}Q_{2}}_{sl(2)}+2s\star
K_{vx}^{Q_{1}-1,Q_{2}}\right)+\sum_{\a=1,2}\left[\frac{A_{v|1}^{\a}}{1-\frac{1}{{\cal
Y}_{+}^{\a}}}y_{v|1}^{\a}\star
s\hat{\star}K_{yQ_{2}}-\frac{A_{w|1}^{\a}}{{\cal
Y}_{+}^{\a}-1}y_{w|1}^{\a}\star s\hat{\star}K_{yQ_{2}}\right.\nn\\
&&\left.+\frac{{\cal Y}_{Q_{1}}\star K_{xv}^{Q_{1}1}}{{\cal Y}_{+}^{\a}-1}\star
s\hat{\star }K_{yQ_{2}}+\frac{{\cal Y}_{Q_{1}}\star K_{Q_{1}y}}{1-\frac{1}{{\cal
Y}_{\pm}^{\a}}}\hat{\star}s\star K_{vx}^{1Q_{2}}-\frac{{\cal Y}_{Q_{1}}\star
K_{Q_{1}y}}{2({\cal Y}_{\pm}^{\a}-1)}\hat{\star}K_{Q_{2}}-\frac{{\cal Y}_{Q_{1}}\star
K_{Q_{1}}}{2({\cal Y}_{+}^{\a}-1)}\hat{\star }K_{yQ_{2}}\right.\nn\\
&&\left.+A_{v|Q_{2}-1}^{\a}y_{v|Q_{2}-1}^{\a}\star s\right]\,,
\label{yQQ'}
\ea 
and analyze it term by term. Since we have already checked in Section (\ref{sec:NLOexpansion}) the matching of the first term with the scalar part of the L\"uscher result, we start by  considering the second term of (\ref{yQQ'}), which can be 
rewritten as 
\be
2{\cal Y}_{Q_{1}}\star s\star K_{vx}^{Q_{1}-1,Q_{2}}=
2{\cal Y}_{Q_{1}}\star K_{Q_{1}-1}\star s\hat{\star} K_{yQ_{2}}
+2{\cal Y}_{Q_{1}}\star \sum_{j=0}^{Q_{1}-2}K_{Q_{2}-Q_{1}+2j+1}\star s\,,
\label{eq:rewrite}
\ee
where we used the property $s\star K_{Q} = K_{Q}\star s$, valid for any $Q$.
Now we consider the terms in the square
brackets of (\ref{yQQ'}), again suppressing the index $\a$.  Using the solution (\ref{tildeyvM}) for
$M=1$ and taking its inverse Fourier transform, we can express the first term as
\ba
\frac{A_{v|1}}{1-\frac{1}{{\cal Y}_{+}}}y_{v|1}\star s\hat{\star
}K_{yQ_{2}}&=&\frac{3}{2-[2]}\Bigg\{\frac{{\cal Y}_{Q_{1}}\star
K_{Q_{1}y}}{2-[2]}\hat{\star }\left(\frac{K_{3}}{3}-K_{1}\right)\nn\\
&+&\frac{1}{3}\frac{{\cal Y}_{Q_{1}}}{Q_{1}}\star
\left[(Q_{1}-1)K_{Q_{1}+1}-(Q_{1}+1)K_{Q_{1}-1}\right]\Bigg\}\star s\hat{\star
}K_{yQ_{2}}\,,
\label{yv1}
\ea 
where the term in the second line can be rewritten, by using  the identity $\left( 
K_{n+1}+K_{n-1}+n \delta_{n,\pm1}\delta \right)\star s = K_{n}$, as
\[\nn \\
-\frac{2}{2-[2]}{\cal Y}_{Q_{1}}\star K_{Q_{1}-1}
\star s\hat{\star}K_{yQ_{2}}+\frac{Q_{1}-1}{Q_{1}(2-[2])}
{\cal Y}_{Q_{1}}\star K_{Q_{1}}\hat{\star}K_{yQ_{2}}\,.\]
The $K_{3}$ contribution in the first line of (\ref{yv1}) cancels, as we have already seen in
the $Q_{1}=Q_{2}=1$ case, with the successive term in (\ref{yQQ'})
\ba
-\frac{A_{w|1}}{{\cal Y}_{+}-1}y_{w|1}\star s\hat{\star
}K_{yQ_{2}}=-\frac{1}{(2-[2])^{2}}
{\cal Y}_{Q_{1}}\star K_{Q_{1}y}\hat{\star
}\left(K_{3}-[3]K_{1}\right)\star s\hat{\star }K_{yQ_{2}}~~\,,
\ea
while the terms with $K_{1}$ give
\be
\frac{[3]-3}{(2-[2])^{2}}{\cal Y}_{Q_{1}}\star K_{Q_{1}y}\hat{\star }K_{1}\star
s\hat{\star }K_{yQ_{2}}\,.
\label{K1sKyQ}
\ee
Summing this contribution to the first two terms in the second line of (\ref{yQQ'}), we obtain
\ba
&&\frac{2}{2-[2]}\left({\cal Y}_{Q_{1}}\star K_{Q_{1}y}\hat{\star }s\star
K_{1}\hat{\star }K_{yQ_{2}}-{\cal Y}_{Q_{1}}\star K_{Q_{1}y}\hat{\star }K_{1}\star
s\hat{\star }K_{yQ_{2}}\right)\nn\\
&&+\frac{[2]}{2-[2]}{\cal Y}_{Q_{1}}\star K_{Q_{1}-1}\star s\hat{\star
}K_{yQ_{2}}+\frac{2}{2-[2]}{\cal Y}_{Q_{1}}\star K_{Q_{1}y}\hat{\star }s\star K_{Q_{2}-1}\,,
\label{bracket}
\ea
where we used the identities $K_{vx}^{1Q}=K_{1}\hat{\star }K_{yQ}+K_{Q-1}$ and
$K_{xv}^{Q1}=K_{Qy}\hat{\star }K_{1}+K_{Q-1}$. \footnote{The latter identity 
is reported in footnote 4 of \cite{Arutyunov:2009ax}; the former can 
be derived analogously using equations (6.19) and (6.39) in \cite{Arutyunov:2009ur}.
The same equations, together with (6.14), can also be used to obtain 
(\ref{eq:rewrite}).}
As already noticed for the case
$Q_{1}=Q_{2}=1$, the first line in the expression above vanishes because $K_{1}\star
s=s\star K_{1}$.
The successive two terms in the second line of (\ref{yQQ'}) give
\ba
-\frac{[2]}{2\pi(2-[2])}{\cal Y}_{Q_{1}}\star i\partial_{u_{1}}\log
a_{1}^{Q_{1}Q_{2}}(u_{1},u_{2}) \,,
\label{eq:othercontribs}
\ea
where we used the identity (\ref{a1identity}) generalized for any 
$Q_{1},Q_{2}$,
\be
K_{Q_{1}y}\hat{\star }K_{Q_{2}}+K_{Q_{1}}\hat{\star }K_{yQ_{2}}=\frac{1}{\pi
i}\partial_{u_{1}}\log\left(\frac{x_{1}^{-Q_{1}}-x_{2}^{+Q_{2}}}{x_{1}^{+Q_{1}}-x_{2}^{-Q_{2}}}
\sqrt{\frac{x_{1}^{+Q_{1}}}{x_{1}^{-Q_{1}}}}\right)=-\frac{1}{\pi i}\partial_{u_{1}}\log
a_{1}^{Q_{1}Q_{2}}(u_{1},u_{2})\,.
\label{a1QQ'}
\ee
Let us turn to the last and most complicated term. Using the inverse
Fourier transform of (\ref{tildeyvM}) for $M=Q_{2}-1$, we can write
it as follows
\ba
A_{v|Q_{2}-1}y_{v|Q_{2}-1}\star s&=&\frac{{\cal
Y}_{Q_{1}}}{Q_{1}Q_{2}}\star
\sum_{k=0}^{Q_{1}-1}k(k-Q_{1})\left[(Q_{2}+1)K_{Q_{2}-Q_{1}+2k-1}-(Q_{2}-1)K_{Q_{2}-Q_{1}+2k+1}\right]\star
s\nn\\
&+&\frac{{\cal Y}_{Q_{1}}\star K_{Q_{1}y}}{Q_{2}(2-[2])}\hat{\star
}\left[(Q_{2}-1)K_{Q_{2}+1}-(Q_{2}+1)K_{Q_{2}-1}\right]\star s\,,
\label{yvQ-1}
\ea
where the second line can be expressed as
\be
-\frac{2}{2-[2]}{\cal Y}_{Q_{1}}\star K_{Q_{1}y}\hat{\star}s\star K_{Q_{2}-1}
+\frac{Q_{2}-1}{Q_{2}(2-[2])}{\cal Y}_{Q_{1}}\star K_{Q_{1}y}\hat{\star}K_{Q_{2}}\,.
\ee
Now, taking into account that in summing over 
$\alpha$ the first term in (\ref{yvQ-1}) gets a factor 2 and 
the other terms get similar coefficients with $q \rightarrow \dot{q}$,
we can sum all the contributions above to get the solution for
$y_{Q_{2}}$ for generic values of $Q_{1},Q_{2}$:
\ba
y_{Q_2}&=&{\cal Y}_{Q_1}\star \Bigg\{K_{sl(2)}^{Q_1Q_2}+2\sum_{j=0}^{Q_1-2}K_{Q_2-Q_1+2j+1}\star s-\frac{\partial_{u_{1}}\log a_{1}^{Q_1Q_2}(u_{1},u_{2})}{\pi i}\nn\\
&-&\sum_{\alpha=\pm}\left[\frac{\partial_{u_{1}}\log
a_{2}^{Q_1Q_2}(u_{1},u_{2})}{2\pi iQ_2(2-[2]_{\alpha})}+\frac{
\partial_{u_{1}}\log a_{2}^{Q_2Q_1}(u_{2},u_{1})^{\star }}{2\pi
iQ_1(2-[2]_{\alpha})}\right]\nn\\
&+&\frac{2}{Q_{1}Q_2}
\sum_{k=0}^{Q_1-1}k(k-Q_1)\left[(Q_2+1)K_{Q_2-Q_1+2k-1}-(Q_2-1)K_{Q_2-Q_1+2k+1}\right]\star
s\Bigg\}\,,~~~~~~
\label{yQ1}
\ea
where we used the following identity
\be
K_{Q_1y}\hat{*}K_{Q_2}
=\frac{1}{2\pi
i}\partial_{u_{1}}\log\left(\frac{x_{1}^{-Q_1}-x_{2}^{+Q_2}}{x_{1}^{+Q_1}-x_{2}^{-Q_2}}
\frac{x_{1}^{+Q_1}-1/x_{2}^{+Q_2}}{x_{1}^{-Q_1}-1/x_{2}^{-Q_2}}\right)\equiv\frac{1}{
2\pi i}\partial_{u_{1}}\log a_{2}^{Q_1Q_2}(u_{1},u_{2})\,,
\ee
its hermitian conjugate (recall that $x(u)^{*}=1/x(u^{*})$
in the mirror kinematics) \footnote{Actually, identities 
(\ref{a1identity}), (\ref{a1QQ'}), and (\ref{a2Q'Q*}) are valid up to vanishing derivatives 
$\partial_{u_{1}}\log \sqrt{\frac{x_{2}^{-}}{x_{2}^{+}}}$,  
$\partial_{u_{1}}\log \sqrt{\frac{x_{2}^{-Q_{2}}}{x_{2}^{+Q_{2}}}}$ 
and $\partial_{u_{1}}\log \frac{x_{2}^{+Q_{2}}}{x_{2}^{-Q_{2}}}$, respectively.}
\be
K_{Q_1}\hat{*}K_{yQ_2}=\frac{1}{2\pi
i}\partial_{u_{1}}\log\left(\frac{x_{1}^{-Q_1}-x_{2}^{+Q_2}}{x_{1}^{+Q_1}-x_{2}^{-Q_2}}
\frac{x_{1}^{-Q_1}x_{2}^{-Q_2}-1}{x_{1}^{+Q_1}x_{2}^{+Q_2}-1}\frac{x_{1}^{+Q_{1}}}{x_{1}^{-Q_{1}}}\frac{x_{2}^{-Q_{2}}}{x_{2}^{+Q_{2}}}\right)=\frac{1}
{2\pi i}\partial_{u_{1}}\log a_{2}^{Q_2Q_1}(u_{2},u_{1})^{*} \,,
\label{a2Q'Q*}
\ee 
and $a_{2}^{Q_1Q_2}(u_{1},u_{2})\, a_{2}^{Q_2Q_1}(u_{2},u_{1})^{*}
=\left[a_{1}^{Q_1Q_2}(u_{1},u_{2})\right]^{-2}$.
Moreover, we can write the sum of the two convolutions involving the universal
kernel $s(u)$ in (\ref{yQ1}) as
\ba
&&\frac{1}{Q_{1}Q_2}
\sum_{k=0}^{Q_1-1}k(k-Q_1)\left[(Q_2+1)K_{Q_2-Q_1+2k-1}-(Q_2-1)K_{Q_2-Q_1+2k+1}\right]\star
s\nn\\
&&+\sum_{j=0}^{Q_1-2}K_{Q_2-Q_1+2j+1}\star s
=\frac{1}{2\pi i Q_{1}Q_2}\partial_{u_{1}}{\cal K}^{Q_{1}Q_{2}}\,,
\ea
where we used the definition (\ref{calK}) of ${\cal K}^{Q_{1}Q_{2}}$.
Remarkably, despite the long computation, the final expression for $y_{Q_2}$ is quite simple and reads
\ba
y_{Q_2}&=&{\cal Y}_{Q_1}\star{1\over 2\pi i}\partial_{u_{1}}\Bigg\{ 
\log S_{sl(2)}^{Q_1Q_2}+\frac{2}{Q_{1} Q_2}{\cal K}^{Q_{1}Q_{2}}
-2\log a_{1}^{Q_1Q_2}(u_{1},u_{2}) \nn \\
&-&\sum_{\alpha=\pm}\frac{1}{(2-[2]_{\alpha})}\left[
\frac{1}{Q_{2}}\log a_{2}^{Q_1Q_2}(u_{1},u_{2})
+\frac{1}{Q_1}\log a_{2}^{Q_{2}Q_1}(u_{2},u_{1})^{\star }
\right]\Bigg\}\,.
\label{TBAQQ'}
\ea
Substituting this result, together with the result (\ref{TBAQLOresult}) for 
${\cal Y}_{Q_{1}}$, into the formula (\ref{energy2}) for the energy correction, 
we obtain
\ba
E_{0}^{(2,2)}  &=&  \sum_{Q_{1},Q_{2}=1}^{\infty}Q_{1}Q_{2}
\int_{-\infty}^{\infty}\frac{d\tilde{p}_{1}}{2\pi}
e^{-L\tilde{\epsilon}_{Q_{1}}(\tilde{p}_{1})}
\int_{-\infty}^{\infty}\frac{d\tilde{p}_{2}}{2\pi}
e^{-L\tilde{\epsilon}_{Q_{2}}(\tilde{p}_{2})} \nonumber \\
&\times & i\partial_{\tilde{p}_{1}}
\Biggl\{ (2-[2]_{\dot{q}})^{2}\left[[3]_{q}\left(-Q_{1}Q_{2}\log a_{1}^{Q_{1}Q_{2}}+{\cal
K}^{Q_{1}Q_{2}}\right) \right.\nn\\
&&\hspace{2.5cm}-[2]_{q}\left(-4Q_{1}Q_{2}\log 
a_{1}^{Q_{1}Q_{2}}-Q_{1}\log a_{2}^{Q_{1}Q_{2}}-Q_{2}\log
a_{2}^{Q_{2}Q_{1}*}+4{\cal K}^{Q_{1}Q_{2}}\right)\nn\\
 && \hspace{2.5cm}+\left.[1]_{q}\left(-5Q_{1}Q_{2}\log a_{1}^{Q_{1}Q_{2}}-2Q_{1}
\log a_{2}^{Q_{1}Q_{2}}-2Q_{2}\log a_{2}^{Q_{2}Q_{1}*} +5{\cal
K}^{Q_{1}Q_{2}}\right)\right]\nn\\
&&\hspace{1cm} + (q \leftrightarrow \dot q) \nn \\
&& \hspace{1cm}+Q_{1}Q_{2}(2-[2]_{q})^{2}(2-[2]_{\dot{q}})^{2}
\log S_{sl(2)}^{Q_{1}Q_{2}}(\tilde{p}_{1},\tilde{p}_{2})
\Bigg\}\,.
\label{NLOTBA}
\ea 
Finally, through the following identifications
\be
a_{1}^{Q_{1}Q_{2}}=(U_{0}U_{1}U_{2})^{-1}\,, \quad a_{2}^{Q_{1}Q_{2}}=U_{0}U_{2}^{2}U_{3}\,,
\quad a_{2}^{Q_{2}Q_{1}*}=U_{0}U_{1}^{2}U_{3}^{-1}\,,
\ee
we find full agreement with the result (\ref{NLOLuscher}) from the L\"uscher computation.

\section{Weak-coupling expansion}\label{sec:weakcoupling}

In this section we calculate the weak-coupling expansion of the ground-state
energy of the twisted AdS/CFT model. 
In order to perform the weak-coupling expansion, we use the parameterization
\[
x^{\pm}(\tilde p)=\frac{(\tilde p-iQ)}{2g}\left(\sqrt{1+\frac{4g^{2}}{Q^{2}+\tilde p^{2}}}\mp1\right)\,,
\]
which follows from (\ref{eq:shortening}) and (\ref{eq:mirrordispersion}).
At leading order in $g$ , and so at weak coupling, we have
\[
x^{-}=\frac{\tilde p-iQ}{g}+O(g)\,, \qquad x^{+}=\frac{g}{\tilde 
p+iQ}+O(g^{3})\,.
\]

\subsection{LO contribution, single wrapping}

The LO correction can be calculated from (\ref{result}) by using the expansion
of the exponential term appearing in $\mathcal{Y}_{Q}$:
\[
e^{-L\tilde{\epsilon}_{Q}(\tilde{p})}=\sum_{j=0}^{\infty}c_{j}
\frac{g^{2(L+j)}}{(\tilde{p}^{2}+Q^{2})^{L+j}}\,.
\]
In particular $c_{0}=1$, while the higher-order terms can be easily
generated with Mathematica. Using the fact that $\frac{1}{n!}f^{(n)}(z)=\oint
\frac{dw}{2\pi i}\frac{f(w)}{(w-z)^{n+1}}$,
we perform the integral in (\ref{result}) by residues 
\begin{equation}
\int_{-\infty}^{\infty}\frac{d\tilde{p}}{2\pi}\frac{1}{(
\tilde{p}^{2}+Q^{2})^{k}}=\left(\begin{array}{c}
2k-2\\
k-1\end{array}\right)\,(2Q)^{1-2k}\,.
\label{eq:singleintegral}
\end{equation}
The summation over $Q$ gives rise to a series of $\zeta$-functions:
\[
E_{0}^{(1)}(L)=-(2-[2]_{q})(2-[2]_{\dot{q}})\sum_{j=0}^{
\infty}c_{j}2^{1-2(L+j)}\left(\begin{array}{c}
2(L+j)-2\\
L+j-1\end{array}\right)\,\zeta_{2(L+j)-3}g^{2(L+j)}\,.
\label{eq:LOwrapping}
\]
This result is exact up to $g^{4L}$ where the NLO L\"uscher correction
starts to play a role. We evaluate the leading $g^{4L}$-order contribution
of the NLO L\"uscher correction in the next subsection.

\subsection{NLO contribution, double wrapping }

The simplest term of the NLO correction comes from (\ref{E021}) and contains
$\mathcal{Y}_{Q}^{2}$. Its contribution at order $g^{4L}$ can
be calculated using eq. (\ref{eq:singleintegral}) to be 
\[
E_{0}^{(2,1)}(L)=(2-[2]_{q})^{2}(2-[2]_{\dot{q}})^{2}2^{-4L}\left(\begin{array}{c}
4L-2\\
2L-1\end{array}\right)\,\zeta_{4L-5}g^{4L}\,.
\label{eq:NLOsimple}
\]

The most complicated term is $E_{0}^{(2,2)}(L).$ We have to evaluate
(\ref{energy2}) based on the solution given in (\ref{TBAQQ'}). The twist dependence comes
in two distinct ways as:
\[
E_{0}^{(2,2)}(L)=(2-[2]_{q})^{2}(2-[2]_{\dot{q}})^{2}
\left[A(L)+B(L)\left(\frac{1}{[2]_{q}-2}+\frac{1}{[2]_{\dot{q}}-2}\right)\right]g^{4L} \,.
\label{eq:NLOhard}
\]
We first calculate $B(L)$ for any value of $L$. The weak-coupling expansion
of the functions $a_{2}^{Q_{1}Q_{2}}$ and $a_{2}^{Q_{2}Q_{1}*}$
are given by
\[
\partial_{\tilde{p}_{1}}\log a_{2}^{Q_{1}Q_{2}}(\tilde{p}_{1},
\tilde{p}_{2})=O(g^{2})\,, \qquad
\partial_{\tilde{p}_{1}}
\log a_{2}^{Q_{2}Q_{1}}(\tilde{p}_{2},\tilde{p}_{1})^{*}=
\frac{2iQ_{1}}{\tilde{p}_{1}^{2}+Q_{1}^{2}}+O(g^{2})\,.
\]
We substitute these results into (\ref{TBAQQ'}) and then into (\ref{energy2}), we perform
the integrals as in (\ref{eq:singleintegral}), and sum up the independent
terms to obtain:
\[
B(L)=-2^{1-4L}\left(\begin{array}{c}
2L-2\\
L-1\end{array}\right)\left(\begin{array}{c}
2L\\
L\end{array}\right)\zeta_{2L-1}\zeta_{2L-3}\,.
\label{eq:BL}
\]
This gives the complete answer for the given $(2-[2]_{q})(2-[2]_{
\dot{q}})(4-[2]_{q}-[2]_{\dot{q}})$
dependence of the double-wrapping correction at leading nonvanishing
order for any $L$. 

We now proceed to calculate $A(L)$. It acquires contributions from
the first line of (\ref{TBAQQ'}), which we denote by $A_{sl(2)}$, $A_{\mathcal{K}}$
and $A_{1}$, respectively, 
\be
A(L) = A_{sl(2)}(L) + A_{\mathcal{K}}(L) + A_{1}(L) \,,
\ee 
where
\ba
\hspace{-0.2in} A_{sl(2)}(L) &=& \sum_{Q_{1},Q_{2}}Q_{1}^{2}Q_{2}^{2}
\int\frac{d\tilde{p}_{1}}{2\pi}e^{-L\epsilon_{Q_{1}}(\tilde{p}_{1})}
\int\frac{d\tilde{p}_{2}}{2\pi}e^{-L\epsilon_{Q_{2}}(\tilde{p}_{2})}
\, i \partial_{\tilde{p}_{1}}\log S_{sl(2)}^{Q_{1}Q_{2}}
(\tilde{p}_{1},\tilde{p}_{2}) \,, \label{defAsl2}\\
A_{\mathcal{K}}(L) &=& 2\sum_{Q_{1},Q_{2}}Q_{1} Q_{2}
\int\frac{d\tilde{p}_{1}}{2\pi}e^{-L\epsilon_{Q_{1}}(\tilde{p}_{1})}
\int\frac{d\tilde{p}_{2}}{2\pi}e^{-L\epsilon_{Q_{2}}(\tilde{p}_{2})}
\, i \partial_{\tilde{p}_{1}} 
\mathcal{K}^{Q_{1}Q_{2}}(\tilde{p}_{1},\tilde{p}_{2})\,, 
\label{defAcalK}\\
A_{1}(L) &=& -2\sum_{Q_{1},Q_{2}}Q_{1}^{2}Q_{2}^{2}
\int\frac{d\tilde{p}_{1}}{2\pi}e^{-L\epsilon_{Q_{1}}(\tilde{p}_{1})}
\int\frac{d\tilde{p}_{2}}{2\pi}e^{-L\epsilon_{Q_{2}}(\tilde{p}_{2})}
\, i \partial_{\tilde{p}_{1}}\log a_{1}^{Q_{1}Q_{2}}
(\tilde{p}_{1},\tilde{p}_{2}) \,. \label{defA1}
\ea 
In order to compute $A_{1}$, we expand
$a_{1}^{Q_{1}Q_{2}}$ to leading order in $g$:
\[
\partial_{\tilde{p}_{1}}\log a_{1}^{Q_{1}Q_{2}}(\tilde{p}_{1},\tilde{p}_{2})=
-\frac{iQ_{1}}{\tilde{p}_{1}^{2}+Q_{1}^{2}}+O(g^{2})\,.
\]
Substituting the result back into (\ref{defA1}) gives 
\[
A_{1}(L)=-2^{1-4L}\left(\begin{array}{c}
2L-2\\
L-1\end{array}\right)\left(\begin{array}{c}
2L\\
L\end{array}\right)\zeta_{2L-2}\zeta_{2L-3}\,.
\label{eq:AL}
\]
Observe that the transcendentality of $A(L)$ and $B(L)$ are different.
It seems the deformation $2-[2]$ carries transcendentality $1$. A similar
effect was observed already in \cite{deLeeuw:2010ed,deLeeuw:2011rw}.

To calculate $A_{sl(2)}$, we have to expand the logarithm of the
dressing factor $\log S_{sl(2)}^{Q_{1}Q_{2}}(\tilde{p}_{1},\tilde{p}_{2})$
in the mirror-mirror kinematics. According to \cite{Arutyunov:2009ur},
it has the structure 
$\log S_{sl(2)}^{Q_{1}Q_{2}}(\tilde{p}_{1},\tilde{p}_{2})=
-\log S_{su(2)}^{Q_{1}Q_{2}}(\tilde{p}_{1},\tilde{p}_{2})-2
\log\Sigma^{Q_{1}Q_{2}}(\tilde{p}_{1},\tilde{p}_{2})$. Hence,
we can write
\[
\frac{1}{2\pi i}\partial_{\tilde{p}_{1}}\log S_{sl(2)}^{Q_{1}Q_{2}}
(\tilde{p}_{1},\tilde{p}_{2})=-K_{Q_{1}Q_{2}}-\frac{1}{\pi i}
\partial_{\tilde{p}_{1}}\log\Sigma^{Q_{1}Q_{2}}(\tilde{p}_{1},\tilde{p}_{2})\,.
\]
Explicitly performing the weak-coupling expansion of (6.14) in 
\cite{Arutyunov:2009kf}, we obtain (see (\ref{eq:dressingweakcoupling}))
\[
i\partial_{\tilde{p}_{1}}\log\Sigma^{Q_{1}Q_{2}}
(\tilde{p}_{1},\tilde{p}_{2})=-\frac{1}{2}\biggr[\psi(1-
\frac{i}{2}(\tilde{p}_{1}+iQ_{1}))
-\psi(1+\frac{1}{2}(i(\tilde{p}_{1}-\tilde{p}_{2})+Q_{1}+Q_{2}))+c.c\biggl] \,,
\]
where $\psi(x)=\partial_{x}(\log\Gamma(x))$ is the polygamma function. 
The $su(2)$ scalar factor results in 
\begin{eqnarray}
K_{su(2)}^{Q_{1}Q_{2}}=K_{Q_{1}Q_{2}} & = & -\frac{1}{4\pi}
\biggr[\psi(\frac{1}{2}(i(\tilde{p}_{2}-\tilde{p}_{1})-Q_{1}+Q_{2}))
+\psi(1+\frac{1}{2}(i(\tilde{p}_{2}-\tilde{p}_{1})-Q_{1}+Q_{2}))\\
 &  & \qquad-\psi(\frac{1}{2}(i(\tilde{p}_{2}-\tilde{p}_{1})+Q_{1}+Q_{2}))
-\psi(1+\frac{1}{2}(i(\tilde{p}_{2}-\tilde{p}_{1})+Q_{1}+Q_{2}))+c.c\biggl] \,.
\nonumber
\end{eqnarray}
Finally,
\begin{eqnarray}
i\partial_{\tilde{p}_{1}}
\mathcal{K}^{Q_{1}Q_{2}} & = & -\frac{1}{8}
\biggl[4(Q_{1}-1)Q_{2}+((Q_{1}-Q_{2})^{2}+(\tilde{p}_{1}-\tilde{p}_{2})^{2})
\times\\
 &  & \bigl(\psi(1+\frac{1}{2}(i(\tilde{p}_{2}-\tilde{p}_{1})-Q_{1}+Q_{2}))-
\psi(\frac{1}{2}(i(\tilde{p}_{2}-\tilde{p}_{1})+Q_{1}+Q_{2}))\bigr)
+c.c.\biggl] \,. \nonumber
\end{eqnarray}
Denoting the contributions to $A_{sl(2)}$ by $A_{\Sigma}$ and 
$A_{su(2)}$, we have that
\be
A_{sl(2)}(L) = A_{\Sigma}(L) + A_{su(2)}(L) \,,
\ee
where
\ba
\hspace{-0.2in} A_{\Sigma}(L) &=& -2\sum_{Q_{1},Q_{2}}Q_{1}^{2}Q_{2}^{2}
\int\frac{d\tilde{p}_{1}}{2\pi}e^{-L\epsilon_{Q_{1}}(\tilde{p}_{1})}
\int\frac{d\tilde{p}_{2}}{2\pi}e^{-L\epsilon_{Q_{2}}(\tilde{p}_{2})}
\, i \partial_{\tilde{p}_{1}}\log \Sigma^{Q_{1}Q_{2}}
(\tilde{p}_{1},\tilde{p}_{2}) \,, \label{defASigma}\\
A_{su(2)}(L) &=& \sum_{Q_{1},Q_{2}}Q_{1}^{2}Q_{2}^{2}
\int\frac{d\tilde{p}_{1}}{2\pi}e^{-L\epsilon_{Q_{1}}(\tilde{p}_{1})}
\int\frac{d\tilde{p}_{2}}{2\pi}e^{-L\epsilon_{Q_{2}}(\tilde{p}_{2})}
2\pi K_{Q_{1}Q_{2}} (\tilde{p}_{1},\tilde{p}_{2}) \,. \label{defAKQQ}
\ea 

Using methods explained in Appendix \ref{sec:details}, we evaluated the integrals by
residues. To demonstrate the structure of the result, we write out
explicitly $A_{\Sigma}$ (see (\ref{eq:sl2part1}) and  (\ref{eq:sl2part2})):
\begin{eqnarray}
A_{\Sigma}(L) & = & -2^{2-2L}\left(
\begin{array}{c}
2L-2  \\
L-1\end{array}
\right)\zeta_{2L-3}\sum_{Q_{1}}\sum_{j=0}^{L-1}\left(\begin{array}{c}
L+j-1\\
j\end{array}\right)\frac{2^{-2L+1}}{(L-1-j)!}(-Q_{1})^{2-L-j}\psi^{(L-j-1)}(Q_{1}+1)
\nonumber \\
 &  & -\sum_{Q_{1},Q_{2}}\sum_{j_{1},j_{2}=0}^{L-1}\left(\begin{array}{c}
L+j_{1}-1\\
j_{1}\end{array}\right)\frac{2^{-2L+2}}{(L-1-j_{1})!}(-Q_{1})^{2-L-j_{1}} 
\nonumber \\
 &  & \qquad\qquad\times \left(\begin{array}{c}
L+j_{2}-1\\
j_{2}\end{array}\right)\frac{2^{-2L+1}}{(L-1-j_{2})!}(-Q_{2})^{2-L-j_{2}}
\psi^{(2L-j_{1}-j_{2}-2)}(Q_{1}+Q_{2}+1) \,.
\label{Asl2}
\end{eqnarray}
These terms can be expressed in terms of multiple zeta values (MZV)
by rewriting%
\footnote{For $n=0$, one has to replace $\zeta(1)$ with $\gamma_{E}$. %
} \[
\psi^{(n)}(Q+1)=(-1)^{n+1}n!(\zeta(n+1)-\sum_{j=1}^{Q}j^{-n-1}) \,,
\label{eq:polygammatoharmonic}
\]
and performing the sums explicitly. We will, however, not pursue this
calculation further here as we did not find an explicit answer for
generic $L$. The integrals can be evaluated similarly for $A_{su(2)}$
and $A_{\mathcal{K}}$ with a similar structural final result, although 
some care must be taken to the $Q_1-Q_2$ dependent term for $Q_1=Q_2$. In
the next subsection, we present explicit results for the smallest nontrivial
length: $L=3$.

\subsection{Specific calculations for $L=3$}

The LO wrapping correction (\ref{eq:LOwrapping}) for $L=2$ is divergent, as we have for
$j=0$ the term $\zeta_{2L-3}=\zeta_{1}$. Similar observations were 
made in \cite{Frolov:2009in,deLeeuw:2011rw}.
We therefore focus now on the first
nontrivial case, namely $L=3$. The LO correction for this case goes as follows: 
\[
E_{0}^{(1)}(3)=-(2-[2]_{q})(2-[2]_{\dot{q}})\left(\frac{3}{16}
\zeta_{3}g^{6}-\frac{15}{16}\zeta_{5}g^{8}+\frac{945}{256}
\zeta_{7}g^{10}-\frac{3465}{256}\zeta_{9}g^{12}+\dots\right) \,.
\]
The simple double-wrapping contribution (\ref{eq:NLOsimple}) at leading order is 
\[
E_{0}^{(2,1)}(3)=(2-[2]_{q})^{2}(2-[2]_{\dot{q}})^{2}\frac{63}{1024}\zeta_{7}g^{12}\,.
\]
In calculating the term $E_{0}^{(2,2)}(3)$, we recall from 
(\ref{eq:NLOhard}) that
\[
E_{0}^{(2,2)}(3)=(2-[2]_{q})^{2}(2-[2]_{\dot{q}})^{2}
\left[A(3)+B(3)\left(\frac{1}{[2]_{q}-2}+\frac{1}{[2]_{\dot{q}}-2}\right)\right]g^{12} \,.
\]
From (\ref{eq:BL}), we have
\[
B(3)=-\frac{15}{256}\zeta_{3}\zeta_{5} \,.
\]

We calculated the contributions to $A(3)$ one by one. The simplest
is 
\[
A_{1}(3)=-\frac{15}{256}\zeta_{3}\zeta_{4}\,,
\label{eq:A1result}
\]
as follows from (\ref{eq:AL}).
In the more complicated terms, we calculated the integrals by residues
as explained in Appendix \ref{sec:details}. Then, in summing up the 
expressions, we employed the following strategies: 
\begin{itemize}
\item We performed the sums analytically by replacing the polygamma functions
with harmonic sums using (\ref{eq:polygammatoharmonic}), and then 
rearranging all the sums into MZVs. These MZVs could then be expressed in
terms of elementary ones, which contained only products of simple
zetas with transcendentality less than or equal to $7$. 
\item Alternatively, for terms involving polygamma functions
depending on $Q_{1}+Q_{2}$, we replaced
the polygamma functions with their integral representations 
\[
\psi^{(n)}(z)=\int_{0}^{\infty}\left(\delta_{n,0}
\frac{e^{-t}}{t}-(-1)^{n}\frac{t^{n}e^{-tz}}{1-e^{-t}}\right)dt \,,
\]
and performed the summations $\sum_{Q_{1},Q_{2}=1}^{\infty}$ explicitly.
The remaining integral over $t$ could be evaluated numerically with very high precision
(100 digits), and the result could be expressed in terms of products of zeta
functions (and the Euler constant $\gamma_{E}$) with the help of the online MZV calculator,
EZ-Face. \footnote{EZ-Face is documented in \cite{Borwein:1999}, and can be accessed at 
{\tt http://oldweb.cecm.sfu.ca/projects/EZFace/index.html}}
\item Finally, for polygamma functions depending on $Q_{1}-Q_{2}$, we evaluated
the sums numerically as $\sum_{Q_{1},Q_{2}=1}^{\infty}=2
\sum_{Q_{1}}^{\infty}\sum_{Q_{2}=1}^{Q1-1}+\sum_{(Q_{1}=Q_{2})=1}^{\infty}$,
and again expressed the result in terms of zeta functions using EZ-Face. 
\end{itemize}
We found the following results: 
\ba
A_{\Sigma}(3) &=& \frac{81}{1024}\zeta_{3}\zeta_{4}+
\frac{21}{512}\zeta_{2}\zeta_{5}-\frac{441}{2048}\zeta_{7} \,, 
\nonumber \\
A_{su(2)}(3) &=& 
-\frac{9}{512}\zeta_{3}\zeta_{4}+\frac{315}{4096}\zeta_{7} \,, 
\nonumber \\
A_{\mathcal{K}}(3)&=&-\frac{9}{256}\zeta_{3}^{2}-\frac{3}{1024}
\zeta_{3}\zeta_{4}-\frac{21}{512}\zeta_{2}\zeta_{5}+\frac{63}{512}\zeta_{7}\,.
\ea
By summing up, we obtain the total $A$ contribution
\ba 
A(3) = A_{\Sigma}(3) + A_{su(2)}(3) + A_{\mathcal{K}}(3) + A_{1}(3) =
-\frac{9}{256}\zeta_{3}^{2}-\frac{63}{4096}\zeta_{7} \,.
\ea
Thus, the total anomalous dimension is 
\begin{eqnarray}
E_{0}(3) & = & E_{0}^{(1)}(3) + E_{0}^{(2,1)}(3) + E_{0}^{(2,2)}(3) + 
\ldots \nonumber \\
& = & -(2-[2]_{q})(2-[2]_{\dot{q}})\left(\frac{3}{16}
\zeta_{3}g^{6}-\frac{15}{16}\zeta_{5}g^{8}+\frac{945}{256}
\zeta_{7}g^{10}-\frac{3465}{256}\zeta_{9}g^{12}+\dots\right)\nonumber \\
 &  & -(2-[2]_{q})(2-[2]_{\dot{q}})\left([2]_{q}+[2]_{\dot{q}}-4)
\right)\frac{15}{256}\zeta_{3}\zeta_{5}g^{12} + \dots \nonumber\\
 &  & +(2-[2]_{q})^{2}(2-[2]_{\dot{q}})^{2}\left(-
\frac{9}{256}\zeta_{3}^{2}+
\frac{189}{4096}\zeta_{7}\right)g^{12}+\dots \, ,
\label{finalresult}
\end{eqnarray}
where we recall that $2-[2]_q=4 \sin( \frac{\gamma_+}{2})^2$ and 
$2-[2]_{\dot q}=4 \sin( \frac{\gamma_-}{2})^2$ 
in terms of the deformation parameters
$\gamma_{\pm}=(\gamma_{3}\pm\gamma_{2})\frac{3}{2}$, as in 
our case $L=3$.

The result (\ref{finalresult}) is indeed the total anomalous dimension, since the vacuum energy 
does not receive any contributions from the asymptotic Bethe ansatz.
Remarkably, even though at intermediate stages of the computation 
there appear terms involving even zeta functions and Euler's 
constant $\gamma_{E}$, all such terms finally cancel.

\section{Conclusion}\label{sec:discuss}

We have computed the NLO finite-volume correction to the vacuum energy
in twisted AdS/CFT by two apparently independent approaches: L\"uscher
(\ref{NLOLuscher}) and TBA (\ref{NLOTBA}).  The fact that both
approaches yield identical results provides a strong consistency check
on the AdS/CFT S-matrices and TBA equations that have been developed
in the literature, as well as on the final result.  This result is
expressed in terms of a double infinite sum of contributions from
the infinitely-many types of massive mirror bound states. Our 
computations check the complete (both horizontal and vertical parts of the)  
$Y$-system,
and go beyond the five-loop calculations presented in \cite{Arutyunov:2010gb,
Balog:2010xa, Balog:2010vf}, which checked at the single wrapping order 
only the vertical part.  

Our result is valid for any value of the coupling constant.  However,
by making a weak-coupling expansion, we have obtained a
prediction (\ref{finalresult}) for the anomalous dimension of the operator ${\rm Tr}
Z^{3}$ in the twisted gauge theory up to six loops.  It should be
possible to check this prediction directly in perturbation
theory by taking into account both single-wrapping and
double-wrapping diagrams.  To our knowledge, this is the first
complete computation of double wrapping in the literature. It may be 
interesting to investigate also the strong-coupling limit.

The key results needed for the NLO L\"uscher computation were the
determinants of the (untwisted) AdS/CFT S-matrices in all the
$su(2)_{L} \otimes su(2)_{R}$ sectors, presented in Tables
\ref{table:general} and \ref{table:special}.  The simplicity of these
results suggests that they may have some group-theoretical
formulation.  In particular, it should be possible to find a general
proof, presumably based on $su(2|2)$ Yangian symmetry.

It would be interesting to extend our analysis of finite-size
corrections in twisted AdS/CFT, which has so far been restricted to
the ground state, to excited states beyond the LO result of \cite{deLeeuw:2011rw}.
It would also be interesting to understand the origin of the divergence of the LO 
and NLO results for $L=2$, which was already 
noticed in similar contexts in \cite{Frolov:2009in,deLeeuw:2011rw}. 
Finally, one can now begin to contemplate triple and higher wrapping.

\section*{Acknowledgments}
We thank Orlando Alvarez, Gleb Arutyunov, J\'anos Balog, Sergey 
Frolov, \'Arp\'ad
Heged\H{u}s, Marius de Leeuw, Christoph Sieg and Stijn van Tongeren
for useful discussions and/or correspondence; and the referees for 
their valuable comments.
CA, DB and RN are grateful for the warm hospitality extended to them at 
ELTE and at the Perimeter Institute during the course of this work. 
This work was supported in part by WCU Grant No. R32-2008-000-101300 (CA),
OTKA 81461 (ZB),
the FCT fellowship SFRH/BPD/69813/2010 and the network UNIFY for travel financial 
support (DB), and by the National Science Foundation under Grant PHY-0854366 
and a Cooper fellowship (RN).

\begin{appendix}

\section{Determinants of S-matrices in the $su(2)_{L} \otimes su(2)_{R}$
sectors}\label{sec:dets}

We describe here how we obtained the results in Tables
\ref{table:general} and \ref{table:special} for $\det 
S^{(Q_{1},Q_{2})}(s_{L},s_{R})$, the determinants of
the AdS/CFT S-matrices in the $su(2)_{L} \otimes su(2)_{R}$ sectors,  which
enter into the NLO L\"uscher computation.  Our straightforward
approach was to explicitly compute these determinants for small values
of $Q_{1}$ and $Q_{2}$ (up to 8), and then infer the general pattern.

For the cases $(Q_{1},Q_{2})=(1,1), (1,2), (2,2)$, we used the explicit 
S-matrices from \cite{Arutyunov:2008zt} to directly compute
the eigenvalues. For the cases $(Q_{1},Q_{2})=(1,Q)$, we used results from 
\cite{Bajnok:2008bm}: from Eq. (56) there, it follows that (up to the overall
factors),
\ba
\det S^{(1,Q)}(\frac{Q-1}{2},1) &=& a_{9}^{9} \,, \nn \\
\det S^{(1,Q)}(\frac{Q}{2},\frac{1}{2})  &=& \frac{1}{Q} 
\det \left( \begin{array}{cc}
a_{5}^{5} & a_{5}^{6} \\
a_{6}^{5} & a_{6}^{6}
\end{array} \right) \,, \nn\\
\det S^{(1,Q)}(\frac{Q-2}{2},\frac{1}{2})  &=& Q \det \left( \begin{array}{cc}
a_{7}^{7} & a_{7}^{8} \\
a_{8}^{7} & a_{8}^{8}
\end{array} \right) \,,  \nn \\
\det S^{(1,Q)}(\frac{Q+1}{2},0) &=& a_{1}^{1} = 1 \,,  \nn \\
\det S^{(1,Q)}(\frac{Q-1}{2},0) &=& 
 \frac{Q+1}{Q-1} \det \left( \begin{array}{ccc}
 a_{2}^{2} & a_{2}^{3} & a_{2}^{4} \\
 a_{3}^{2} & a_{3}^{3} & a_{3}^{4} \\
 a_{4}^{2} & a_{4}^{3} & a_{4}^{4}  
 \end{array} \right) \,,  \nn \\
\det S^{(1,Q)}(\frac{Q-3}{2},0) &=& \frac{Q-1}{2}  a_{10}^{10} \,.
\label{BJ}
\ea
One can verify using the explicit values of $a_{i}^{j}$ 
that \footnote{We note a couple of typos in
appendix B of \cite{Bajnok:2008bm}: $a_{3}^{2}$ should not have the
factor $x^{+}z^{+}$ in the denominator; and $a_{3}^{3}$ is missing an
overall minus sign.}
\ba
a_{9}^{9} &=& \frac{1}{Q} 
 \det \left( \begin{array}{cc}
 a_{5}^{5} & a_{5}^{6} \\
 a_{6}^{5} & a_{6}^{6}
 \end{array} \right)  =  U_{0} U_{1} U_{2} 
\,, \nn \\
Q \det \left( \begin{array}{cc}
 a_{7}^{7} & a_{7}^{8} \\
 a_{8}^{7} & a_{8}^{8}
 \end{array} \right) &=&  \frac{Q+1}{Q-1} \det \left( \begin{array}{ccc}
 a_{2}^{2} & a_{2}^{3} & a_{2}^{4} \\
 a_{3}^{2} & a_{3}^{3} & a_{3}^{4} \\
 a_{4}^{2} & a_{4}^{3} & a_{4}^{4}  
 \end{array} \right) = U_{0}^{2} U_{1} U_{2}^{3} U_{3} 
\,, \nn \\
\frac{Q-1}{2}  a_{10}^{10} &=& 
U_{0} U_{2}^{2} U_{3}  \,, 
\label{BJnum}
\ea 
where the notation is defined in (\ref{detnotation}).  It is then easy
to see that the expressions in Table \ref{table:special} are
consistent with the results (\ref{BJ}), (\ref{BJnum}).

For general values of $(Q_{1},Q_{2})$, we made use of the formalism
developed in \cite{Arutyunov:2009mi}.  As an example, let us consider
the case $(Q_{1},Q_{2}) = (2,3)$.  Since the state of a single $Q$-particle (the
$4Q$-dimensional totally symmetric representation of $su(2|2)$) has 
the $su(2)_{L} \otimes su(2)_{R}$ decomposition
\be
V^{\frac{Q}{2}}\times V^{0} + V^{\frac{Q-1}{2}}\times V^{\frac{1}{2}} 
+ V^{\frac{Q-2}{2}}\times V^{0} \,,
\label{eq:1pdecomposition}
\ee
the decomposition of the corresponding 2-particle states can be obtained
from the tensor product
\be
\left( V^{1} \times V^{0} + V^{\frac{1}{2}} \times V^{\frac{1}{2}} 
+ V^{0} \times V^{0} \right) \otimes 
\left( V^{\frac{3}{2}} \times V^{0} + V^{1} \times V^{\frac{1}{2}} 
+ V^{\frac{1}{2}} \times V^{0} \right) \,,
\label{eq:2pdecomposition}
\ee 
where in this appendix we denote by $\times$ the tensor product of 
the $su(2)_{L}$ and $su(2)_{R}$ representations.
For concreteness, let us focus on the computation of $\det S^{(2,3)}(1, \frac{1}{2})$.
The tensor product in (\ref{eq:2pdecomposition}) can be decomposed,
by the Clebsch-Gordan theorem, into a sum of irreducible 
representations of $su(2)_{L} \otimes su(2)_{R}$. In this 
decomposition, there appear four
representations with $(s_{L},s_{R}) = (1, \frac{1}{2})$, which are 
the relevant ones for computing this determinant. These four representations
come from the following channels:
\ba
&1:& \qquad \left( V^{1} \times V^{0}\right) \otimes \left( V^{1} 
\times V^{\frac{1}{2}}\right) \nn \\
&2:& \qquad \left( V^{\frac{1}{2}} \times V^{\frac{1}{2}}\right) \otimes 
\left( V^{\frac{1}{2}} \times V^{0}\right) \nn \\
&3:& \qquad \left( V^{0} \times V^{0}\right) \otimes 
\left( V^{1} \times V^{\frac{1}{2}}\right) \nn \\
&4:& \qquad \left( V^{\frac{1}{2}} \times V^{\frac{1}{2}}\right) \otimes 
\left( V^{\frac{3}{2}} \times V^{0}\right) \,.
\label{eq:channels}
\ea 
The corresponding highest-weight states 
$|\psi_{I}^{(Q_{1},Q_{2})}(s_{L},s_{R})\rangle$
with $s_{L} = m_{L} = 1$ and $s_{R} = m_{R} = \frac{1}{2}$ are given 
(up to an overall normalization factor) by 
\ba
|\psi_{1}^{(2,3)}(1, \frac{1}{2})\rangle & \propto & |0,1\rangle^{\rm II}_{2} -
|1,0\rangle^{\rm II}_{2} \,, 
\nn \\
|\psi_{2}^{(2,3)}(1, \frac{1}{2})\rangle &\propto & |0,1\rangle^{\rm II}_{3} 
\,, 
\nn \\
|\psi_{3}^{(2,3)}(1, \frac{1}{2})\rangle &\propto & |1,0\rangle^{\rm II}_{4} 
\,, 
\nn \\
|\psi_{4}^{(2,3)}(1, \frac{1}{2})\rangle & \propto & |0,1\rangle^{\rm II}_{1} - 
|1,0\rangle^{\rm II}_{1} \,, 
\label{hwstates}
\ea
respectively, where the states $|k,l\rangle^{\rm II}_{i}$ are defined in 
\cite{Arutyunov:2009mi}. It is convenient to introduce a basis 
$|e_{i}\rangle$
of these so-called type-II states with $N \equiv k+l = 1$: 
\ba
|e_{1}\rangle & = & |0,1\rangle^{\rm II}_{1} \,,\nn \\
|e_{2}\rangle & = & |0,1\rangle^{\rm II}_{2} \,,\nn \\
|e_{3}\rangle & = & |0,1\rangle^{\rm II}_{3} \,,\nn \\
|e_{4}\rangle & = & |1,0\rangle^{\rm II}_{1} \,,\nn \\
|e_{5}\rangle & = & |1,0\rangle^{\rm II}_{2} \,,\nn \\
|e_{6}\rangle & = & |1,0\rangle^{\rm II}_{4} \,.
\ea 
Although these states are orthogonal, they are not
normalized.\footnote{We are grateful to M. de Leeuw for pointing this 
out to us.} Indeed, defining
\be
n_{i} \equiv \langle e_{i}| e_{i} \rangle \,,
\ee
it readily follows from the definitions of the states \cite{Arutyunov:2009mi}
that here $n_{i} = (2,2,1,6,2,2)$. An orthonormal basis $|\tilde e_{i}\rangle$
is therefore given by
\be
| \tilde e_{i} \rangle \equiv \frac{1}{\sqrt{n_{i}}} | e_{i} \rangle \,, \qquad 
\langle \tilde e_{i}| \tilde e_{j} \rangle = \delta_{ij} \,.
\ee 

The S-matrix acts as 
\be
S |e_{i}\rangle = \sum_{j}|e_{j}\rangle U_{ji} \,.
\ee
Numerical values for the coefficients $U_{ji}$ can be 
computed using formulas in \cite{Arutyunov:2009mi},
for given numerical values of momenta $p_{1}, p_{2}$, coupling 
constant $g$, and representations $Q_{1}, Q_{2}$.\footnote{We note that
version 1 in the arXiv of \cite{Arutyunov:2009mi} contains a number of
typos, most of which are corrected in the journal. However, some typos
remain in the latter. In particular, in (5.14): $\bar{Q}_{ij} = b_i d_j 
- b_j d_i$. Also,  in $A^{-1}$ in (5.17): in the (2,2)
element of the big matrix, $c_1^-$ should be instead $c_1^+$ ;
and in the (2,1) matrix element, the sign in front of 
$[M+(l_1-l_2)/2]$ should be plus instead of minus. Finally, 
in (A.8), the formulas for $b_{1}, \ldots , b_{4}$ should have sign 
plus instead of minus; and the formulas for $d_{2}$ and $d_{3}$ should 
not have $i$ in the denominator. We are grateful to G. Arutyunov and 
M. de Leeuw for correspondence on these points.} Hence, we can obtain 
the corresponding coefficients $\tilde U_{ji}$ in the normalized basis
\ba
\tilde U_{ji} &\equiv& \langle \tilde e_{j}| S | \tilde e_{i} \rangle 
\nn \\
&=& \sqrt{\frac{n_{j}}{n_{i}}}U_{ji} = \sum_{k,l} M_{jk} U_{kl} 
M^{-1}_{li} \,,
\ea
where we have introduced the diagonal matrix $M_{ij} \equiv \sqrt{n_{i}} 
\delta_{ij}$. A useful check is that the matrix $\tilde U_{ji}$ 
(unlike $U_{ji}$) is unitary.

We use (\ref{hwstates}) to express the highest-weight states
$|\psi_{I}^{(Q_{1},Q_{2})}(s_{L},s_{R})\rangle$ 
in terms of the normalized basis
\be
|\psi_{I}^{(Q_{1},Q_{2})}(s_{L},s_{R})\rangle = \sum_{i} | \tilde e_{i} \rangle
c_{i I} \,, \qquad 
c_{i I} \equiv \langle \tilde e_{i}|
\psi_{I}^{(Q_{1},Q_{2})}(s_{L},s_{R})\rangle \,,
\ee
where the states themselves are normalized,
\be
\langle
\psi_{I}^{(Q_{1},Q_{2})}(s_{L},s_{R})|\psi_{J}^{(Q_{1},Q_{2})}(s_{L},s_{R}
)\rangle = \delta_{IJ} \,.
\ee 

We can finally construct the $S$-matrix in the $(s_{L},s_{R})$ sector,
\be
S_{IJ}^{(Q_{1},Q_{2})}(s_{L},s_{R}) \equiv 
\langle \psi_{I}^{(Q_{1},Q_{2})}(s_{L},s_{R})| S\,
|\psi_{J}^{(Q_{1},Q_{2})}(s_{L},s_{R})\rangle =
\sum_{i,j} c_{i I}^{*} \tilde U_{ij} c_{j J} \,.
\label{eq:Ssector}
\ee 
Another useful check is that the matrix $S_{IJ}^{(Q_{1},Q_{2})}(s_{L},s_{R})$ is
also unitary. 
Computing numerically the determinant of this matrix \footnote{We use the
convention that the determinant of 
a number (i.e., a $1 \times 1$ matrix) is the number itself.}
\be
\det S^{(Q_{1},Q_{2})}(s_{L},s_{R}) \equiv  \det \left( 
S_{IJ}^{(Q_{1},Q_{2})}(s_{L},s_{R}) \right) \,,
\ee
we find for the case in question (namely, $(Q_{1},Q_{2}) = (2,3)$ and 
$(s_{L},s_{R})= (1, \frac{1}{2})$) that the result coincides with 
$\left(U_{0} U_{1} U_{2} \right)^{4} S_{3}^{2}$, in agreement with 
Table \ref{table:general}. Other cases $(Q_{1},Q_{2})$ and other sectors 
$(s_{L},s_{R})$ can be treated in a similar way. Note 
that sectors with $s_{R}=1,\frac{1}{2},0$ are constructed with 
states of type I, II, III, respectively.  After some 
effort to accumulate results for sufficiently many cases,
the general pattern summarized in Tables \ref{table:general} and \ref{table:special} became evident.

Before closing this section, it may be worthwhile to frame the problem 
that we have addressed here in a more general context. Consider an S-matrix (solution 
of the Yang-Baxter equation) that is invariant under a group $G$, which 
here is $su(2)_{L} \otimes su(2)_{R}$. As is well known (see e.g. 
\cite{Kulish:1981gi, Alishauskas:1986}), the S-matrix is a matrix 
$S_{ab}$ defined in the tensor product of two vector spaces $V_{a} 
\otimes V_{b}$ in which representations $\Pi_{a}$ and $\Pi_{b}$ of 
$G$ act,  \footnote{The representations
$\Pi_{a}$ and $\Pi_{b}$ need not be irreducible 
representations of $G$. Indeed, in the AdS/CFT case, they 
are sums of irreducible representations, as in (\ref{eq:1pdecomposition}).}
\be
\left[ \Pi_{a}(g) \otimes \Pi_{b}(g) \,, S_{ab} \right] = 0 \,, 
\qquad g \in G \,.
\label{eq:Smatrixinvariance}
\ee
The representation space decomposes into a sum of irreducible 
representations of $G$ parameterized by highest weights $\Lambda_{k}$,
which here are $(s_{L},s_{R})$,
\be
V_{a} \otimes V_{b} = \sum_{k} V(\Lambda_{k}) \,.
\label{eq:CG}
\ee
Since the S-matrix is $G$-invariant (\ref{eq:Smatrixinvariance}), it
has the corresponding spectral resolution
\be
S_{ab} = \sum_{k} \rho_{k} P_{\Lambda_{k}} \,,
\label{eq:spectralresolution}
\ee
where $P_{\Lambda_{k}}$ is a projector onto the irreducible subspace 
$V(\Lambda_{k})$. 

In the seminal work \cite{Kulish:1981gi} on the construction of
rational $S$-matrices, it was essential to assume that the
Clebsch-Gordan series (\ref{eq:CG}) is multiplicity free (i.e., a
given irreducible representation appears at most once), in which case
$\rho_{k}$ in (\ref{eq:spectralresolution}) is a scalar.  For AdS/CFT,
the decomposition (\ref{eq:CG}) is unfortunately {\em not}
multiplicity free: the Clebsch-Gordan series contains multiple
irreducible representations, as we have seen in the example (\ref{eq:channels}).
Hence, $\rho_{k}$ becomes an $r \times r$ matrix, where $r$ is the
multiplicity of the corresponding irreducible representation with
highest weight $\Lambda_{k}$.  In the AdS/CFT case, $\rho_{k}$ is the
matrix that we have defined in (\ref{eq:Ssector}).  The problem of
explicitly determining this matrix can be quite complicated even for
rational $S$-matrices, see e.g. \cite{Alishauskas:1986}.  In the
present work, we have restricted to the problem of computing its
determinant.

\section{Details of the weak-coupling expansion}\label{sec:details}

\subsection{Weak coupling expansion of the dressing phase}

The dressing phase in the mirror-mirror kinematics is given by \cite{Arutyunov:2009kf} 
\begin{eqnarray}
-i\log\Sigma_{Q_{1}Q_{2}}(y_{1},y_{2}) & = & \Phi(y_{1}^{+},y_{2}^{+})
-\Phi(y_{1}^{+},y_{2}^{-})-\Phi(y_{1}^{-},y_{2}^{+})+\Phi(y_{1}^{-},y_{2}^{-})\nonumber\\
 &  & +\frac{1}{2}\left[-\Psi(y_{1}^{+},y_{2}^{+})+\Psi(y_{1}^{+},y_{2}^{-})
-\Psi(y_{1}^{-},y_{2}^{+})+\Psi(y_{1}^{-},y_{2}^{-})\right]\nonumber\\
 &  & -\frac{1}{2}\left[-\Psi(y_{2}^{+},y_{1}^{+})+\Psi(y_{2}^{+},y_{1}^{-})
-\Psi(y_{2}^{-},y_{1}^{+})+\Psi(y_{2}^{-},y_{1}^{-})\right]\\
 &  & +\frac{1}{i}\log\left[\frac{i^{Q_{1}}\Gamma(Q_{2}-\frac{i}{2}g(y_{1}^{+}
+\frac{1}{y_{1}^{+}}-y_{2}^{+}-\frac{1}{y_{2}^{+}}))}{i^{Q_{2}}\Gamma(Q_{1}+
\frac{i}{2}g(y_{1}^{+}+\frac{1}{y_{1}^{+}}-y_{2}^{+}-
\frac{1}{y_{2}^{+}}))}
\left(\frac{1-\frac{1}{y_{1}^{+}y_{2}^{-}}}{1-\frac{1}{y_{1}^{-}y_{2}^{+}}}\right)
\sqrt{\frac{y_{1}^{+}y_{2}^{-}}{y_{1}^{-}y_{2}^{+}}}\right] \,, \nonumber
\end{eqnarray}
where 
\begin{equation}
\Psi(x_{1},x_{2})=i\oint_{C_{1}}\frac{dw_{2}}{2\pi i}\frac{1}{w_{2}-x_{2}}
\log\frac{\Gamma(1+i\frac{g}{2}(x_{1}+x_{1}^{-1}-w_{2}-w_{2}^{-1}))}{
\Gamma(1-i\frac{g}{2}(x_{1}+x_{1}^{-1}-w_{2}-w_{2}^{-1}))} \,,
\label{eq:Psi}
\end{equation}
and for $\Phi(x_{1},x_{2})$ we just note that it starts in any kinematics
at least with $g^{2}$. We calculate the O(1) expansion of the phase 
(\ref{eq:Psi}).
Using the property $\Psi(x_{1},x_{2})=\Psi(x_{1},0)-\Psi(x_{1},x_{2}^{-1})$,
being valid if $\vert x_{2}\vert\neq1$, and that for $\vert x_{2}\vert>1$
it starts at $g^{2}$, it is easy to see that we need to calculate
$\Psi(x_{1},x_{2})\equiv\Psi(x_{1},0)$ for $\vert x_{2}\vert<1$,
i.e. for $x^{+}_{2}$. Since we are interested in the derivative of the expanded functions
with respect to the first argument $\partial_{1}$, we need to expand $-\frac{1}{2}(
\Psi(y_{1}^{+},0)+\Psi(y_{1}^{-},0))$
only. Rescaling the integration variable $w_{2}$ by $g$ and evaluating
the leading residue for small $g$, we obtain
\[
\Psi(y_{1}^{+},0)=i\log\frac{\Gamma(1+\frac{i}{2}(\tilde{p}_1 +iQ_1))}{\Gamma(1-
\frac{i}{2}(\tilde{p}_1+iQ_1))}+\dots\,,\qquad
\Psi(y_{1}^{-},0)=i\log\frac{\Gamma(1+
\frac{i}{2}(\tilde{p}_1-iQ_1))}{\Gamma(1-\frac{i}{2}(\tilde{p}_1-iQ_1))}+\dots
\]
The logarithmic derivative of the whole dressing phase is then
\be
-\frac{1}{\pi i}\partial_{\tilde{p}_{1}}\log\Sigma^{Q_{1}Q_{2}}(\tilde{p}_{1},
\tilde{p}_{2}) =  \frac{1}{2\pi}\biggr[-\psi(1-\frac{i}{2}(
\tilde{p}_{1}+iQ_{1}))+\psi(1+\frac{1}{2}(i(\tilde{p}_{1}-
\tilde{p}_{2})+Q_{1}+Q_{2}))+c.c\biggl] \,,
\label{eq:dressingweakcoupling}
\ee
where c.c. denotes complex conjugate, and 
we used that $\psi(-\frac{i}{2}(\tilde{p}-iQ))+c.c=
\psi(1-\frac{i}{2}(\tilde{p}+iQ))+c.c$
for integer $Q$.

\subsection{Performing the integrals by residues}

We demonstrate here how we performed the integrals by 
evaluating $A_{\Sigma}$ (\ref{defASigma}). In view of the result (\ref{eq:dressingweakcoupling}), we
start by evaluating the term with 
$\psi(1-\frac{i}{2}(\tilde{p}_{1}+iQ_{1})) + c.c.$.
Its contribution factorizes for the indices $1,2$ into a product of 
two factors. The more complicated factor is 
\[
\sum_{Q_{1}}Q_{1}^{2}\int_{-\infty}^{\infty}\frac{d\tilde{p}_{1}}{2\pi}\frac{1}{(
\tilde{p}_{1}^{2}+Q_{1}^{2})^{L}}\left[\psi(1-\frac{i}{2}(\tilde{p}_{1}+iQ_{1}))+
\psi(1+\frac{i}{2}(\tilde{p}_{1}-iQ_{1}))\right]\,.
\]
Let us analyze the pole structure of the integrand. Additionally to the
two ``kinematical'' poles at $\tilde{p}=\pm iQ$, the polygamma function
has poles for $\psi(-n)$ if $n\geq0$. These poles are located on
the lower half plane (LHP) for the first and on the upper half plane
(UHP) for second polygamma function: 
\[
\frac{1}{2}(Q_{1}+2\mp 
i\tilde{p}_{1})=-n\quad\longrightarrow\quad\tilde{p}_{1}=\mp i(2(n+1)+Q_{1})\,.
\]
We now use the trick in \cite{Bajnok:2010ud} of
exploiting the reality of the integrand to rewrite the integral as
\[
2\, \Re e\sum_{Q_{1}}Q_{1}^{2}\int_{-\infty}^{\infty}\frac{d\tilde{p}_{1}}{2\pi}
\frac{1}{(\tilde{p}_{1}^{2}+Q_{1}^{2})^{L}}\left[\psi(1-\frac{i}{2}
(\tilde{p_{1}}+iQ_{1}))\right] \,,
\]
and close the contour on the UHP. In so doing, we have to pick up
the residue at $\tilde p_{1}=iQ_{1}$ only: 
\begin{eqnarray}
\lefteqn{2i\sum_{Q_{1}}Q_{1}^{2}\frac{\partial_{\tilde{p}_{1}}^{L-1}}{(L-1)!}\frac{\psi(1-
\frac{i}{2}(\tilde{p}_{1}+iQ_{1}))}{(\tilde{p}_{1}+iQ_{1})^{L}}\vert_{\tilde{p}_{1}=iQ_{1}}}\nonumber\\
&& = -\sum_{Q_{1}}\sum_{j=0}^{L-1}\left(\begin{array}{c}
L+j-1\\
j\end{array}\right)\frac{2^{-2L+2}}{(L-1-j)!}(-Q_{1})^{2-L-j}\psi^{(L-j-1)}(Q_{1}+1) \,.
\end{eqnarray}
We now note that the $Q_{2}$-dependent terms give
\[
\sum_{Q_{2}=1}^{\infty}Q_{2}^{2}\int_{-\infty}^{\infty}\frac{d\tilde{p}_{2}}{2\pi}
\frac{1}{(\tilde{p}_{2}^{2}+Q_{2}^{2})^{L}}=\sum_{Q_{2}=1}^{\infty}\left(\begin{array}{c}
2L-2\\
L-1\end{array}\right)Q_{2}^{3-2L}\, 2^{1-2L}=2^{1-2L}\left(\begin{array}{c}
2L-2\\
L-1\end{array}\right)\zeta_{2L-3}\,.
\]
Hence, the factorizing contribution to $A_{\Sigma}$, which we denote
by $A_{\Sigma}^{(1)}$, is given by
\[
A_{\Sigma}^{(1)}=-2^{1-2L}\left(\begin{array}{c}
2L-2\\
L-1\end{array}\right)\zeta_{2L-3}\sum_{Q_{1}}\sum_{j=0}^{L-1}\left(\begin{array}{c}
L+j-1\\
j\end{array}\right)\frac{2^{-2L+2}}{(L-1-j)!}(-Q_{1})^{2-L-j}\psi^{(L-j-1)}(Q_{1}+1) \,.
\label{eq:sl2part1}
\]
 
Let us concentrate now on the nonfactorizing contributions, which we 
denote by $A_{\Sigma}^{(2)}$. Using again the reality trick, we can write 
\[
A_{\Sigma}^{(2)}=-2\sum_{Q_{1},Q_{2}}Q_{1}^{2}Q_{2}^{2}\int\frac{d
\tilde{p}_{1}}{2\pi}\int\frac{d\tilde{p}_{2}}{2\pi}\frac{1}{(
\tilde{p}_{2}^{2}+Q_{2}^{2})^{L}(\tilde{p}_{1}^{2}+Q_{1}^{2})^{L}}
\psi\left(1+\frac{1}{2}(Q_{1}+Q_{2}-i(\tilde{p}_{2}-\tilde{p}_{1}))\right)\,,
\]
and close the $\tilde{p}_{2}$ integration contour on the UHP. By
picking up the only residue at $\tilde{p}_{2}=iQ_{2}$, the result
is 
\begin{eqnarray}
\lefteqn{-2i\sum_{Q_{2}}Q_{2}^{2}\frac{\partial_{
\tilde{p}_{2}}^{L-1}}{(L-1)!}\frac{\psi(1+\frac{1}{2}(Q_{1}+Q_{2}-i(
\tilde{p}_{2}-\tilde{p}_{1})))}{(\tilde{p}_{2}+iQ_{2})^{L}}\vert_{
\tilde{p}_{2}=iQ_{2}}} \nonumber \\
&& = \sum_{Q_{2}}\sum_{j=0}^{L-1}\left(\begin{array}{c}
L+j-1\\
j\end{array}\right)\frac{2^{-2L+2}}{(L-1-j)!}(-Q_{2})^{2-L-j}
\psi^{(L-j-1)}(Q_{2}+1+\frac{1}{2}(Q_{1}+i\tilde{p}_{1})) \,.
\end{eqnarray}
The next integral we close on the lower half plane and pick up the
residue at $-iQ_{1}$: 
\begin{eqnarray}
A_{\Sigma}^{(2)} & = & -\sum_{Q_{1},Q_{2}}\sum_{j_{1},j_{2}=0}^{L-1}\left(\begin{array}{c}
L+j_{1}-1\\
j_{1}\end{array}\right)\frac{2^{-2L+2}}{(L-1-j_{1})!}(-Q_{1})^{2-L-j_{1}} \nonumber \\
 &  & \times \left(\begin{array}{c}
L+j_{2}-1\\
j_{2}\end{array}\right)\frac{2^{-2L+1}}{(L-1-j_{2})!}(-Q_{2})^{2-L-j_{2}}
\psi^{(2L-j_{1}-j_{2}-2)}(Q_{1}+Q_{2}+1) \,.
\label{eq:sl2part2}
\end{eqnarray}
Adding the two terms $A_{\Sigma}=A_{\Sigma}^{(1)}+A_{\Sigma}^{(2)}$
gives the result we presented in (\ref{Asl2}).

\end{appendix}


\begin{thebibliography}{10}

\bibitem{Tseytlin:2009zz}
C.~Kristjansen, (ed.~), M.~Staudacher, (ed.~), and A.~Tseytlin, (ed.~),
  ``{Gauge-string duality and integrability: Progress and outlook},''
\href{http://dx.doi.org/10.1088/1751-8121/42/25/250301}{{\em J. Phys.}
  {\bfseries A42} (2009) 250301}.

\bibitem{Beisert:2010jr}
N.~Beisert, C.~Ahn, L.~F. Alday, Z.~Bajnok, J.~M. Drummond, {\em et al.},
  ``{Review of AdS/CFT Integrability: An Overview},''
  \href{http://arxiv.org/abs/1012.3982}{{\ttfamily arXiv:1012.3982 [hep-th]}}.
  * Temporary entry *.

\bibitem{Staudacher:2010jz}
M.~Staudacher, ``{Review of AdS/CFT Integrability, Chapter III.1: Bethe
  Ans\"atze and the R-Matrix Formalism},''
  \href{http://arxiv.org/abs/1012.3990}{{\ttfamily arXiv:1012.3990 [hep-th]}}.
  * Temporary entry *.

\bibitem{Ahn:2010ka}
C.~Ahn and R.~I. Nepomechie, ``{Review of AdS/CFT Integrability, Chapter III.2:
  Exact world-sheet S-matrix},''
  \href{http://arxiv.org/abs/1012.3991}{{\ttfamily arXiv:1012.3991 [hep-th]}}.
  * Temporary entry *.

\bibitem{Sieg:2010jt}
C.~Sieg, ``{Review of AdS/CFT Integrability, Chapter I.2: The spectrum from
  perturbative gauge theory},''
  \href{http://arxiv.org/abs/1012.3984}{{\ttfamily arXiv:1012.3984 [hep-th]}}.
  * Temporary entry *.

\bibitem{Janik:2010kd}
R.~A. Janik, ``{Review of AdS/CFT Integrability, Chapter III.5: L\'uscher
  corrections},'' \href{http://arxiv.org/abs/1012.3994}{{\ttfamily
  arXiv:1012.3994 [hep-th]}}. * Temporary entry *.

\bibitem{Ambjorn:2005wa}
J.~Ambjorn, R.~A. Janik, and C.~Kristjansen, ``{Wrapping interactions and a new
  source of corrections to the spin-chain / string duality},''
  \href{http://dx.doi.org/10.1016/j.nuclphysb.2005.12.007}{{\em Nucl. Phys.}
  {\bfseries B736} (2006) 288--301},
\href{http://arxiv.org/abs/hep-th/0510171}{{\ttfamily arXiv:hep-th/0510171}}.

\bibitem{Arutyunov:2007tc}
G.~Arutyunov and S.~Frolov, ``{On String S-matrix, Bound States and TBA},''
  \href{http://dx.doi.org/10.1088/1126-6708/2007/12/024}{{\em JHEP} {\bfseries
  12} (2007) 024},
\href{http://arxiv.org/abs/0710.1568}{{\ttfamily arXiv:0710.1568 [hep-th]}}.

\bibitem{Bajnok:2010ke}
Z.~Bajnok, ``{Review of AdS/CFT Integrability, Chapter III.6: Thermodynamic
  Bethe Ansatz},'' \href{http://arxiv.org/abs/1012.3995}{{\ttfamily
  arXiv:1012.3995 [hep-th]}}. * Temporary entry *.

\bibitem{Gromov:2009tv}
N.~Gromov, V.~Kazakov, and P.~Vieira, ``{Exact Spectrum of Anomalous Dimensions
  of Planar N=4 Supersymmetric Yang-Mills Theory},''
  \href{http://dx.doi.org/10.1103/PhysRevLett.103.131601}{{\em Phys.Rev.Lett.}
  {\bfseries 103} (2009) 131601},
  \href{http://arxiv.org/abs/0901.3753}{{\ttfamily arXiv:0901.3753 [hep-th]}}.

\bibitem{Bombardelli:2009ns}
D.~Bombardelli, D.~Fioravanti, and R.~Tateo, ``{Thermodynamic Bethe Ansatz for
  planar AdS/CFT: A Proposal},''
  \href{http://dx.doi.org/10.1088/1751-8113/42/37/375401,
  10.1088/1751-8113/42/37/375401}{{\em J.Phys.A} {\bfseries A42} (2009)
  375401}, \href{http://arxiv.org/abs/0902.3930}{{\ttfamily arXiv:0902.3930
  [hep-th]}}.

\bibitem{Gromov:2009bc}
N.~Gromov, V.~Kazakov, A.~Kozak, and P.~Vieira, ``{Exact Spectrum of Anomalous
  Dimensions of Planar N = 4 Supersymmetric Yang-Mills Theory: TBA and excited
  states},'' \href{http://dx.doi.org/10.1007/s11005-010-0374-8}{{\em
  Lett.Math.Phys.} {\bfseries 91} (2010) 265--287},
  \href{http://arxiv.org/abs/0902.4458}{{\ttfamily arXiv:0902.4458 [hep-th]}}.

\bibitem{Arutyunov:2009ur}
G.~Arutyunov and S.~Frolov, ``{Thermodynamic Bethe Ansatz for the AdS$_5$
  $\times$ S$^5$ Mirror Model},''
  \href{http://dx.doi.org/10.1088/1126-6708/2009/05/068}{{\em JHEP} {\bfseries
  05} (2009) 068},
\href{http://arxiv.org/abs/0903.0141}{{\ttfamily arXiv:0903.0141 [hep-th]}}.

\bibitem{Arutyunov:2009ux}
G.~Arutyunov and S.~Frolov, ``{Simplified TBA equations of the AdS(5) x S**5
  mirror model},'' \href{http://dx.doi.org/10.1088/1126-6708/2009/11/019}{{\em
  JHEP} {\bfseries 0911} (2009) 019},
  \href{http://arxiv.org/abs/0907.2647}{{\ttfamily arXiv:0907.2647 [hep-th]}}.

\bibitem{Gromov:2010kf}
N.~Gromov and V.~Kazakov, ``{Review of AdS/CFT Integrability, Chapter III.7:
  Hirota Dynamics for Quantum Integrability},''
  \href{http://arxiv.org/abs/1012.3996}{{\ttfamily arXiv:1012.3996 [hep-th]}}.
  * Temporary entry *.

\bibitem{Arutyunov:2009ax}
G.~Arutyunov, S.~Frolov, and R.~Suzuki, ``{Exploring the mirror TBA},''
  \href{http://dx.doi.org/10.1007/JHEP05(2010)031}{{\em JHEP} {\bfseries 1005}
  (2010) 031}, \href{http://arxiv.org/abs/0911.2224}{{\ttfamily arXiv:0911.2224
  [hep-th]}}.

\bibitem{Arutyunov:2010gb}
G.~Arutyunov, S.~Frolov, and R.~Suzuki, ``{Five-loop Konishi from the Mirror
  TBA},'' \href{http://dx.doi.org/10.1007/JHEP04(2010)069}{{\em JHEP}
  {\bfseries 1004} (2010) 069},
  \href{http://arxiv.org/abs/1002.1711}{{\ttfamily arXiv:1002.1711 [hep-th]}}.

\bibitem{Arutyunov:2011uz}
G.~Arutyunov and S.~Frolov, ``{Comments on the Mirror TBA},''
  \href{http://dx.doi.org/10.1007/JHEP05(2011)082}{{\em JHEP} {\bfseries 1105}
  (2011) 082}, \href{http://arxiv.org/abs/1103.2708}{{\ttfamily arXiv:1103.2708
  [hep-th]}}. * Temporary entry *.

\bibitem{Cavaglia:2010nm}
A.~Cavaglia, D.~Fioravanti, and R.~Tateo, ``{Extended Y-system for the
  $AdS_5/CFT_4$ correspondence},''
  \href{http://dx.doi.org/10.1016/j.nuclphysb.2010.09.015}{{\em Nucl.Phys.}
  {\bfseries B843} (2011) 302--343},
  \href{http://arxiv.org/abs/1005.3016}{{\ttfamily arXiv:1005.3016 [hep-th]}}.

\bibitem{Cavaglia:2011kd}
A.~Cavaglia, D.~Fioravanti, M.~Mattelliano, and R.~Tateo, ``{On the
  $AdS_5/CFT_4$ TBA and its analytic properties},''
  \href{http://arxiv.org/abs/1103.0499}{{\ttfamily arXiv:1103.0499 [hep-th]}}.
  * Temporary entry *.

\bibitem{Balog:2011nm}
J.~Balog and A.~Hegedus, ``{$AdS_5\times S^5$ mirror TBA equations from
  Y-system and discontinuity relations},''
  \href{http://arxiv.org/abs/1104.4054}{{\ttfamily arXiv:1104.4054 [hep-th]}}.

\bibitem{Balog:2011cx}
J.~Balog and A.~Hegedus, ``{Quasi-local formulation of the mirror TBA},''
  \href{http://arxiv.org/abs/1106.2100}{{\ttfamily arXiv:1106.2100 [hep-th]}}.
  * Temporary entry *.

\bibitem{Hegedus:2009ky}
A.~Hegedus, ``{Discrete Hirota dynamics for AdS/CFT},''
  \href{http://dx.doi.org/10.1016/j.nuclphysb.2009.09.012}{{\em Nucl.Phys.}
  {\bfseries B825} (2010) 341--365},
  \href{http://arxiv.org/abs/0906.2546}{{\ttfamily arXiv:0906.2546 [hep-th]}}.

\bibitem{Gromov:2010km}
N.~Gromov, V.~Kazakov, S.~Leurent, and Z.~Tsuboi, ``{Wronskian Solution for
  AdS/CFT Y-system},'' \href{http://dx.doi.org/10.1007/JHEP01(2011)155}{{\em
  JHEP} {\bfseries 1101} (2011) 155},
  \href{http://arxiv.org/abs/1010.2720}{{\ttfamily arXiv:1010.2720 [hep-th]}}.

\bibitem{Gromov:2010vb}
N.~Gromov, V.~Kazakov, and Z.~Tsuboi, ``{PSU(2,2|4) Character of Quasiclassical
  AdS/CFT},'' \href{http://dx.doi.org/10.1007/JHEP07(2010)097}{{\em JHEP}
  {\bfseries 1007} (2010) 097},
  \href{http://arxiv.org/abs/1002.3981}{{\ttfamily arXiv:1002.3981 [hep-th]}}.

\bibitem{Suzuki:2011dj}
R.~Suzuki, ``{Hybrid NLIE for the Mirror $AdS_5 x S^5$},''
  \href{http://dx.doi.org/10.1088/1751-8113/44/23/235401}{{\em J. Phys.}
  {\bfseries A44} (2011) 235401},
\href{http://arxiv.org/abs/1101.5165}{{\ttfamily arXiv:1101.5165 [hep-th]}}.

\bibitem{Lunin:2005jy}
O.~Lunin and J.~M. Maldacena, ``{Deforming field theories with U(1) $\times$
  U(1) global symmetry and their gravity duals},'' {\em JHEP} {\bfseries 05}
  (2005) 033,
\href{http://arxiv.org/abs/hep-th/0502086}{{\ttfamily arXiv:hep-th/0502086}}.

\bibitem{Frolov:2005dj}
S.~Frolov, ``{Lax pair for strings in Lunin-Maldacena background},'' {\em JHEP}
  {\bfseries 05} (2005) 069,
\href{http://arxiv.org/abs/hep-th/0503201}{{\ttfamily arXiv:hep-th/0503201}}.

\bibitem{Beisert:2005if}
N.~Beisert and R.~Roiban, ``{Beauty and the twist: The Bethe ansatz for twisted
  $\mathcal{N}$ = 4 SYM},'' {\em JHEP} {\bfseries 08} (2005) 039,
\href{http://arxiv.org/abs/hep-th/0505187}{{\ttfamily arXiv:hep-th/0505187}}.

\bibitem{Frolov:2005iq}
S.~A. Frolov, R.~Roiban, and A.~A. Tseytlin, ``{Gauge-string duality for
  (non)supersymmetric deformations of $\mathcal{N}$ = 4 super Yang-Mills
  theory},'' \href{http://dx.doi.org/10.1016/j.nuclphysb.2005.10.004}{{\em
  Nucl. Phys.} {\bfseries B731} (2005) 1--44},
\href{http://arxiv.org/abs/hep-th/0507021}{{\ttfamily arXiv:hep-th/0507021}}.

\bibitem{Zoubos:2010kh}
K.~Zoubos, ``{Review of AdS/CFT Integrability, Chapter IV.2: Deformations,
  Orbifolds and Open Boundaries},''
  \href{http://arxiv.org/abs/1012.3998}{{\ttfamily arXiv:1012.3998 [hep-th]}}.
  * Temporary entry *.

\bibitem{Ananth:2007px}
S.~Ananth, S.~Kovacs, and H.~Shimada, ``{Proof of ultra-violet finiteness for a
  planar non-supersymmetric Yang-Mills theory},''
  \href{http://dx.doi.org/10.1016/j.nuclphysb.2007.04.005}{{\em Nucl. Phys.}
  {\bfseries B783} (2007) 227--237},
\href{http://arxiv.org/abs/hep-th/0702020}{{\ttfamily arXiv:hep-th/0702020}}.

\bibitem{Arutyunov:2010gu}
G.~Arutyunov, M.~de~Leeuw, and S.~J. van Tongeren, ``{Twisting the Mirror
  TBA},'' \href{http://dx.doi.org/10.1007/JHEP02(2011)025}{{\em JHEP}
  {\bfseries 1102} (2011) 025},
  \href{http://arxiv.org/abs/1009.4118}{{\ttfamily arXiv:1009.4118 [hep-th]}}.

\bibitem{Ahn:2010yv}
C.~Ahn, Z.~Bajnok, D.~Bombardelli, and R.~I. Nepomechie, ``{Finite-size effect
  for four-loop Konishi of the $\beta$-deformed N=4 SYM},''
  \href{http://dx.doi.org/10.1016/j.physletb.2010.08.056}{{\em Phys.Lett.}
  {\bfseries B693} (2010) 380--385},
  \href{http://arxiv.org/abs/1006.2209}{{\ttfamily arXiv:1006.2209 [hep-th]}}.

\bibitem{Ahn:2010ws}
C.~Ahn, Z.~Bajnok, D.~Bombardelli, and R.~I. Nepomechie, ``{Twisted Bethe
  equations from a twisted S-matrix},''
  \href{http://dx.doi.org/10.1007/JHEP02(2011)027}{{\em JHEP} {\bfseries 1102}
  (2011) 027}, \href{http://arxiv.org/abs/1010.3229}{{\ttfamily arXiv:1010.3229
  [hep-th]}}.

\bibitem{Gromov:2010dy}
N.~Gromov and F.~Levkovich-Maslyuk, ``{Y-system and $\beta$-deformed N=4
  Super-Yang-Mills},''
  \href{http://dx.doi.org/10.1088/1751-8113/44/1/015402}{{\em J.Phys.A}
  {\bfseries A44} (2011) 015402},
  \href{http://arxiv.org/abs/1006.5438}{{\ttfamily arXiv:1006.5438 [hep-th]}}.

\bibitem{Reshetikhin:1990ep}
N.~Reshetikhin, ``{Multiparameter quantum groups and twisted quasitriangular
  Hopf algebras},'' \href{http://dx.doi.org/10.1007/BF00626530}{{\em
  Lett.Math.Phys.} {\bfseries 20} (1990) 331--335}.

\bibitem{deLeeuw:2011rw}
M.~de~Leeuw and S.~J. van Tongeren, ``{Orbifolded Konishi from the Mirror
  TBA},'' \href{http://dx.doi.org/10.1088/1751-8113/44/32/325404}{{\em
  J.Phys.A} {\bfseries A44} (2011) 325404},
  \href{http://arxiv.org/abs/1103.5853}{{\ttfamily arXiv:1103.5853 [hep-th]}}.

\bibitem{Beccaria:2011qd}
M.~Beccaria and G.~Macorini, ``{Y-system for $Z_S$ Orbifolds of N=4 SYM},''
  \href{http://dx.doi.org/10.1007/JHEP06(2011)004}{{\em JHEP} {\bfseries 1106}
  (2011) 004}, \href{http://arxiv.org/abs/1104.0883}{{\ttfamily arXiv:1104.0883
  [hep-th]}}.

\bibitem{Bajnok:2004jd}
Z.~Bajnok and A.~George, ``{From defects to boundaries},''
  \href{http://dx.doi.org/10.1142/S0217751X06025262}{{\em Int.J.Mod.Phys.}
  {\bfseries A21} (2006) 1063--1078},
  \href{http://arxiv.org/abs/hep-th/0404199}{{\ttfamily arXiv:hep-th/0404199
  [hep-th]}}.

\bibitem{Dorey:2004xk}
P.~Dorey, D.~Fioravanti, C.~Rim, and R.~Tateo, ``{Integrable quantum field
  theory with boundaries: The Exact g function},''
  \href{http://dx.doi.org/10.1016/j.nuclphysb.2004.06.045}{{\em Nucl.Phys.}
  {\bfseries B696} (2004) 445--467},
  \href{http://arxiv.org/abs/hep-th/0404014}{{\ttfamily arXiv:hep-th/0404014
  [hep-th]}}.

\bibitem{Bajnok:2004tq}
Z.~Bajnok, L.~Palla, and G.~Takacs, ``{Finite size effects in quantum field
  theories with boundary from scattering data},''
  \href{http://dx.doi.org/10.1016/j.nuclphysb.2005.03.021}{{\em Nucl.Phys.}
  {\bfseries B716} (2005) 519--542},
  \href{http://arxiv.org/abs/hep-th/0412192}{{\ttfamily arXiv:hep-th/0412192
  [hep-th]}}.

\bibitem{Balog:2001sr}
J.~Balog and A.~Hegedus, ``{Virial expansion and TBA in O(N) sigma models},''
  \href{http://dx.doi.org/10.1016/S0370-2693(01)01307-7}{{\em Phys. Lett.}
  {\bfseries B523} (2001) 211--220},
\href{http://arxiv.org/abs/hep-th/0108071}{{\ttfamily arXiv:hep-th/0108071}}.

\bibitem{Klassen:1990dx}
T.~R. Klassen and E.~Melzer, ``{The Thermodynamics of purely elastic scattering
  theories and conformal perturbation theory},''
  \href{http://dx.doi.org/10.1016/0550-3213(91)90159-U}{{\em Nucl.Phys.}
  {\bfseries B350} (1991) 635--689}.

\bibitem{Zamolodchikov:1977nu}
A.~B. Zamolodchikov and A.~B. Zamolodchikov, ``{Relativistic Factorized S
  Matrix in Two-Dimensions Having O(N) Isotopic Symmetry},''
  \href{http://dx.doi.org/10.1016/0550-3213(78)90239-0,
  10.1016/0550-3213(78)90239-0}{{\em Nucl.Phys.} {\bfseries B133} (1978) 525}.

\bibitem{Zamolodchikov:1992zr}
A.~B. Zamolodchikov and A.~B. Zamolodchikov, ``{Massless factorized scattering
  and sigma models with topological terms},''
  \href{http://dx.doi.org/10.1016/0550-3213(92)90136-Y}{{\em Nucl.Phys.}
  {\bfseries B379} (1992) 602--623}.

\bibitem{Gromov:2008gj}
N.~Gromov, V.~Kazakov, and P.~Vieira, ``{Finite Volume Spectrum of 2D Field
  Theories from Hirota Dynamics},''
  \href{http://dx.doi.org/10.1088/1126-6708/2009/12/060}{{\em JHEP} {\bfseries
  0912} (2009) 060}, \href{http://arxiv.org/abs/0812.5091}{{\ttfamily
  arXiv:0812.5091 [hep-th]}}.

\bibitem{Arutyunov:2006yd}
G.~Arutyunov, S.~Frolov, and M.~Zamaklar, ``{The Zamolodchikov-Faddeev algebra
  for AdS$_5$ $\times$ S$^5$ superstring},'' {\em JHEP} {\bfseries 04} (2007)
  002,
\href{http://arxiv.org/abs/hep-th/0612229}{{\ttfamily arXiv:hep-th/0612229}}.

\bibitem{Arutyunov:2009mi}
G.~Arutyunov, M.~de~Leeuw, and A.~Torrielli, ``{The Bound State S-Matrix for
  AdS$_5$ $\times$ S$^5$ Superstring},''
  \href{http://dx.doi.org/10.1016/j.nuclphysb.2009.03.024}{{\em Nucl. Phys.}
  {\bfseries B819} (2009) 319--350},
\href{http://arxiv.org/abs/0902.0183}{{\ttfamily arXiv:0902.0183 [hep-th]}}.

\bibitem{Beisert:2007ds}
N.~Beisert, ``{The S-Matrix of AdS/CFT and Yangian Symmetry},'' {\em PoS}
  {\bfseries SOLVAY} (2006) 002,
\href{http://arxiv.org/abs/0704.0400}{{\ttfamily arXiv:0704.0400 [nlin.SI]}}.

\bibitem{Arutyunov:2009kf}
G.~Arutyunov and S.~Frolov, ``{The Dressing Factor and Crossing Equations},''
  \href{http://dx.doi.org/10.1088/1751-8113/42/42/425401}{{\em J.Phys.A}
  {\bfseries A42} (2009) 425401},
  \href{http://arxiv.org/abs/0904.4575}{{\ttfamily arXiv:0904.4575 [hep-th]}}.

\bibitem{Arutyunov:2008zt}
G.~Arutyunov and S.~Frolov, ``{The S-matrix of String Bound States},''
  \href{http://dx.doi.org/10.1016/j.nuclphysb.2008.06.005}{{\em Nucl. Phys.}
  {\bfseries B804} (2008) 90--143},
\href{http://arxiv.org/abs/0803.4323}{{\ttfamily arXiv:0803.4323 [hep-th]}}.

\bibitem{Arutyunov:2009zu}
G.~Arutyunov and S.~Frolov, ``{String hypothesis for the AdS$_5$ $\times$ S$^5$
  mirror},'' \href{http://dx.doi.org/10.1088/1126-6708/2009/03/152}{{\em JHEP}
  {\bfseries 03} (2009) 152},
\href{http://arxiv.org/abs/0901.1417}{{\ttfamily arXiv:0901.1417 [hep-th]}}.

\bibitem{deLeeuw:2010ed}
M.~de~Leeuw and T.~Lukowski, ``{Twist operators in N=4 beta-deformed theory},''
  \href{http://dx.doi.org/10.1007/JHEP04(2011)084}{{\em JHEP} {\bfseries 04}
  (2011) 084},
\href{http://arxiv.org/abs/1012.3725}{{\ttfamily arXiv:1012.3725 [hep-th]}}.

\bibitem{Frolov:2009in}
S.~Frolov and R.~Suzuki, ``{Temperature quantization from the TBA equations},''
  \href{http://dx.doi.org/10.1016/j.physletb.2009.06.069}{{\em Phys. Lett.}
  {\bfseries B679} (2009) 60--64},
\href{http://arxiv.org/abs/0906.0499}{{\ttfamily arXiv:0906.0499 [hep-th]}}.

\bibitem{Borwein:1999}
J.~Borwein, D.~Bradley, D.~Broadhurst, and P.~Lisonek, ``{Special values of
  multiple polylogarithms},'' {\em Trans. Amer. Math. Soc.} {\bfseries 353}
  (2001) 907--941, \href{http://arxiv.org/abs/9910045}{{\ttfamily arXiv:9910045
  [math]}}.

\bibitem{Balog:2010xa}
J.~Balog and A.~Hegedus, ``{5-loop Konishi from linearized TBA and the XXX
  magnet},'' \href{http://dx.doi.org/10.1007/JHEP06(2010)080}{{\em JHEP}
  {\bfseries 1006} (2010) 080},
  \href{http://arxiv.org/abs/1002.4142}{{\ttfamily arXiv:1002.4142 [hep-th]}}.

\bibitem{Balog:2010vf}
J.~Balog and A.~Hegedus, ``{The Bajnok-Janik formula and wrapping
  corrections},'' \href{http://dx.doi.org/10.1007/JHEP09(2010)107}{{\em JHEP}
  {\bfseries 1009} (2010) 107},
  \href{http://arxiv.org/abs/1003.4303}{{\ttfamily arXiv:1003.4303 [hep-th]}}.

\bibitem{Bajnok:2008bm}
Z.~Bajnok and R.~A. Janik, ``{Four-loop perturbative Konishi from strings and
  finite size effects for multiparticle states},''
  \href{http://dx.doi.org/10.1016/j.nuclphysb.2008.08.020}{{\em Nucl. Phys.}
  {\bfseries B807} (2009) 625--650},
\href{http://arxiv.org/abs/0807.0399}{{\ttfamily arXiv:0807.0399 [hep-th]}}.

\bibitem{Kulish:1981gi}
P.~Kulish, N.~Reshetikhin, and E.~Sklyanin, ``{Yang-Baxter Equation and
  Representation Theory. 1.},''
  \href{http://dx.doi.org/10.1007/BF02285311}{{\em Lett.Math.Phys.} {\bfseries
  5} (1981) 393--403}.

\bibitem{Alishauskas:1986}
S.~Alishauskas and P.~Kulish, ``{Spectral resolution of su(3)-invariant
  solutions of the Yang-Baxter equation},'' {\em J.Sov.Math.} {\bfseries 35}
  (1986) 2563--2574.

\bibitem{Bajnok:2010ud}
Z.~Bajnok and O.~el~Deeb, ``{6-loop anomalous dimension of a single impurity
  operator from AdS/CFT and multiple zeta values},''
  \href{http://dx.doi.org/10.1007/JHEP01(2011)054}{{\em JHEP} {\bfseries 1101}
  (2011) 054}, \href{http://arxiv.org/abs/1010.5606}{{\ttfamily arXiv:1010.5606
  [hep-th]}}.

\end{thebibliography}

\end{document}